\documentclass{JHEP3}
\usepackage{epsfig}
\newcommand{\be}{\begin{equation}}
\newcommand{\ee}{\end{equation}}
\newcommand{\bea}{\begin{eqnarray}}
\newcommand{\eea}{\end{eqnarray}}
\newcommand{\IR}{\mathbb{R}} 

\def\IZ{\relax\ifmmode\hbox{Z\kern-.4em Z}\else{Z\kern-.4em Z}\fi}
\newcommand{\IS}{{\bf S}} \newcommand{\IT}{{\bf T}}

\newcommand{\non}{\nonumber \\}

\def\half{{1 \over 2}} 

\def\del{{\partial}}

 \def\bh{{\bar h}}

\def\wttriangle{\widetilde{\triangle}}

\def\al{\alpha} 
  \def\eps{\epsilon}
\def\teps{\tilde{\epsilon}}

\def\trho{{\tilde \rho}} \def\hrho{{\hat \rho}}
 \def\sig{\sigma}
\def\hlambda{{\hat \lambda}}

\def\presub{\vspace{.5cm} \noindent}

\def\bi{\begin{itemize}} \def\ei{\end{itemize}}

\def\Schw{Schwarzschild }

\def\({\left(} \def\){\right)}
\def\[{\left[} \def\]{\right]}
\preprint{{\tt hep-th/0411240}}

\title{ \center{The Phase Transition between Caged Black Holes and Black
Strings -- A Review}}

\author{Barak Kol \\
 Racah Institute of Physics\\
 Hebrew University \\
 Jerusalem 91904,
 Israel\\
{\tt barak\_kol@phys.huji.ac.il}}

\abstract{Black hole uniqueness is known to fail in higher
dimensions, and the multiplicity of black hole phases leads to
phase transitions physics in General Relativity. The black-hole
black-string transition is a prime realization of such a system
and its phase diagram has been the subject of considerable study
in the last few years. The most surprising results seem to be the
appearance of critical dimensions where the qualitative behavior
of the system changes, and a novel kind of topology change.
Recently, a full phase diagram was determined numerically,
confirming earlier predictions for a merger of the black-hole and
black string phases and giving very strong evidence that the
end-state of the Gregory-Laflamme instability is a black hole (in
the dimension range $5 \le D \le 13$).
 Here this progress is reviewed, illustrated with figures,
put into a wider context, and the still open questions are
listed.}

\begin{document}

\newpage

\begin{flushright}
To Dorit, Inbal and Neta,  \\
my wife and daughters \hspace*{0.3cm}
\end{flushright}

\section{Introduction}


In this introduction we will present the general background for
studying \emph{General Relativity in higher dimensions} and the
novel field of \emph{phase transitions in General Relativity}. We
will list the systems known to exhibit phase transitions, and take
the opportunity to discuss the rotating black ring before we
proceed to concentrate on our system of choice, \emph{the
black-hole black-string transition}.

\presub {\bf Why study GR in higher dimensions?}
 There are several good reasons to study General Relativity (GR)
in higher dimensions, namely $D>4$, where $D$ is the total
space-time dimension. From a theoretical point of view there is
nothing in GR that restricts us to $D=4$. On the contrary, the
theory is independent of $D$, and {\it $D$ should be considered as
a parameter}. It is common practice in theoretical physics to
explore large regions of parameter space of a theory in order to
enhance its understanding, rather than restrict to the
experimental values and GR should be no exception. For example, in
the study of gauge theories it is standard to consider various
possibilities for the gauge group and matter content which differ
from the standard model.

Additional reasons to study higher dimensional GR include
\emph{string theory} and the phenomenological scenario of
\emph{``large extra dimensions''}, as we proceed to discuss.
String theory has a ``built-in'' preference for higher dimensional
spacetimes with 10 (the ''critical dimension'') or 11 dimensions,
where the extra dimensions must be compactified. This preference
originates in the cancellation of the conformal (quantum) anomaly
in 10d which is necessary for the consistency of weakly coupled
string theories. The ``large extra dimensions'' scenario (which is
presumably string theory inspired)
 stresses the following important realizations: that
to date gravity is measured only down to $1\mu$--1mm range (which
is an ``astronomically'' poor resolution relative to the one we
have for other forces), that it is quite consistent to assume the
existence of a compact dimension(s) smaller than the experimental
bound and that the situation can be rectified only by improving
gravitational and accelerator experiments.


\presub {\bf The novel feature - non-uniqueness of black objects.}
 Often when we generalize a problem to allow for an arbitrary dimension
the qualitative features do not change and thus the generalization
does not produce ``new physics'', even if the quantitative
expressions are different. However, in GR we do find qualitative
changes. If we roughly divide the field of General Relativity into
black holes, gravitational waves and cosmology, we find a
qualitative change in the first of these categories: one of the
basic properties of 4d black holes changes, namely {\it black hole
uniqueness}.\footnote{Other qualitative differences include the
disappearance of stable circular orbits for $D>4$ (in Newtonian
gravity), the absence of propagating gravitational waves in $D<4$,
and the Belinskii-Khalatnikov-Lifshitz (BKL) analysis of the
approach to a space-like singularity, where there is a critical
dimension $D^*_{BKL}=10$, such that for $D>D^*_{BKL}$ the system
becomes non-chaotic (see the review \cite{CosmoBilliard} and
references therein).}

Here we should digress to make the distinction between two closely
related black hole notions: ``no hair'' and ``uniqueness'' (see
for example \cite{Mazur}). \emph{``No hair''} denotes the feature
that the space of black hole solutions has a small dimension
usually parameterized by asympototically measured quantities
(mass, angular momentum and electric charge, for example) much
like macroscopic thermodynamical variables. In this respect a
black hole strongly contrasts with a non-black-hole star which
typically has a much larger number of characteristics such as its
internal matter ingredients each with its own equation of state
and spatial distribution possibly resulting in an unbounded number
of independent multipoles for mass, charge and angular momentum.
Whether this ``no-hair'' property continues to hold for higher
dimensional black holes could depend on the way one chooses to
generalize it. If one generalizes ``no hair'' to mean that the
solutions are determined in term of a small number of (not
necessarily conserved) asymptotic data then it continues to hold
in higher dimensions as far as we know. However, if one would
choose the more restrictive definition which requires conserved
charges then this property fails in higher dimensions as was
demonstrated in the generalized rotating black ring
\cite{Emparan-with-hair,BenaWarnerRing,EEMRdipoleRing}, whose
parameters include some non-conserved dipole charges.

\emph{``Uniqueness''} on the other hand is the more specialized
statement that a choice of all of these asymptotic black hole
parameters selects a unique black hole rather than a discrete set.
In other words, that only a single branch of solutions exists. In
4d uniqueness was proven to hold, namely that given the mass,
charge and angular momentum (satisfying some inequalities to
ensure the existence of a solution with no naked singularities)
there is a unique black hole. However, the proof relies heavily on
properties which are special to 4d: Hawking's proof that the
horizon topology has to be $\IS^2$ and the simplifying gauge
choices of Weyl-Papapetrou and Ernst (see \cite{uniqueness} for
references to original papers and reviews and for a speculative
generalization of uniqueness to higher dimensions). See
\cite{HelfgottOzYanay,GallowaySchoen} for a determination of the
allowed horizon topologies in certain $D>4$.

The breakdown of black hole uniqueness in higher dimensions
implies the coexistence of several phases with the same asymptotic
charges on a non-trivial phase diagram. \emph{Phase transitions}
between the various phases should occur as parameters are changed.
As always one may define the order of the phase transition. It
could be a first order transition in which case it is triggered
non-perturbatively by a competition of entropies between two
phases which are separated by a finite distance in configuration
space, or it could be of second or higher order, in which case it
is triggered by perturbative tachyons and the transition is smooth
(see subsection \ref{transition-order-subsection}).

Such first order transitions would be accompanied by an
exceptional release of energy, sometimes called a {\it
thunderbolt},\footnote{This term was introduced by
\cite{HawkingStewart} for a certain gravitational shock wave in
the presence of a naked singularity and seems appropriate for the
system under study as well.} simply since the total mass of the
final state must be lower than or equal to that of the initial
state and the excess energy must be lost through radiation.
Moreover, exact mass equality is highly unlikely, but rather a
loss of mass is natural as spacetime would undergo violent changes
including sometimes the roll-down of a tachyonic mode.

\presub {\bf The two systems.}
 To date we know of two systems with higher dimensional non-uniqueness
 resulting in non-trivial phase transition physics
 \bi
 \item The rotating ring
 \item The black-hole black-string transition
\ei

The latter was chosen for a thorough study of its phase structure
which is the subject of this review, presumably since it is
somewhat simpler to analyze on account of the smaller number of
metric functions and its higher degree of symmetry. Before
proceeding to analyze it in detail, we discuss the other example,
the black ring.

\begin{figure}[t!]
\centering \noindent
\includegraphics[width=10cm]{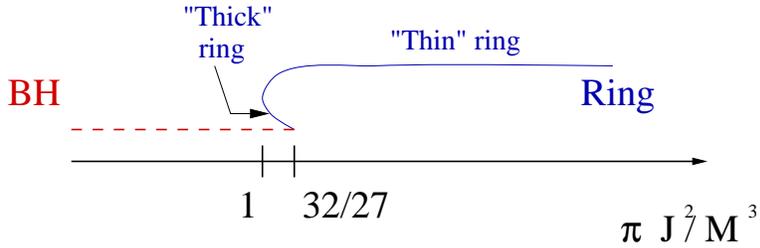}
\caption[]{In asymptotically flat 5d spacetime uniqueness is
violated by the co-existence of the rotating black hole and the
rotating black ring. The figure shows the range of existence of
each phase on the dimensionless angular momentum axis. Note that
for $1 \le \pi\, J^2/M^3 \le 32/27$ three phases co-exist.}
\label{ring-range}
\end{figure}

The rotating black ring lives in the flat (and topologically
trivial) 5d background. Spherical rotating black holes solutions
in higher dimensions, which generalize the 4d Kerr solution were
already found in 1986 by Myers and Perry \cite{MyersPerry}. These
solutions have an $\IS^3$ horizon topology in 5d and display a
maximal angular momentum (at fixed mass). In 2001 Emparan and
Reall \cite{EmparanReall-ring} discovered a beautiful solution,
the ring, with horizon topology $\IS^2 \times \IS^1$, and with an
angular momentum which is bounded from below, but not from above
(at fixed mass). Figure \ref{ring-range} shows the regions on the
angular momentum axis which are occupied by the various phases,
and one notes that there is a middle region where three phases
coexist -- one black hole and two rings. Unlike the black string,
rotating ring solutions are known only in 5d, presumably due to
the special property that in 5d the centrifugal potential and the
(Newtonian) gravitation potential have the same ($1/r^2$)
$r$-dependence.

For some time it was not clear whether the black ring is stable,
and despite some recent findings the issue is not settled yet. In
2004 charged rings were shown to be BPS \cite{EEMRsusyRing} and
hence plausibly ``super-stable'' (that is, non-perturbatively
stable). Although the stability of the original non-BPS ring of
\cite{EmparanReall-ring} is still undetermined (however, see
\cite{ArcioniLozano} for interesting partial results on
stability), there is comfort in knowing that some of its closest
relatives which share many of their outstanding properties are
plausibly stable. Soon after, several groups made progress in
obtaining larger families of ring solutions, both BPS
\cite{ConcentricRing1,BenaWarnerRing,EEMRdipoleRing,ConcentricRing2}
and non-BPS \cite{ElvangEmparanFigueras}, all of them restricted
to 5d. Moreover, the inclusion of non-conserved dipole moments in
\cite{EEMRdipoleRing} demonstrates the ``no-hair'' principle in
higher dimensions must be generalized at least to allow for
non-conserved quantities.

As the recent discovery of families of black rings demonstrates,
the black ring may hold further surprises. In particular, we do
not know the full parameter space for rings, and we know close to
nothing about the associated phase transitions. Thus rotating
rings constitute a promising and active field of research.

\presub {\bf Outline}. At this point we set aside the topic of
black rings until the discussion section and we turn in the next
section to the other example for non-uniqueness, the black-hole
black-string transition which is the main topic of this review. In
section \ref{set-up} the physical set-up is described and the
questions of interest are formulated. In section \ref{qualitative}
we describe the analytic considerations that culminate in
subsection \ref{predicted-phase-diag} to a certain suggestive
qualitative form of the phase diagram which is compared there with
numerical data. Section \ref{solutions} describes the quantitative
tools that were employed in order to obtain solutions, including
both numerical and analytic methods. Finally, related work is
described in section \ref{related} and we conclude with a summary
of the results and a discussion of open questions in section
\ref{summary}.


\section{Set-up and formulation of questions}
\label{set-up}

\subsection{Background metric and phases}

{\bf Background metric}. We consider a background with extra
compact dimensions. In such a background one expects to find
several phases of black objects depending on the relative size of
the object and the relevant length scales in the compact
dimensions. For simplicity we discuss here pure GR (the only field
is the metric) with no cosmological constant. Thus the backgrounds
considered are of the form $\IR^{d-1,1} \times X^p$ where $X^p$ is
any $p$-dimensional compact Ricci-flat manifold, $d$ is the number
of extended spacetime dimensions, and the total spacetime
dimension is $D=d+p$.

The simplest compactifying manifold is a single compact dimension
$X=\IS^1$, and accordingly that was the $X$ considered in most of
the research so far. $\IS^1$ was chosen not only for its
simplicity but also since while more involved $X$ will have
several phases of black objects, the phase transition physics
between any two specific phases is expected to be essentially
similar (generically) to the $\IS^1$ case. Some research was
devoted to $X=\IT^p$, the $p$-dimensional torus
\cite{LargeD-GL,torus}, and we shall discuss it later. Other
possibilities for $X$ include K3 and Calabi-Yau threefolds, as
well as Ricci-flat spaces $X$ which are not supersymmetric.

Thus we consider a background with a single compact dimension of
size $L$, namely $\IR^{d-1,1} \times \IS^1$, and $D=d+1 \ge 5$
(the lower bound on $D$ is set in order to avoid spacetimes with 2
or less extended spatial dimensions where the presence of a
massive source is inconsistent with asymptotic flatness, see
\cite{Myers-4dcaged,Korotkin_Nicolai,FrolovFrolov} for a limited
analogue in $D=4$).

\begin{figure}[t!]
\centering \noindent
\includegraphics[width=7cm]{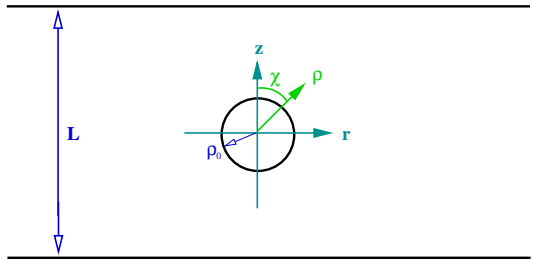}
\caption[]{Definition of coordinates. For backgrounds with a
single compact dimension the essential geometry is 2d after
suppressing time and angular coordinates in the extended
dimensions.  The cylindrical coordinates $(r,z)$ are defined such
that $z \sim z+L$ is the coordinate along the compact dimension
and $r$ is the radial coordinate in the extended spatial
directions. For black holes we define another set of local
coordinates $(\rho,\chi)$, defined only for $\rho \le L/2$, which
are radial coordinates in the 2d plane with origin at the center
of the BH.} \label{coordinates}
\end{figure}

\presub {\bf Black objects}. The non-rotating black objects in
which we are interested are static and spherically symmetric.
Thus, the essential geometry is 2d after suppressing time and the
angular coordinates in the extended dimensions. Our coordinates
are defined in figure \ref{coordinates}.

Such solutions are characterized by 3 dimensionful parameters
$M,\, L$ and $G_N$ where $M$ is the mass of the black object
(measured in the asymptotic $d$ dimensional spacetime, and the
detailed expression is given in \ref{asymp_to_charges}), and $G_N$
is Newton's constant. These define a single dimensionless
parameter \footnote{We are dealing with classical GR, and thus we
{\it do not} set $\hbar=1$.} \be
 \mu := {G_N\, M \over L^{D-3}} \label{def-mu}~.\ee
Alternatively one may use a different parameterization of the
solutions such as replacing $M$ by $\beta$, the inverse
temperature.\footnote{More precisely in order to avoid using
$\hbar$ we define here $\beta=2 \pi/\kappa$, namely the period of
the Euclidean time direction, and $\kappa$ is the surface
gravity.} Correspondingly one may define another dimensionless
parameter \be
 \mu_\beta \propto {\beta \over L} ~, \label{def-mu-beta} \ee
 where the proportionality factor may be chosen later by
 convenience. In thermodynamic terms, the parameter (\ref{def-mu}) or
(\ref{def-mu-beta}) is the \emph{``control parameter''} of the
system, and the choice between these two depends on whether one
prefers the micro-canonical or canonical ensembles, respectively.

\begin{figure}[t!]
\centering \noindent
\includegraphics[width=8cm]{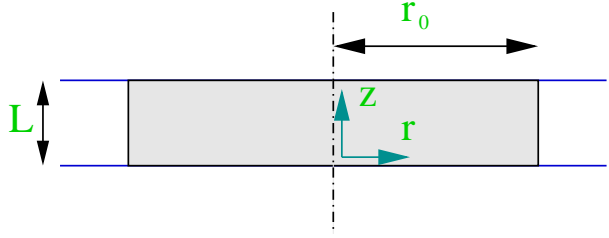}
\caption[]{The uniform black string. $r_0$ is its \Schw radius.}
\label{uniform-string-figure}
\end{figure}

\begin{figure}[t!]
\centering \noindent
\includegraphics[width=6cm]{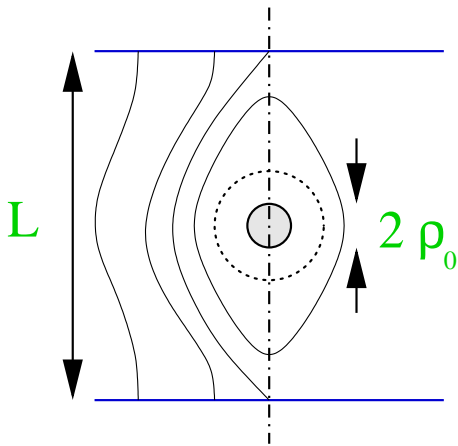}
\caption[]{A caged black hole (BH). Newtonian equipotential lines
are shown.} \label{BH-figure}
\end{figure}

In this background one expects at least two phases of black object
solutions: when $\mu \ll 1$, namely the size of the black object
is small (compared to $L$ the size of the extra dimension) one
expects the region near the object to closely resemble a
$D$-dimensional black hole, while as one increases the mass one
expects that at some point the black hole will no longer fit in
the compact dimension and a black string, whose horizon winds
around the compact dimension will be formed. The precise distinction between these two phases
is give by\\
\underline{Definition}:
 We distinguish between the \emph{black hole} (BH) and the \emph{black
 string} according to their \emph{horizon topology} which is either spherical ---
$\IS^{D-2}$ or cylindrical
--- $\IS^{d-2} \times X$, respectively. \\
 These phases are illustrated in figures
\ref{uniform-string-figure},\ref{BH-figure},\ref{GL-setup-figure}(b).
 We shall sometimes refer to such a black hole localized in a compact dimension as a
\emph{``caged black hole''}.

\presub {\bf Applications.} Before proceeding to discuss the
phases in more detail, let us mention some applications that
contribute to its importance, beyond its considerable intrinsic
value. In String Theory it has attracted continued interest,
particularly regarding the thermodynamic phase diagram for various
gravity theories and/or field theories \cite{IMSY,ABKR} which are
related by dualities to the higher dimensional origin of brane
solitons (such as the M-theory origin of string branes), where the
physics is significantly affected by the question whether they are
localized in the compact dimensions or wrap them. Another field of
application is black holes on brane-worlds \cite{CHR,EHM}, a
problem closely related to the one discussed here, only the
background in which the black objects live includes not only an
extra dimension but also a ``phenomenological'' brane localized in
that dimension and carrying the fields of the standard model.

\vspace{0.5cm}

We now proceed to discuss the two phases with more detail.\\
\noindent {\bf The black string.} We can readily write down
solutions which describe {\it uniform black strings} (see figure
\ref{uniform-string-figure}) \be
 ds^2 = ds^2_{\mbox{Schw}} + ds^2_X \label{string-metric} \ee
 where $ds^2_X$ is the metric on $X$, which for our central example, an $\IS^1$
 parameterized by the coordinate $z$, is just \be
 ds^2_X=dz^2 ~,\label{string-metric2} \ee
 and $ds^2_{\mbox{Schw}}$ is the $d$-dimensional \Schw black hole
 (also known as Schwarzschild-Tangherlini \cite{Tangherlini}), which is given
 by \be
 ds^2_{\mbox{Schw}} =
 -f(\rho)\,dt^{2}+\frac{1}{f(\rho)}\,d\rho^{2}+\rho^{2}\,d\Omega_{d-2}^{2}
 \label{uniform-string}
 \ee
 where \be
 f(\rho) = 1-\frac{\rho_{0}^{d-3}}{\rho^{d-3}} ~,\ee
  $d\Omega_{d-2}^{2}$ is the metric on the sphere $S^{d-2}$ \be
 d\Omega_{d-2}^{2} = d\chi^{2}+\sin^{2}\chi \,d \theta_{1}^{2}+...+
 (\sin^{2}\chi\,\sin^{2}\theta_{1}...\sin^{2}\theta_{d-4})\,d\theta_{d-3}^{2}~.\ee
 $\rho_0$ is related to the black hole mass, $M$, via
\cite{MyersPerry} \be
 \rho_{0}^{d-3} = \frac{16\,\pi\, G_{d}\,M}{(d-2)\,\Omega_{d-2}}
 ~, \label{rho-m} \ee
 where $G_d$ is the $d$-dimensional Newton constant,
and $ \Omega_{d-1}=d { \pi^{d/2} \over (d/2)!} =
\frac{2\,\pi^{\frac{d}{2}}}{\Gamma(\frac{d}{2})}$
 is the area of a unit sphere $S^{d-1}$. The relation between $r_0$ and the inverse
 temperature $\beta$ is \be
 \beta = {2\, \pi \over \kappa}={4\, \pi \over f'(r_0)}={4\, \pi\, r_0 \over d-3}~.\ee

These metrics are Ricci flat as a result of being a direct product
of Ricci flat metrics. They are called ``uniform'' for being a
direct product with $X$ (moreover, for $\IS^1$ the full metric is
 $z$-independent). Later we will encounter also
\emph{non-uniform strings} (see figure \ref{GL-setup-figure}(b)).
Note that for general $p>1$ (namely dim$(X)$) these metrics
actually describe $p$-branes rather than strings.

The uniform black string solution is valid for any $r_0$ (and
fixed $X$). However, we shall soon see that for ``thin'' enough
strings, namely small enough $r_0$, an instability develops.

\presub {\bf Caged black holes.}
 One expects localized black hole (BH) solutions to exist (see figure \ref{BH-figure}),
intuitively obtained by constructing a black hole locally without
ever being ``aware'' of the compactness of the some of the
dimensions, at least as long as the black hole is much smaller
than the compact dimensions (and the number of extended spacetime
dimension is $d \ge 4$ to avoid problems with asymptotics) .

As the black hole grows it will start feeling the presence of the
compact dimensions and it will deform accordingly. At some
critical $\mu$ one may expect that the black hole will be too
large to fit into $X$, and so the mass of this phase will be
bounded from above.

Unlike the uniform black string there is no explicit metric that
we can write down. This situation was confronted by two methods:
an analytic perturbative expansion
\cite{Harmark4,dialogue,KSSW1,dialogue2} and numerical analysis
\cite{KPS1,KPS2,KudohWiseman1,KudohWiseman2}. Both techniques will
be described in section \ref{solutions}, and here we only note
that the analytic method is useful for small black holes (actually
$\mu_\beta$ is the small parameter for the perturbation series),
while for large black holes, where the interesting phase
transition physics occurs the numerical methods are essential. The
existence of both techniques created a healthy feedback where both
methods were used to test and improve each other.

\subsection{Gregory-Laflamme instability}

\begin{figure}[t!]
\centering \noindent
\includegraphics[width=8cm]{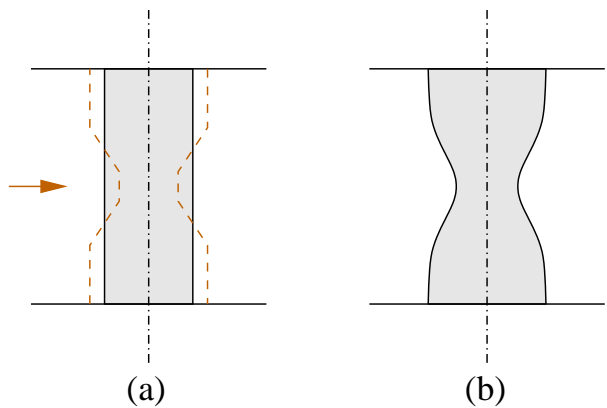}
\caption[]{(a) The Gregory-Laflamme instability. (b) A non-uniform
black string.} \label{GL-setup-figure}
\end{figure}

Gregory and Laflamme (GL, 1993 \cite{GL1})  discovered that the
uniform black string solution (\ref{uniform-string}) develops a
$z$-dependent metric-instability below a certain critical mass
\cite{GL1} (see figure \ref{GL-setup-figure}). By a
``metric-instability'' one means that when one analyzes the
spectrum of frequencies-squared for small perturbations around
this background, a negative eigenvalue is found.\footnote{Such a
metric instability is also known as a ``tachyon'', where the
latter term is used in a more general sense than the ``usual'' 4d
tachyonic field. While one usually considers 4d tachyonic fields
$\phi$, whose Lagrangian behaves as $\sim \half \left[
\dot{\phi}^2-(\vec{\nabla}^2 \phi)^2+m_T^{~2}\, \phi^2 \right] $,
where $m_T^{~2}>0$ and $\vec{\nabla}$ stands for a 3d spatial
gradient, one also generalizes it to arbitrary spatial dimension,
including spatial dimension 0, which is the case here, when we
take $\phi$ to be the amplitude of the GL mode and $m_T^{~2}
\equiv \Omega^2$, where $\Omega$, the inverse decay time, is to be
defined shortly in
figure 6.}

In hindsight, this instability makes a lot of sense. In general,
gravity has a tendency to clump matter. For example, a uniform
distribution of gravitating matter (``gas'') is know to be
unstable against the formation of inhomogeneities (the so-called
``Jeans instability''): when an inhomogeneity forms the denser
regions exert a stronger gravitational pull on their neighborhood,
thereby triggering an unstable positive feedback. Similarly here,
a long enough string ``wants'' to develop inhomogeneities (if it
is short enough then it gets stabilized by the energetic costs of
spatial gradients). Another perspective is to recall that the
\Schw black hole has negative specific heat (black hole
thermodynamics). While this is not enough to de-stabilize a single
black hole, it should certainly destabilize a homogeneous
collection of black holes, namely a black string, which could
increase its entropy by re-distributing its mass non-uniformly.
This intuition is the basis for the Gubser-Mitra Correlated
Stability Conjecture \cite{GubserMitra,GubserMitra-detail} which
states that a homogeneous black brane is (classically)
perturbatively unstable if and only if the dimensionally reduced
black hole is thermodynamically unstable
(semi-classically)\footnote{See \cite{Reall-onGM} for proofs of
certain aspects of this conjecture.}. From now on we continue to
discuss only perturbative instabilities.

\begin{figure}[h!]
\centering \noindent
\includegraphics[width=10cm]{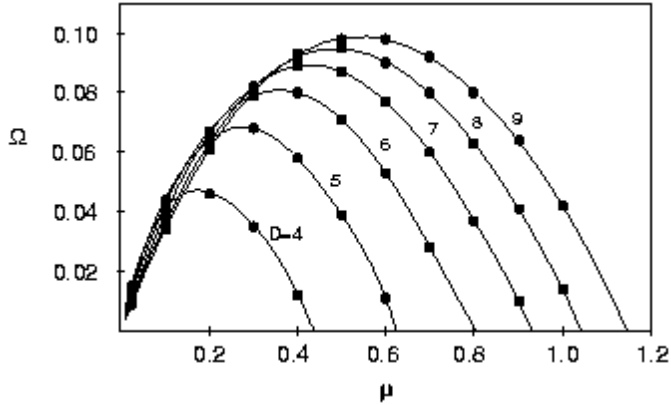}
\caption[]{Characteristic inverse decay times $\Omega$ (giving the
perturbation an $e^{\Omega\, t}$ time dependence) as a function of
the perturbation wavenumber $\mu: = \pi\, r_0/L=r_0\, k/2$
(proportional to $\mu_\beta$ in our notation) for $4 \le D \le 9$
where $D$ is the extended space-time dimension ($d$ in our
notation) in backgrounds with an $\IS^1$ compactification.
$k_{GL}$, the critical Gregory-Laflamme wave-numbers, are the
maximal wavenumbers for which the instability exists -- namely,
the intersection points with the horizontal axis. The bold points
correspond to value calculated numerically and the lines have been
traced to guide the eye (all lines converge to the origin
$\Omega=\mu=0$ which is a non-physical gauge mode). Reproduced
from \cite{GL1,GLcharged}.} \label{GL-figure}
\end{figure}

The main results of GL are summarized in figure \ref{GL-figure}
(taken from \cite{GL1,GLcharged}) which depicts the inverse decay
times as a function of $r_0/L$ for total spacetime dimensions $5
\le D \le 10$.

From figure \ref{GL-figure} we see that the tachyonic mode appears
for wavenumbers $k$ (at fixed $r_0$) which are lower than \emph{a
critical wavenumber} $k_{GL}$ (which depends on $d$). In order to
find $k_{GL}(d)$ it is not necessary to look at perturbations in
$D$ dimensions, but rather it suffices to find the negative mode
of the Euclidean $d$ dimensional \Schw black hole. This mode,
discovered by Gross, Perry and Yaffe \cite{GPY} (in the 4d case)
and hence denoted here by $h_{GPY}$ satisfies \be
 L_{Schw}\, h_{GPY} = -\lambda_{GPY}\, h_{GPY} \ee
 where $L_{Schw}$ is the Lichnerowicz operator for perturbations
in the \Schw background and $-\lambda_{GPY}$ is the negative
eigenvalue. Given $h_{GPY}$ the marginally tachyonic mode is given
by \cite{GL0,Reall-onGM}\footnote{Footnote 4 of \cite{Reall-onGM}
explains ``This relationship (eq. \ref{k-lambda} - BK) was noted
in \cite{GL0}, although the connection with classical stability
was not appreciated.''}\bea
 h=h_{GPY}\, \exp{(i\, k_{GL}\, z)} \non
 k_{GL} := \sqrt{\lambda_{GPY}} \label{k-lambda}\eea

In \cite{SorkinD*} the critical GL lengths were obtained for \Schw
black holes in various dimensions (see table \ref{table_kGL}).
From these the high $d$ asymptotic form was extracted and later
proven analytically in \cite{LargeD-GL} to be \be
 k_{GL} \simeq \sqrt{d}\, {1 \over r_0} ~.\label{larged-GL} \ee
This means that for large $d$ the black string becomes unstable at
a compactification length $L_{GL}=2\pi/k_{GL} \sim r_0/\sqrt{d}$
when it is quite ``fat'' (namely $r_0 \gg L_{GL}$) and indicates
that such a string would not decay into a black hole which would
not ``fit'' inside the extra dimension.

At $k=k_{GL}$ the GL mode is marginally tachyonic, namely a
zero-mode. Morse theory arguments strongly suggest\footnote{The
phrase ``strongly suggest'' is used conservatively
  due to possible subtleties in the argument which are indicated in subsection \ref{morse-subsection}
  and were not explored in full rigor. However, it is the author's opinion that Morse theory
  arguments essentially \emph{guarantee} the existence of the non-uniform branch.}
that this zero-mode produces a branch of solutions emanating from
the GL point describing \emph{non-uniform strings} due to the
$z$-dependence of the GL mode.

\begin{table}[h!]
\centering \noindent
\begin{tabular}{|c||c|c|c|c|c|c|c|c|}\hline
  d& 4& 5 & 6&7&8&9&10&11 \\ \hline
$k_{GL}$  & .876& 1.27& 1.58& 1.85&2.09& 2.30 &2.50 &2.69
\\ \hline
\hline
  d &12&13&14&15&19&29&49&99 \\ \hline
$k_{GL}$  & 2.87 &3.03&3.19 &3.34&3.89 &5.06& 6.72& 9.75
\\ \hline
\end{tabular}
\caption[]{Numerically computed static
  mode wavenumbers $k_{GL}$ in units of $r_0^{-1}$ as a function of $d$,
  the number of extended space-time dimensions \cite{LargeD-GL,SorkinD*}.}
\label{table_kGL}
\end{table}

\presub {\bf The end-state.}
 Whenever one discovers a perturbative
tachyon indicating a decay, the question of its end-state is
naturally raised. As the end-state configuration often lies away
from the initial configuration a perturbative analysis does not
suffice and one needs global information regarding all stable
static solutions which is more difficult to obtain.

Gregory and Laflamme believed the end-state to be the black hole,
both since that was the only other phase they knew about and since
comparing entropies at small $\mu$ one notices that the black hole
phase has superior entropy in this regime. Indeed for small $\mu$,
where the black hole is well approximated by a $D$ dimensional
spherical black hole, the entropies scale  as \bea
 S_{BH} \sim r_0^{~D-2} \sim \mu^{{D-2 \over D-3}} \non
 S_{St} \sim r_0^{~d-2} \sim \mu^{{d-2 \over d-3}} ~,\eea
in units where $L=G_d=G_D=1$. It is seen that the exponent
$(D-2)/(D-3)=1+1/(D-3)$ is a monotonically decreasing function of
its argument and hence the exponent is smaller for $D>d$ (the
black hole) resulting in a larger area.

More recently Horowitz and Maeda \cite{HM} showed that the black
string horizon cannot pinch in finite ``horizon time'' (namely,
finite affine parameter along the horizon generators). They
interpreted that as an indication that surprisingly a black hole
could not be the end-state of decay and predicted instead the
existence of a stable non-uniform string phase that would serve as
an end-state. The argument for pinching in infinite horizon time
relied on assuming the increasing area theorem for an event
horizon and applying it to an area element at the ``waist'' -- the
inward collapsing region of the event horizon. The extension to
the claim on the end-state involved estimates on why infinite
horizon time should imply infinite asymptotic time (time for an
asymptotic observer). While these claims stimulated much of the
research reported here, and while numerical evidence lends support
for ``pinching in infinite horizon time'' (see subsection
\ref{time-evolution}), strong evidence against the end-state being
a non-uniform string will be described as this review proceeds,
implying that the end-point is actually the black hole phase as
originally argued by Gregory and Laflamme (at least in dimension
$D \le 13$). See the summary section for a more complete
discussion.

\subsection{Issues}
\label{issues}

Let us formulate some major issues or questions regarding this
phase transition. These issues may be roughly divided into two
groups: static and time evolution.

The static issues include \bi
 \item End-state of decay.
 \item Qualitative form of the phase diagram
 including the determination of all static phases.
 \item Detailed quantitative data on the phase diagram: the
domain of existence of each phase and the determination of
critical points. \ei

During the last couple of years there was significant progress on
the static issues, resulting also in the surprising discoveries of
critical dimensions and a topology change. The deepest issues
belong however to the time evolution \bi
 \item The spacetime structure, namely determination of the
 Penrose diagram, or an appropriate generalization thereof.
 \item A naked singularity and a violation of Cosmic Censorship.

It is plausible that as the black string pinches a naked
singularity is formed, naively because the singularity which
``originally'' winds the compact dimension gets ``broken'',
perhaps at the event of pinching. Another argument
 comes from the clash between the arguments of \cite{HM} and results on
the system's phase diagram \cite{TopChange,KudohWiseman2}.
Possibilities include a problem with the assumption that there are
no singularities strictly outside the horizon and an infinite
duration with respect to ``horizon-time'' (horizon affine
parameter) while the asymptotic-time duration is finite.  Note
that the initial conditions in this case are generic, unlike known
examples of naked singularities.
 \item A thunderbolt and quantum gravity.

The decay is accompanied by a release of energy (after all, a
tachyon is involved) in the form of radiation (see
\cite{explosive}). It is plausible that this radiation pulse is
classically singular (a ``thunderbolt''), perhaps due to its
origin from the naked singularity. In such a case it is quite
plausible that some knowledge of quantum gravity will be necessary
in order to understand this outgoing radiation. \ei

While there was much progress on the static issues, there was
practically none on the time evolution issues, and these remain
unsolved.

\newpage
\section{Qualitative features}
\label{qualitative}

In order to understand the phase transition physics and to resolve
the issue of the end-state it suffices to map out all static and
stable solutions of the system, since the end-state is certainly
static and stable.
 But actually, Morse theory arguments will lead us to consider all static solutions
whether stable or not, in order to take advantage of a ``phase
conservation rule'' which is a powerful qualitative constraint on
the form of the phase diagram. So we seek the phase diagram of all
static solutions as a function of $\mu$, and throughout this
review we will restrict ourselves to static aspects of the
system.\footnote{Except for subsection \ref{time-evolution} where
a simulated time evolution is described, and subsections
\ref{issues},\ref{open-questions} where the open questions are
discussed.}

In this section we seek to determine the phase diagram
qualitatively, and the quantitative aspects will be described in
the next section.

\subsection{Order parameter}
\label{order-param}

We wish to define an order parameter $\hlambda$ such that the
uniform string will have $\hlambda=0$ and the emerging non-uniform
branch (from the GL point) will have $\hlambda \neq 0$, namely
$\hlambda$ should be a measure of non-uniformity\footnote{We use
the notation $\hlambda$ for the general discussion of an order
parameter, to distinguish it from the closely related perturbation
parameter around uniform strings, which we denote by $\lambda$,
that will be introduced later and which also satisfies that
$\lambda = 0$ if and only if the string is uniform.}. Actually, it
is desirable to have both the black strings and the black holes at
finite values of $\hlambda$, motivated by the expectation for a
merger of the two due to Morse theory arguments as will be
explained in subsection \ref{morse-subsection}.

It turns out that an asymptotic analysis of the metric and the
associated charges furnishes physically meaningful candidates
\cite{HO2,KPS1}. However, it should be noted that the central
discussion on the qualitative form of the phase diagram that will
culminate in subsection \ref{predicted-phase-diag} will be
independent of this choice of the order parameter, and the
discussion here is intended mainly to avoid any unnecessary
vagueness that tends to lead to concerns, such as the very
existence of an appropriate order parameter which is finite on
both black-hole and black-string.

For concreteness, we take the compactification manifold to be
$X=\IS^1$ throughout this subsection. Far away from the black
object the leading behavior of the radial coordinate $r$ is
well-defined (by comparison with the flat geometry) and thus, as
usual, it is possible to read the (ADM) mass of the object by the
asymptotic $r$ behavior of the metric functions. One such
asymptotic constant can be measured from the fall-off of $g_{tt}$
and for a spherical hole in a flat (and topologically trivial)
background this would be the only independent asymptotic constant,
and it would be proportional to the mass. Here there is one more
asymptotic constant:
 the metric becomes $z$-independent ($z$-dependent modes are
massive from the lower dimensional point of view and hence they
decay like $\exp(-2 \pi n\, r/L), ~ n \in \IZ$) and thus it is
sensible to perform a dimensional reduction asymptotically. After
dimensional reduction $g_{zz}$, the size of the extra dimension,
turns into a scalar field. Thus we expect \emph{two asymptotic
charges -- the mass and the scalar charge}. (The latter is
non-conserved, but is conventionally called ``a charge'',
presumably since it is the coefficient for the leading $1/r^{d-3}$
asymptotic fall-off of a field just like the electric charge can
be read from the fall-off of the electro-static potential. It is
precisely this property of being the leading term in the
asymptotic region which we are interested in.)

One can define the asymptotic charges from either the higher
dimensional or from the dimensionally reduced points of view. In
the higher dimension the metric defines two  asymptotic constants
\footnote{Exactly two independent ones as discussed above.} $a,\,
b$ \bea
 -g_{tt} &=& 1-{2 \, a \over r^{d-3}} \non
 g_{zz}  &=& 1+{2 \, b \over r^{d-3}} \label{def-ab} ~.\eea

From the lower dimension point of view the definition for $b$
conforms with the standard definition of a scalar charge: One
defines the scalar field from the $g_{zz}$ component of the metric
through $e^{2\, \Phi} :=g_{zz}$, and the scalar charge
$\Lambda_\Phi$ through the asymptotic behavior of $\Phi$ as
$\Phi=\Lambda_\Phi/r^{d-3}$ (we are not careful here to fix
constants in any particular way). Thus from (\ref{def-ab}) we see
that $\Lambda_\Phi \equiv b$.

We identify the total mass from the dimensionally reduced metric
which is gotten by a Weyl rescaling of the metric $g_{tt} \to
\tilde{g}_{tt}=g_{zz}^{1/(D-3)}\, g_{tt}$ and therefore
$\tilde{a}=a-b/(D-3)$, where $\tilde{a}$ is defined by
$-\tilde{g}_{tt} = 1-2 \, \tilde{a}/ r^{d-3}$. Identifying $2\,
\tilde{a}$ with $r_0^{~d-3}$ and using (\ref{rho-m}) we get \be
G_d\, M = {\Omega_{D-3}\over 8\, \pi} [ (D-3)\, a-b ] ~.
\label{dim-red-mass} \ee

In addition to the mass, the asymptotic constants $a,\, b$ can be
used to express another \emph{physical charge, the tension
$\tau$}. In a satisfying analogy with the well-known first law of
gas thermodynamics \be dE=T\, dS - P\, dV ~,\ee
 where $E,\, T,\, S,\, P,\, V$ are the energy, temperature, entropy,
pressure and volume, respectively, $\tau$ is defined here through
\be
 dM= T\, dS + \tau\, dL ~. \label{differential-first} \ee
Namely, the tension is defined to be the thermodynamic conjugate
to $L$, the size of the extra dimension (see
\cite{TraschenFoxTension,TownsendZamaklarTension} for earlier and
equivalent definitions of tension).

The thermodynamic charges are related to the asymptotic constants
$a,\, b$ through \be 
  \left[ \begin{array}{c} M \\ \tau\, L \\  \end{array} \right] =
  {\Omega_{D-3} \over 8 \pi} \left[ \begin{array}{cc}
  D-3 & -1 \\
  1   & -(D-3) \\
   \end{array} \right] \,
 \left[ \begin{array}{c} a \\ b \\ \end{array} \right]
  \label{asymp_to_charges} ~.\ee
This relation may derived either through the thermodynamic
definitions or from the ``method of equivalent
sources''\footnote{This is our own notation for this known method
-- see for instance \cite{MyersPerry}, but we shall not attempt
complete referencing for it.}, as we proceed to explain. The
thermodynamic definition is fully specified by the gravitational
(Gibbons-Hawking) action $I=-\beta F$ where $\beta$ is the inverse
temperature and $F$ is the free energy.\footnote{The gravitational
action is defined in \ref{Itotal} and discussed around it.}
Alternatively, in the ``method of equivalent sources'' one
imagines that the asymptotic fields were generated by a weak
stress-energy source and uses the linearized equations to infer
the integrated stress-energy charges from the metric asymptotics.
Note that the current expression for the mass
(\ref{asymp_to_charges}) coincides with the mass read off the
dimensionally reduced metric (\ref{dim-red-mass}).

Let us gain some insight into the behavior of the tension and the
scalar charge.  For the uniform string metric
(\ref{string-metric},\ref{string-metric2}) $b=0$ from its
definition (\ref{def-ab}), and hence $\tau\, L=M/(D-3)$ from
(\ref{asymp_to_charges}). For the small black hole, on the other
hand, one finds from the Newtonian approximation
(\ref{Newton2},\ref{thephi}) that $\tau=0$ and $b=a/(D-3)$. More
precisely, to leading order the tension is proportional to $M^2$:
$\tau L= (D-3)\, \zeta(D-3)/2 ~(\rho_0/L)^{D-3}\, M$
\cite{Harmark4,dialogue2}. See table
\ref{order-param-table-for-phases} for a summary.

\begin{table} \begin{center}
\begin{tabular}{c|c|c}
              & Uniform string & Small black hole \\
\hline
Scalar charge & 0              & $G_d M/ (D-3)$ \\
Tension       & $M\, L^{-1}/(D-3)$        & 0       \\
 \end{tabular} \end{center}
\caption{Values for some possible order parameters for both
``extreme'' phases}
 \label{order-param-table-for-phases}
\end{table}

Inverting (\ref{asymp_to_charges}) we obtain \be
  \left[ \begin{array}{c} a \\ b \\  \end{array} \right] =
  {8\pi \over \Omega_{D-3} (D-4)\,(D-2)} \left[ \begin{array}{cc}
  D-3 & -1 \\
  1   & -(D-3) \\
   \end{array} \right] \,
 \left[ \begin{array}{c} M \\ \tau\, L \\ \end{array} \right] \label{charges_to_asymp} ~.\ee
Looking at the expression for the scalar charge $b$ we see that
the mass tends to increase it, namely mass ``wants to generate
more space'' for itself, while tension ``wants'' to contract the
extra dimension. Thus we may say that for the uniform string
($b=0$) the tension has exactly the correct value to cancel the
tendency of mass to expand the extra dimension.

In fact we empirically find that for all black hole and black
string solutions their $b,\, \tau$ parameters lie between the
uniform string and the small BH. This is only partly understood.
Positive tension $\tau>0$, was proven in analogy with the positive
mass theorem \cite{TraschenTension>0,STI-Tension>0}. However, it
is not clear so far why $b>0$ holds. Actually, when one considers
also bubbles (for instance \cite{EHObubbles}) then $b$ is no
longer positive. While $b>0$ would correspond to the bound $\tau\,
L/M<1/(D-3)$ it was argued in \cite{HO2} (see also
\cite{HOtension}) that the completely general bound is higher,
namely $\tau\, L/M<(D-3)$. This bound is set by the bubble and is
consistent with the Strong Energy Condition $T_{00} - 1/(D-2)\,
T^\mu_\mu\, g_{00} >0$. \footnote{The metric signature convention
is ``mostly plus''.}

We may now {\it define the order parameter}. Since $b$ is zero
exactly for the uniform string we can use some multiple of it. The
natural choice is a dimensionless scalar charge, being either
$b/(G_d\, M)$ for the micro-canonical ensemble or $b /\beta^{d-3}$
for the canonical ensemble. The dimensionless scalar charge has
the additional advantage of placing all phases at finite values:
not only is the uniform string at $b=0$ but also the small black
hole is at finite value, namely $b/M=1/(D-3)$, as can be seen from
table \ref{order-param-table-for-phases}. Alternatively, one may
choose a dimensionless tension, such as $\tau/M$, as an order
parameter that vanishes not for uniform strings, but rather for
black holes, and is closely related to $b$ and $M$
(\ref{asymp_to_charges}).

Here we note that the differential form of the first law of black
hole thermodynamics (\ref{differential-first}) may be integrated
using the scale invariance of GR. Namely, when one scales the
lengths $L \to e^\al\, L$ the mass and area (entropy) scale
according to $M \to e^{(D-3)\al}\, M,\, S \to e^{(D-2)\al}\, S$.
Taking the differentials for these transformation with respect to
$\al$ at $\al=1$ and substituting into (\ref{differential-first})
yields \be
 (D-3)\, M = (D-2)\, T\, S + \tau\, L \label{integrated-first} ~,\ee
which is a useful formula known as \emph{``the integrated first
law''} or ``Smarr's formula'' (shown in the current context in
\cite{HO2,KPS1}).

An interesting property of a phase diagram with this order
parameter is that intersections in the phase diagram are
constrained due to the first law \cite{HO2}.

\subsection{Order of phase transition}
\label{transition-order-subsection}

In general, one of the basic properties of any phase transition is
its order. In this subsection we first review black hole
thermodynamics in the Gibbons-Hawking formalism and the general
Landau-Ginzburg theory and then we summarize the results for the
system under study.

\presub {\bf The gravitational free energy}. We will use the
standard semi-classical\footnote{
 From a practical point of view
all the computations with this action are classical. $\hbar$
enters only in the dictionary between the variables of $F$ such as
$\beta,A$ (the periodicity of Euclidean time and the area), and
the thermodynamic variables such as $T^{-1},S$ (the inverse
temperature and the entropy). For example $S=A/(4\, G\, \hbar)$.}
 Gibbons-Hawking \cite{GibbonsHawking-action} gravitational free
energy given by the gravitational action $I(g_{\mu\nu};\,
\beta)=-\beta\, F$ (to be described below) evaluated on a
``Euclidean section'' of the metric. Since our solutions are
static,\footnote{``Static'' means by convention ``non-rotating and
time-independent'' or more formally, invariance not only under
time translations but also under time reversal, so that
$g_{ti}=0,\, i \ne t$. Equivalently, there exist hypersurfaces
such that the Killing vector field $\del_t$ is orthogonal to them,
namely the $t=const$ hypersurfaces.} it is straightforward to
obtain a ``Euclidean section'' simply by taking the transformation
$t \to i\, t$. In a standard way, requiring the absence of conical
singularities at the horizon fixes the period of Euclidean time to
be $\beta=2 \pi/\kappa$, where $\kappa$ is the surface gravity
which is constant over the horizon by the zeroth law.\footnote{The
zeroth law is derived by imposing the constraint $G_{ni}=0$, where
$G$ is the Einstein tensor, $n$ is the coordinate normal to the
horizon and $x^i$ are the coordinates tangent to the horizon,
excluding $t$, the time.}

The gravitational action $I$ is given by the standard
Einstein-Hilbert action with an additional boundary term $I_\del$
\be
 I=I_{EH}+ I_\del \label{Isum} \ee
 such that $I$ is stationary on solutions with respect to
variations of the metric which preserve the boundary metric
\cite{GibbonsHawking-action,York-action}.  \be
 I_{EH}={1 \over 16 \pi G} \int_{\cal M} R ~, \label{IEH} \ee
where $R$ is the Ricci scalar and $I_\del$ can be defined by
either of the following equivalent
definitions\footnote{\cite{GibbonsHawking-action} uses definition
(\ref{Idel1}), and the equivalence with definition (\ref{Idel2})
is used in a computation. I believe that the third and last
definition is implied by the relation between the boundary
conditions and the action.} \bi
 \item  \be 8 \pi G\, I_\del := \int_{\del {\cal M}} K \label{Idel1} \ee where $K$
 is the trace of the second fundamental form for the embedding of the boundary in the manifold.
 \item  \be 8 \pi G\, I_\del := \del_n V_\del,\label{Idel2} \ee the derivative
 of the boundary $(D-1)$-volume with respect to a (proper length) shift in the
 normal direction.
 \item $I_\del$ is the boundary term obtained through integration by parts of
$I_{EH}$ such that $I$ contains only first derivatives of the
metric and no second derivatives, namely $(\del g)(\del g) \in
I,\, g \del^2 g \notin I$ where $g$ denotes here a generic metric
element. \ei

In asymptotically flat spaces $I_\del$ as defined above
(\ref{Idel1}) diverges and must be regularized. The standard
regularization \cite{GibbonsHawking-action} is done by measuring
$I$ relative to flat space. One chooses a large cutoff $r=R$,
where in our case the boundary is $\IS^{d-2}_R \times \IS^1_{L(R)}
\times \IS^1_{\beta(R)}$, where $L(R),\, \beta(R)$ are the periods
of the $z,\, t$ directions, respectively\footnote{
 For large $r$ the black
object metric is virtually $z$ independent, as we discuss above
\ref{def-ab}.}. Next one subtract the action of a flat space with
the same boundary, which in our case is $\IR^{d-1} \times
\IS^1_{L(r)} \times \IS^1_{\beta(R)}$, namely \be
 8 \pi G \, I_0=\del_r \left. 
  \( \Omega_{d-2}\, r^{d-2}\, L(R)\, \beta(R)
 \) \right|_R= (d-2)\, \Omega_{d-2}\,  L(R)\, \beta(R) R^{d-3} \label{defI0} ~. \ee
 Finally one takes $R \to \infty$.

Combining this with (\ref{Isum},\ref{IEH},\ref{Idel1}), the free
energy $F=F(g_{\mu\nu};\beta)$ is finally given by \be
 - \beta\, F = I = {1 \over 16\, \pi\, G} \int_{\cal M} R + {1 \over 8\, \pi\,
 G} \int_{\del {\cal M}} (K-K_0) \label{Itotal} \ee
which includes a bulk integral over $R$, the Ricci scalar and a
boundary integral over $K-K_0$,  where $K$ is the trace of the
second fundamental form on the boundary, and $K_0$ is the same
quantity for the reference flat space geometry.

\presub {\bf Landau-Ginzburg theory}. In a phase transition some
derivative of the free energy is discontinuous and goes through a
jump. The order of a phase transition is defined to be the order
of this derivative. A first order transition is between two phases
which are separated in configuration space and hence have
different entropies (and other thermodynamic variables) and are
therefore exothermic (involving latent heat) while for second
order and higher the phases are continuously connected and there
is no finite release of energy.

The Landau-Ginzburg theory of phase transitions \cite{LL} tells us
how to infer the order of the transition from the local behavior
of the free energy near the critical point $F=F(\lambda;\mu)$,
where the $\lambda$ variables parameterize (a selected subset of)
the configuration space, and may act as order parameters, and the
$\mu$ variables are the control parameters, for instance the
temperature. In our case, the dimensionless control parameter may
be chosen as $\mu \equiv \mu_\beta \propto \beta/L$. Geometrically
it is the ratio of the two asymptotic periods, while physically it
is a dimensionless inverse temperature. Since the control
parameter is essentially the temperature, we interpret the
thermodynamics as taking place in the canonical ensemble, and
accordingly, the name ``free energy'' is fitting.
 The relevant configuration variable is $\lambda$, the amplitude of
the marginally tachyonic GL mode (\ref{k-lambda}), so symbolically
the perturbation is \be
   h \sim \lambda\, \exp(i\, k\, z)\, h_{GPY} ~. \label{def-lambda} \ee

\begin{figure}[t!] \centering \noindent
\includegraphics[width=7cm]{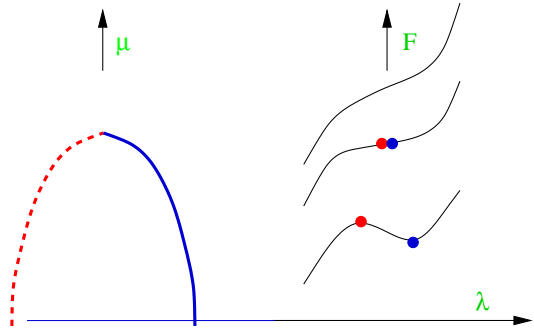}
\caption[]{The most generic phase transition -- first order
resulting from a non-zero cubic term in $F$. On the right we see a
sequence of free energy functions, parameterized by $\mu$, with
their extrema (phases) highlighted. On the left the phases are
extracted into a phase diagram. Stable (unstable) phases are
denoted by solid (dashed) lines.} \label{1tran}
\end{figure}

A generic phase transition is of first order as depicted in figure
\ref{1tran}. \footnote{See a discussion of the Hawking-Page
transition in subsection \ref{morse-subsection}.}
 However, in our case $F$ possesses a
certain parity symmetry (which is non-generic) which opens the
possibility for higher order transitions, as we proceed to
explain.
 We note from (\ref{def-lambda}) that $\lambda$ is complex and
its phase is related to translations in the $z$ direction. Since
the action is invariant under $z$-translations we have \be
   F=F(\mu,|\lambda|^2) ~,\ee
 and the non-uniform phase spontaneously breaks this symmetry.
 From now on, without loss of generality, we
consider $\lambda$ to be real and omit the absolute value
notation, so that $F$ is an even function, as claimed. This
corresponds to fixing the $z$-translations in (\ref{def-lambda})
through $h \sim \lambda\, \cos(k\, z)\, h_{GPY}$ with a real
$\lambda$, and the sign reversal now amounts to shifting $z$ by
half a period.

Note that $\lambda$ can be related to the order parameter defined
in the previous section, the scalar charge $b$. The latter being
invariant under $z$-translations must be a function of
$|\lambda|^2$ as well. Since they both vanish for the uniform
string we conclude that \be
   b \propto |\lambda|^2
 \ee
(there is a genericity assumption made here which is confirmed by
calculations).

\begin{figure}[t!] \centering \noindent
\includegraphics[width=7cm]{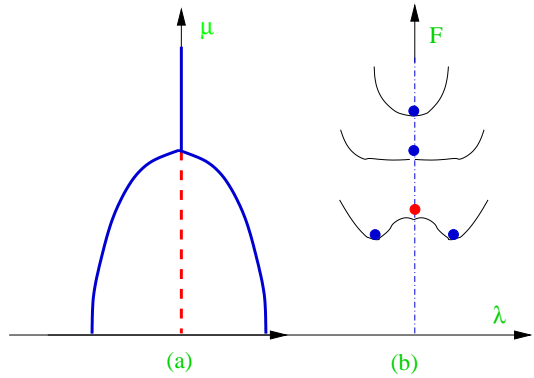}
\caption[]{For an even free energy the order of transition could
be first or higher depending on a sign. This figure shows a second
order transition (or higher) since the free energy has a minimum
at the critical point. Same conventions as in figure \ref{1tran}.}
\label{2tran2nd}
\end{figure}

\begin{figure}[t!] \centering \noindent
\includegraphics[width=7cm]{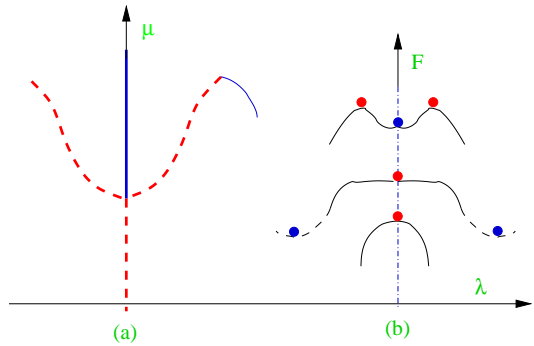}
\caption[]{An even free energy results in a first order transition
if it has a maximum (or more precisely a non-minimum) at the
critical point.  Same conventions as in figure \ref{1tran}.}
\label{2tran1st}
\end{figure}

Having a marginally tachyonic mode appear at some critical value
$\mu_c$ means that the quadratic term in $\lambda$ has a zero at
$\mu_c$, namely \be
 F = F_1\, (\mu-\mu_c)\, \lambda^2 + \dots \ee
 where $F_1$ is some positive constant (assuming the phase is stable for
$\mu>\mu_c$ as is our case). Now the Landau-Ginzburg theory
suggests to expand the free energy at $\mu_c$ to higher orders in
$\lambda$ and to test whether the free energy has a minimum or
not\footnote{In the general case, where $F$ may also have negative
modes, the precise criterion for whether the transition is first
order is whether the $\mu$ of the new phase increases. Note that
in the expansion one may need to incorporate the back-reaction.
These issues are further discussed below}. If $F$ has a minimum at
$\mu_c$, as in figure \ref{2tran2nd}, then for $\mu<\mu_c$ two
stable minima are created close by at small $\lambda$, the system
will continuously evolve into these new phases and the transition
must be second order (or higher). If however $F$ has a ``direction
of descent''\footnote{A non-standard term which we use to mean
``some direction in configuration space where $F$ decreases''.} at
$\mu_c$, as in figure \ref{2tran1st}, that means that since $F$
must be bounded from below there must be some other minimum at
some finite value of $\lambda$, whose free energy is lower than
the $\lambda=0$ phase. Therefore the system underwent already a
first order transition at some higher value of $\mu$ where the
free energies of both phases were equal.

Therefore Landau-Ginzburg theory instructs us (in our case, where
$F$ is even) to expand \be
 F = F_1\, (\mu-\mu_c)\, \lambda^2 + F_2\,  \lambda^4 + \dots \ee
 and to \emph{determine the sign of $F_2$, the quartic coefficient at
$\mu_c$}. Positive (negative) $F_2$ implies a second (first) order
transition. Of course if $F_2$ happens to vanish one needs to
compute higher orders, but this did not happen in this system.

Actually, so far we neglected all other modes except for the GL
mode. When these are brought into consideration one finds it is
required to compute first the (quadratic) back-reaction of the GL
mode and incorporate it into the computation of the quartic term
in the action. The reason for taking the back-reaction is that
what we really want to know is whether the emergent phase from the
critical point goes up or down in $\mu$ in the phase diagram,
indicating a first or second order transition. To see that we
denote the extra modes by $y^i$ and expand the free energy as
follows \footnote{The index $i$ runs over all the extra modes,
namely the additional configuration variables. In our case it
should really be the continuous variable $r$ and sums should be
replaced by integrals, but we keep this notation for conceptual
clarity. We expended $F$ up to quartic order in $\lambda$,
incorporating the fact that $y^i$ will be second order in
$\lambda$.} \be
 F(\lambda,y^i;\mu) = F_1\, (\mu-\mu_c)\, \lambda^2 + F_{2\lambda}\,  \lambda^4 +
 F_{2i}\, \lambda^2 y^i + K_{ij}\, y^i\, y^j + \dots \label{def-Fy} \ee
The equation of motion for the $y^i$ is $0={d \over dy^i}F=2
K_{ij} y^j+ F_{2i}\, \lambda^2=0$. We denote the solution by $y^i=
\lambda^2\, B^i$, where $B^i$ stands for back-reaction.
 The equation of motion for $\lambda$ is $0={d \over d\lambda}F=2\, F_1\,
(\mu-\mu_c)\, \lambda + 4\, F_{2\lambda}\,  \lambda^3 +
 2\, F_{2i}\, \lambda y^i$.
 After diving by $\lambda$ and using the equations of motion
for $y^i$ we find that
 $ -2 F_1 (\mu-\mu_c) =4 \lambda^2 \( F_{2\lambda} - K_{ij}\,
 B^i\, B^j \).$
 Namely, in order to determine $\mbox{sgn}(\mu-\mu_c)$ and from that
the order of the transition we need to compute \be
 F_2 := F_{2\lambda} - K_{ij}\, B^i\, B^j \ee
 where $B^i$ is the back-reaction that solves the $y^i$ equations
of motion. Note that $F_2$ is computed without deviating from
criticality $\mu=\mu_c$.

\begin{figure}[t!]
\centering \noindent
\includegraphics[width=10cm]{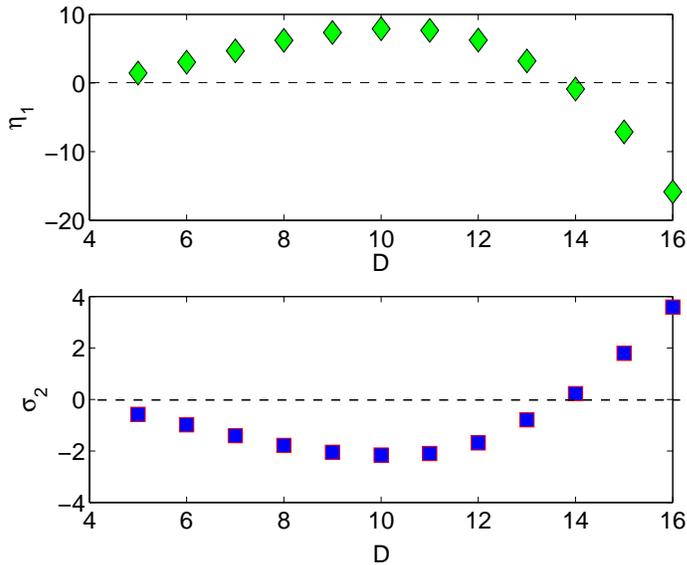}
\caption[]{The first correction to the mass $\eta_1$ and to
 entropy (relative to a uniform string with the same mass)
 $\sigma_2$ for the non-uniform string branch
emanating from the GL point for backgrounds $\IR^{D-2,1} \times
\IS^1$ -- see \cite{SorkinD*} for precise definitions.
 From our general discussion we know that there is only one sign to
be determined (that of $F_2$), and indeed the signs of $\eta_1,\,
\sigma_2$ are correlated. At the critical dimension $D^*=13.5$
there is a change of sign indicating the transition becomes second
order for higher $D$. Adapted with permission from
\cite{SorkinD*}.} \label{sorkin-critical-dim-figure}
\end{figure}

 \presub {\bf Results}. The determination of the order
was first carried out by Gubser \cite{Gubser} for the background
$\IR^{3,1} \times \IS^1$ (see also Wiseman's improvement
\cite{Wiseman1}). Part of the original motivation there was to
find a second order phase transition and hence together with it a
branch of stable non-uniform strings emanating from the GL point,
such as those predicted by \cite{HM}. However, the transition was
found to be first order. Sorkin \cite{SorkinD*} generalized the
method to the backgrounds $\IR^{D-2,1} \times \IS^1$ and found a
surprising critical dimension
\begin{center}
\fbox{\vspace*{.5cm} \begin{tabular}{cc}
 $D \le 13$ &  first order \\
 $D \ge 14$ &  second order ~.\\ \end{tabular} \vspace*{.5cm}} \end{center}
 The critical dimension $D^*=``13.5"$
 \footnote{Of course dimensions are integral, and the notation
means only that the change in the order happens between $D=13$ and
$D=14$.}
 is demonstrated in figure \ref{sorkin-critical-dim-figure}.
Kudoh and Miyamoto \cite{KudohMiyamoto} observed that the critical
dimension depends on the ensemble: Sorkin's critical dimension
holds for the micro-canonical ensemble, while in the canonical
ensemble \footnote{A black hole embedded in a heat bath.} the
critical dimension is lower by one.
 In subsection \ref{results}, ``results'', we comment on the relation of these
results with the predictions of \cite{HM}.

Due to the importance of the critical dimension it is a good idea
to develop some intuition about it. First, as discussed around
equation (\ref{larged-GL}) for high $D$ the critical GL string is
quite ``fat'' and we expect that the string will not decay
(directly) into the black hole, which would be ``too big to fit''
inside the extra dimension (see also \cite{Park} for a similar
argument involving $\mu_S$, which is defined below). Another
indicator comes from comparing $\mu_{GL}$ with $\mu_S$ the value
of the approximate equal area point: $S_{St}(\mu_S)=S_{BH}(\mu_S)$
where $S_{St}$ is the entropy of the uniform string and $S_{BH}$
is BH area approximated by the small black hole expression. For a
first order transition we expect $\mu_S > \mu_{GL}$ but that holds
only for $D < 12.5$.

\presub {\bf Method}. Gubser's method \cite{Gubser} was to
perturbatively follow the branch of non-uniform strings emanating
from the GL critical point. The perturbation parameter is
$\lambda$, the amplitude for the GL mode (\ref{def-lambda}), and
in order to determine the order of the transition is was necessary
to compute some metric functions up to the third order in the
perturbation parameter. As we mentioned above, to determine the
order there is a somewhat more efficient method, namely it is
actually enough to compute only the second order back-reaction and
substitute into the quartic part of the action (see appendix A in
\cite{TopChange} and \cite{torus}). However, the longer
computation naturally yields additional results not included in
the shorter one.

 In practice, when computing the back-reaction we consider a
continuum of modes, and therefore the discrete index $i$ in
(\ref{def-Fy}) and the discussion around it is replaced by the
continuous variable $r$, and the ``field''$y$ is replaced by the
all the fields in the problem. Moreover, inverting the linear
operator $K$ to solve for $y$ means obtaining the fields by
solving a second order ODE (obtained from linearizing the
equations of motion) with a source term quadratic in the
perturbation (the first order mode).

\subsection{Morse theory}
\label{morse-subsection}


When one turns to consider the question of the end-point for
decay, or more generally of finding the phase diagram of all
static phases, one is at first discouraged by the lack of any
knowledge regarding the non-uniform strings and the big black
holes. The most interesting question is to determine the
qualitative features of the phase diagram. For that purpose one
needs qualitative tools, and one such tool was given in
\cite{TopChange} under the name ``Morse theory'', fulfilling the
intuition that generically a phase persists as the system is
``deformed'' by changing a parameter, and that the disappearance
of a phase should require some special circumstances that are
worth elucidating.

Indeed solutions of the Einstein equations are extrema of the
gravitational action in the space of metrics, and as such are
generically stable under perturbations. The topological theory of
extrema of functions is well-known and is called ``Morse theory''
and it includes the specification of the allowed transitions.
\footnote{Some readers may be familiar with the way Morse theory
measures global properties of manifolds (Homology), but here we
need a different aspect of the theory -- local invariants of
extrema under deformation of the function.}

For other qualitative tools in the study of thermodynamics in the
astrophysical context of self-gravitating systems see the review
\cite{KatzThermo}, and especially the closely related Poincar\'{e}
method to determine the perturbative stability of phases just by
looking at a certain kind of a phase diagram.

Actually, there is a subtlety in the identification of extrema of
the action with solutions of the equations of motion due to gauge
(diffeomorphism) redundancy, which we would like to mention. It is
certainly true that solutions are extrema of the action. However
we wish to consider the action as a function of metrics up to
gauge invariance, namely as a function of the gauge-fixed metric.
Therefore the extremum equations should be supplemented by the
gauge-fixing constraints. Nevertheless, in this case
 it was found \cite{Wiseman1} that the constraints are actually
implied by the extremum requirement through a combination of
properly chosen boundary conditions and the constraints' Bianchi
identities. It could be that this is true more generally.

\presub {\bf A lightning review of Morse theory} (see section
(3.2) of \cite{TopChange} for a somewhat longer introduction). For
functions of one variable the ways in which an extrema can
disappear are clear \bi
 \item Annihilation. A maximum and a minimum can coalesce under continuous deformation
 and disappear into a monotonous function. We call this the basic
 1d vertex
\footnote{Here we introduce the term ``vertex'' to mean the event
when two or more extrema of a function coincide as a deformation
parameter is varied. Such an event looks like a collision of
phases in a phase diagram and the name comes from the analogy with
the Feynman diagram vertex at the collision point between two or
more particle world-lines.}
  of ``annihilation'' -- see figure \ref{1tran}.
 \item Run-away. An extrema of a function $f(x)$ may run away to infinity either
 in $x$ or in $f$ during a finite range of deformation parameter.
 In this paper we shall find ``annihilation'' explanations for changing phases, and thus we
 will not need to resort to ``run-away'' explanations, even though
 they are certainly a logical possibility.
 \ei

When one considers a function of several variables,
$f=f(\vec{x})$, one may get a simple generalization of the basic
1d vertex by adding $n$ spectating negative directions as well as
$N-n-1$ positive directions, namely $f(\vec{x})=f_0(x_0;\mu) -
\sum_{i=1}^n f_i\, x_i^{~2} + \sum_{j=n+1}^{N-1} f_j\, x_j^{~2}$,
where $f_0(x_0;\mu)$ is a 1d function such as the one depicted in
figure \ref{1tran}, and $f_i, ~1 \le i \le N-1$ are positive
constants. Now, the minimum in figure \ref{1tran} turns into an
extremum with $n$ negative modes, while the maximum has $n+1$
negative directions. Therefore we obtain, what we call ``the basic
vertex''
\begin{center}
 \fbox{Basic vertex: two extrema with $n$ and $n+1$ negative directions
 may annihilate.}
 \end{center}
While for generic extrema the Hessian is non-degenerate and $n$ is
well-defined, one may wish to know the rules for more general
vertices. Indeed, Morse theory can be phrased as saying that the
most general vertex is a coincidence of several of these basic
vertices, thereby justifying our use of the adjective ``basic''.

\presub {\bf The conclusion and some reservations}. We see that a
stable phase ($n=0$) is allowed to disappear at the expense of
``annihilating'' with an unstable phase with one negative mode
($n=1$). The latter in turn can disappear by annihilating against
either $n=0$ or $n=2$ phases and so one. This is exactly the
``{\it phase conservation rule}'' \cite{TopChange} which we were
seeking. It sets a strong qualitative constraint on the existence
of phases. However, the price to be paid is that all phases must
now be mapped out, not only the stable ones.

As shown in figures \ref{2tran2nd},\ref{2tran1st} this rule
determines the stability of the non-uniform string emanating from
the GL point. One may ask where this phase might end. From the
phase conservation rule we conclude that the simplest way to
satisfy it, without requiring any additional phases, is that
\emph{the non-uniform string phase would annihilate against the
black hole phase} \cite{TopChange} at a point on the phase diagram
that we call \emph{``the merger''}. Taking into account the
existence of a critical dimension we find that depending on the
dimension we get two possible behaviors: for $5 \le D < D^*$ the
non-uniform black string, which is unstable for small
non-uniformity, annihilates with the black hole (which is stable
when it is small) while for $D > D^*$ a stable non-uniform string
transforms into a stable BH, and $D^*=12.5$ ($13.5$) in the
canonical (micro-canonical ensemble). This is our main conclusion
from Morse theory and we stress again that it seems to be the
simplest scenario, but others cannot be excluded.

The rigor of the prediction of a phase merger, even if intuitive
and clear, is questionable due to the following observation. In
the next section we will see that the Euclidean versions of the
black hole and the string have {\it different topologies} and
hence their metrics would be expected to live in different,
disconnected spaces of metrics, and it wouldn't make sense for
phases to move from one space to the other. Nevertheless, we shall
take the prediction above seriously and look whether these two
spaces of metrics are in some sense glued together. Indeed we
shall find a continuous transition (and in finite distance)
between the two spaces, which we view as an important confirmation
for the consistency of the picture. However, the way in which the
spaces are glued is still poorly understood, and the gluing may
very well be non-smooth as well as involve the infinite
dimensionality of the space of metrics in an essential way, in
which case the validity of the Morse theory argument is not
self-evident. At this point, I consider the conclusion above to be
essentially correct and justified if not rigorously {\it a priori}
then {\it a posteriori} by the agreement of the predicted and the
numerically computed phase diagrams.

\presub {\bf An example: the Hawking-Page transition.} The reader
familiar with the phase transition between thermal Anti-de-Sitter
(AdS) and large black holes in AdS, known as the ``Hawking-Page
transition'' \cite{HawkingPage}, may benefit from applying the
Morse theory ideas above to that context.

In the Hawking-Page transition in its canonical ensemble setting,
one considers a space-time with a negative cosmological constant
$\Lambda$ whose boundary is that same as that of thermal AdS,
namely a large spatial sphere of Radius $R$ times a compact circle
of Euclidean time of size $\beta\, \sqrt{-\Lambda}\, R$. The
dimensionless control parameter is
$\beta_\Lambda:=\beta\,\sqrt{-\Lambda}$, and as usual
$\beta/\hbar$ is the inverse temperature. In 4d Hawking and Page
found that several phases exist in this system as follows. For
$\beta_\Lambda>\beta_0={1 \over 2\, \pi}$ the only phase which
exists is thermal AdS, AdS filled with a thermal gas of radiation.
For $\beta_0>\beta_\Lambda>\beta_1={1 \over \sqrt{3}\, \pi}$ two
additional phases show up, the small and large AdS black holes
(inside a thermal bath), the large (small) black hole being
thermodynamically stable (unstable) due to its positive (negative)
specific heat. However, the black holes' free energy is inferior
to that of thermal AdS. Finally at $\beta_\Lambda<\beta_1$ the
large black hole dominates.\footnote{Another transition is
expected at $\beta_2 \propto \hbar^{1/4}$ which is of quantum
nature and will not be discussed here.}

\begin{figure}[t!]
\centering \noindent
\includegraphics[width=10cm]{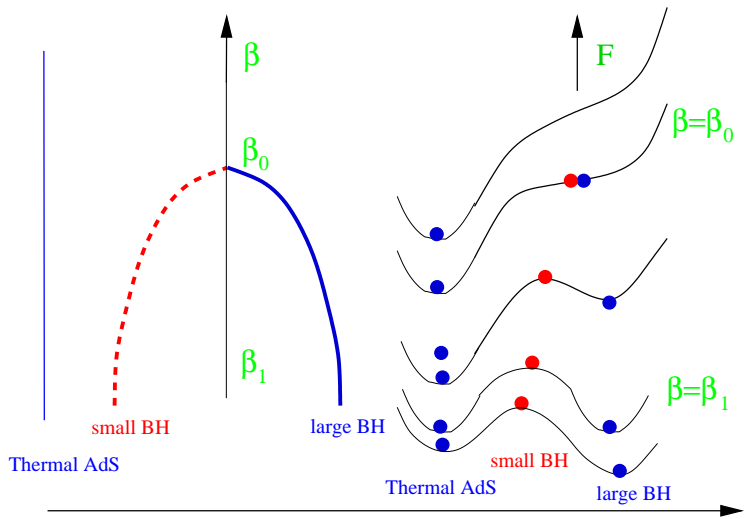}
\caption[]{The free energy and phase diagram for the Hawking-Page
transition \cite{HawkingPage} in AdS space, which is a first order
phase transition. At $\beta=\beta_0$ the unstable small black hole
phase meets the stable large black hole phase and they
``annihilate'' according to our ''basic vertex'' rule.}
\label{HPtran}
\end{figure}

In the current context, this phase transition, which is a first
order phase transition is described in figure \ref{HPtran}
(compare with the general first order transition of figure
\ref{1tran}). For $\beta>\beta_0$ there is a single minimum for
the free energy which is thermal AdS. At $\beta=\beta_0$ two
phases are ``pair created'' (or ``annihilated'' if one approached
$\beta_0$ from below) through our ``basic vertex'': a stable large
black hole and an unstable small black hole, but thermal AdS still
has lower free energy. Here we are assuming the Gubser-Mitra
Correlated Stability Conjecture
\cite{GubserMitra,GubserMitra-detail} to infer the perturbative
stability or instability from the sign of the specific heat and
the associated thermodynamic stability of the black holes. Then at
$\beta=\beta_1$ a first order phase transition occurs when the
free energies of thermal AdS and the large BH become equal.
Finally, for smaller $\beta$ the large BH dominates.

\subsection{Merger point}
\label{merger-subsection}

In the last subsection we saw how Morse theory makes it plausible
that the non-uniform string phase merges with the BH phase. We
shall first encounter a problem for this picture, namely
topological differences between the two phases. However, we are
familiar with some topology changes such as the flop and the
conifold, and one of the surprising results of the research on
this system is the emergence of a novel type of topology change,
called the ``merger'' transition in \cite{TopChange}.

\begin{figure}[t!] \centering \noindent
\includegraphics[width=9cm]{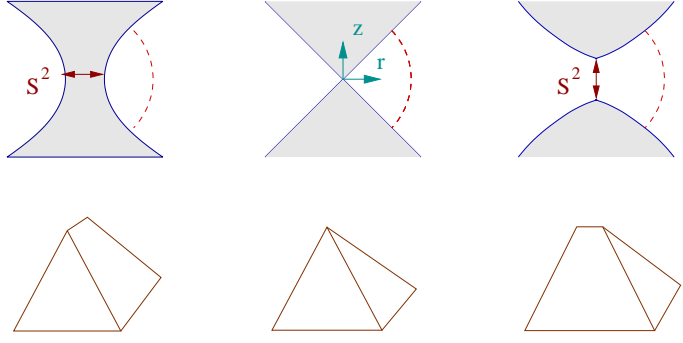}
\caption[]{The merger transition. Shaded regions are inside the
horizon and the dashed line is a boundary far away. The singular
configuration is a cone over $\IS^2 \times \IS^2$.} \label{merger}
\end{figure}

\presub {\bf A topology change}. Intuitively the transition from
black string to black hole involves a region where the horizon
becomes thinner and thinner as a parameter is changed until it
pinches and the horizon topology changes. We call this region
``the waist'' and this process is described in the upper row of
figure \ref{merger} using the $(r,z)$ coordinates defined in
figure \ref{coordinates}. It is important to remember that all
metrics under consideration are static and that they change as we
change an external parameter, not time. Since the metrics are
static we may as well consider their Euclidean versions (this
point was discussed in the second paragraph of section
\ref{transition-order-subsection}).

We shall now demonstrate that this merger transition involves a
local topology change of the Euclidean manifold. Let us zoom in
around a very thin waist, whereby all the scales of the problem
such as $GM$ and $L$ are very large and irrelevant, identifying
what may be called the local geometry at the waist. Consider a
co-dimension 1 surface within the local geometry but far away from
the waist, such as the one denoted by a dashed line in all three
geometries in the top row of figure \ref{merger}. \footnote{In
other words, if we denote by $\rho_w$ the radius of the (minimal)
angular sphere ($S^{D-3}$) at the waist, then we wish to consider
the fixed $\rho$ surface where $\rho_w \ll \rho \ll L$.} Actually,
 this surface is the asymptotic boundary of the local geometry.

\begin{figure}[t!]
\centering \noindent
\includegraphics[width=8cm]{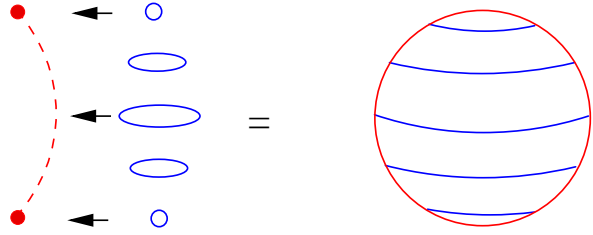}
\caption[]{An illustration of the fibration of an $\IS^2$: circles
fibred over an interval such that the fiber shrinks at the edges
(left) are equivalent to a surface of a sphere (right). In this
way the dashed intervals in the top row of figure 12 actually
represent spheres, once the circles of Euclidean time are taken
into account.} \label{S2fib}
\end{figure}

The topology of this asymptotic boundary is given by
$\IS^{D-3}_{\theta} \times \IS^2_{r,z,t}$ as we proceed to
explain. The angular piece, $\IS^{D-3}_{\theta}$, is obvious while
the $\IS^2_{r,z,t}$ requires explanation. One should remember that
in our figures, such as figure \ref{merger}, we suppress not only
the angular coordinates but also the time variable. In the
Euclidean continuation the Euclidean time must be periodic in
order to avoid a conical singularity at the horizon, and moreover,
the proper size of this circle vanishes at the horizon ($g_{tt}=0$
at the horizon). Thus the circle fibration of Euclidean time
\footnote{which is topologically trivial since $g_{ti}=0$ for all
$i \ne t$.} over an interval (the dashed line in figure
\ref{merger}), such that the fibred circle shrinks
 on the edges \footnote{More formally, contracts to a point.}
{\it exactly produces a topological $\IS^2$}. This is just like
the fibration of the surface of the earth by latitude lines, which
shrink at the poles (see figure \ref{S2fib}).

Naturally, the topology of this boundary surface is constant as
local changes occur near the waist. Next we notice that in the
black-hole phase the $\IS^{D-3}_{\theta}$ is contractible (onto
the exposed $r=0$ axis), while in the string phase the
$\IS^2_{r,z,t}$ is contractible. Therefore {\it the local topology
of (Euclidean) spacetime is changing}, not only the horizon
topology. Thus, {\it the topology change can be modelled by a
``pyramid''} (just like the conifold transition) -- see the lower
row of figure \ref{merger}: the rectangular basis of the pyramid
denotes the asymptotic boundary $\IS^{D-3}_{\theta} \times
\IS^2_{r,z,t}$ where each edge represents one of the sphere
factors, and the pyramid's truncated apex encodes which one of the
spheres is contractible and which one remains of finite size and
is non-contractible. By the nature of topology, in order to change
it there must be at least one singular solution along the way
(with at least one singular point). The simplest possibility,
(which is also realized in the conifold) is  to assume that the
{\it the singular geometry is the cone over} $\IS^{D-3} \times
\IS^2$.

It is suggestive that the local space-time topology change is also
accompanied by a change in the global topology. I believe this is
the case for the following reason. The black string solution is
contractible to $\IS^{D-3}_\theta \times \IS^1_z$ (after
contracting the $r,t$ cigar), and thus its elementary
non-contractible cycles are $\IS^{D-3}_\theta$ and $\IS^1$. For
the black hole on the other hand the non-contractible cycles seem
to be $\IS^{D-2}_{\chi\theta},\, \IS^2_{z,t}$ and $\IS^1_z$: the
$\IS^{D-2}_{\chi\theta}$ is the horizon, the $\IS^1_z$ is the
compact $z$ direction and the $\IS^2_{z,t}$ is the $r=0$ axis
connecting the poles together with the time fibration. It is seen
that there are several topological differences between the
geometries, for instance, the black hole has a 2d topologically
non-trivial cycle while the string does not.

\presub {\bf Cones}. Figure \ref{merger} encodes the topological
nature of the merger. But is it really true that this can be
realized with Ricci-flat metrics?

It is easy to write down a Ricci flat metric for the singular
solution, the cone (see the middle of figure \ref{merger}).
Actually this can be done for a somewhat more general cone, the
cone over $\IS^m \times \IS^n$ with no additional ``cost''. The
metric is \be
 ds^2 = d\rho^2 + {\rho^2 \over D-2}
 \left[ (m-1)\, d\Omega^2_{\IS^m} + (n-1)\, d\Omega^2_{\IS^n}
 \right] ~,\ee
where the $\rho$ coordinate measures the distance from the tip of
the cone, $D=m+n+1$ is as usual the total spacetime dimension and
the constant pre-factors are essential for Ricci-flatness. Note
that $\rho=0$ is the singular tip of the cone, unless $m=0$ (or
$n=0$) when it becomes the smooth origin of $\IR^D$ in spherical
coordinates.

In order to exhibit ``smooth cone'' metrics which approach the
singular cone from both sides of the transition (see right and
left portions of figure \ref{merger}) one may use the following
ansatz \be \label{def_cone_ansatz_gen}
 ds^2 = d\rho^2 + e^{2\, a(\rho)}\, d\Omega^2_{\IS^m} + e^{2\, b(\rho)}\,
  d\Omega^2_{\IS^n} \ee
  with boundary conditions at $\rho \to 0$ \bea
  a(\rho=0) &=& a_0 \non
  b(\rho) &=& \log(\rho) \label{smooth-cone-bc} \eea
such that $\IS^m$ becomes non-contractible while $\IS^n$ joins
with $\rho$ to make a smooth neighborhood of the origin of
$\IR^{n+1}$.

Using (\ref{Rfibration}) for the Ricci scalar of a fibration  and
after integration by parts one gets the action for $a,b$ \be
 S= {\Omega_m\, \Omega_n \over 16\, \pi\, G_D} \int d\rho\, e^{m\, a
 + n\, b}\, \left[ m(m-1)\, (a'^2+e^{-2\, a}) + n(n-1)\, (b'^2 + e^{-2\,
 b}) + 2\, m\, n\, a'\, b' \right] ~~.\ee
Einstein's equations are \bea \label{abeom}
 a'' &=& (m-1)\, e^{-2\, a} -m\, a'^{~2} - n\, a'\, b' \non
 b'' &=& (n-1)\, e^{-2\, b} -n\, b'^{~2} - m\, a'\, b' \non
 0   &=& m\,(a'^{~2} + a'') + n\,(b'^{~2} + b''). \eea
These equations are very similar to the equations encountered in
the Belinskii-Khalatnikov-Lifshitz (BKL) analysis of the approach
to a space-like singularity (see the recent excellent
``Cosmological Billiard" review \cite{CosmoBilliard}). Although
the general (and often singular) qualitative behavior of these
equations as $\rho \to 0$ for arbitrary initial conditions was not
obtained in \cite{TopChange},\footnote{One can prove useful
theorems for the evolution of the volume factor $U=e^{m\, a + n\,
b}$ using the geometric analogue of the ``c-theorem'' (I thank J.
Maldacena for pointing this). From (\ref{abeom}) we have
$(\log(U))''=-m\, a'^2 - n\, b'^2 \le 0 $ Hence if $U$ (or
equivalently $\log(U)$) is somewhere decreasing it must continue
to decrease monotonically. At the same time $U'' = \left( m(m-1)\,
e^{-2\, a} + n(n-1)\, e^{-2\, b} \right) U \ge 0$ which together
with the previous result guarantees that $U$ is monotonic.} it was
checked that for the boundary condition (\ref{smooth-cone-bc})
$a(\rho), \, b(\rho)-\log(\rho)$ can be expanded in a Taylor
series around $\rho=0$ and the recurrence equations for the Taylor
coefficients could be solved without encountering an obstruction.

\begin{figure}[t!] \centering \noindent
\includegraphics[width=5cm]{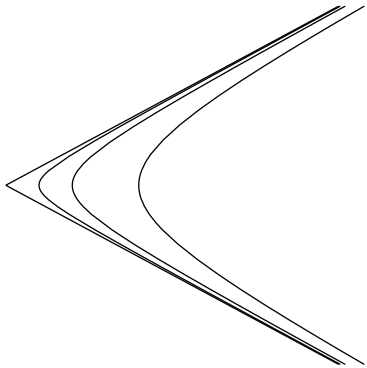}
\caption[]{A single smoothed cone solution can be scaled down to
provide a continuous family of metrics which approach the cone.}
\label{ScaledCones}
\end{figure}

Once a single smoothed cone solution is available, constructing a
full family that approaches the singular cone is a matter of
simply rescaling it.
 As illustrated in figure \ref{ScaledCones}, since away from the
smoothed tip the smoothed cone asymptotes to a cone, a geometry
which is scale invariant, then after rescaling there is a natural
way to identify the asymptotic cones, thereby specifying the way
to take the limit over the family of rescaled metrics.

Altogether we succeeded in realizing the local topology change
encoded in figure \ref{merger} by a family of Ricci-flat metrics.
I consider this non-trivial property to be \emph{strong evidence
for the merger picture} as presented in subsection
\ref{morse-subsection}.

I would like to stress some of the \emph{assumptions involved} in
locally modelling by cones \bi
 \item The singular solution has a single singular point.
 \item The singular solution is continuously self similar (CSS).
 \ei
Both assumptions are reasonable and minimal: there could be more
than a single singularity, but there is at least one, and the
local singular solution must forget the ``long distance'' scales
and hence it would be scale-free, and the simplest way to obtain
that is if the solution is self-similar. Continuous
self-similarity would be the simplest possibility and in
\cite{TopChange} it was found to lead to a local isometry
enhancement where the $\chi$ coordinate (see figure
\ref{coordinates}) conspires with $t$ to make a round $\IS^2$.
Indeed, a Numeric study \cite{KolWiseman}, limited by numerical
resolution, found consistent evidence for this cone structure.
More recently it was claimed in \cite{scaling} that in a certain
range of dimensions ($5 \le D < 10$) continuous self-similarity
(CSS) is in fact (spontaneously) broken into discrete
self-similarity (DSS) which requires to replace the cones by
certain ``wiggly cones'' as local models of the merger point.

{\bf Tachyons on cones}. It turns out that the cones may have
tachyons and that their existence surprisingly depends on a
critical dimension $D^*_{\mbox{merger}}=10$ as we proceed to show
-- see also \cite{TopChange} and references therein including
\cite{KlebanovRangamaniWitten-private}.

The dangerous mode is a function $\eps(\rho)$ which inflates
slightly one of the spheres while shrinking the other. The ansatz
for the perturbation is \be \label{stab_ansatz_gen}
 ds^2 = d\rho^2 + {\rho^2 \over D-2}
 \left( (m-1)\, e^{2\, \eps/m}\, d\Omega^2_{\IS^m}
  +(n-1)\, e^{-2\, \eps/n}\, d\Omega^2_{\IS^n} \right)~. \ee
{\it A priori} one could start with two separate scale functions,
one for each sphere, but the $G_{\rho\rho}=0$ constraint relates
them as above (after ignoring the trivial perturbation which
represents $\rho$-translation) -- for more details see
\cite{TopChange} eq. (6.6) and below.

The quadratic part of the action, disregarding an overall constant
is \be \label{eps_action}
 I \sim \int \rho^{D-1}\, d\rho \left[ {2\,(D-2) \over \rho^2}\, \eps^2 -
\eps'^{~2} \right] ~, \ee
 and through the change of variables $\hrho^{-1} = (D-2)\, \rho^{D-2}$
it can be recast to have a canonical kinetic term \be
 \label{eps_action2}
 I \sim \int d\hrho
 \left[ {2 \over (D-2) \, \hrho^2}\, \eps^2 - \eps'^{~2} \right] ~.\ee

The equation of motion for $\eps$, namely the zero mode equation,
is \bea
 [ -\hat{\del}^2+V(\hrho) &]& \eps = 0 \label{zm} \\
  V(\hrho) &=& - {2 \over (D-2)\, \hrho^2} ~, \label{zm-pot} \eea
The solutions are \bea
 \eps &=& \rho^{s_{1,2}} \non
 s_{1,2} &=& {D-2 \over 2}\,\left(-1 \pm i
\sqrt{{8 \over D-2}-1}\right) ~. \label{linearized-exponents} \eea

The expression (\ref{linearized-exponents}) for $s_{1,2}$, the
characteristic exponents, reveals a \emph{critical dimension}
\cite{TopChange} \be
 D^*=10 \label{crit-dim} \ee
 such that for $D<D^*$ $s$ are complex while for $D>D^*$ they are real.
Complex characteristic exponents (for $D<D^*$) imply that (the
real part of) the zero mode has infinitely many nodes (zeroes)
equally spaced in $\log(\rho)$ with ``log-period'' $\Delta_e=2
\pi/\Im(s)$. The presence of infinitely many nodes for the zero
mode implies the presence of infinitely many tachyons, just like
the number of nodes of the zero-energy solution to a
Schr\"{o}dinger equation counts the number of negative energy
states. In \cite{scaling} these log-periodicity and tachyons (for
$D<D^*$) were interpreted as an indication for a spontaneous
breaking of continuous self-similarity (CSS) into discrete
self-similarity (DSS). For $D>D^*$ on the other hand, $\Im(s)=0$
and there are no nodes nor tachyons.

Note that $D^*$ is the total spacetime dimension, and it is
independent of $m$ and $n$ separately. When we wish to distinguish
this critical dimension from others we shall denote it by
$D^*_{\mbox{merger}}$.

One can view the zero-mode equation (\ref{zm}) in a wider context
by considering the eigenvalue problem \be
 [ -\hat{\del}^2+V(\hrho) ] \eps = \lambda \eps \label{zm-eigen} \ee
 where $V$ is the same as in eq. (\ref{zm-pot}), $\lambda$ is the
 eigenvalue,\footnote{This $\lambda$ is unrelated to the $\lambda$
 introduces in subsection \ref{transition-order-subsection} to
represent the amplitude of the perturbation, and will not be used
elsewhere in this review.}
 and we note that (\ref{zm-eigen}) reduces to (\ref{zm}) upon
setting $\lambda=0$.

The eigenvalue problem (\ref{zm-eigen}) is in Schr\"{o}dinger
form, and we may apply known results. For potentials of the
general form $V=-c/r^\al$ it is well-known that while the
classical energy is unbounded from below, the quantum problem may
have a ground state as long as the potential is not ``too
negative''. For instance, for $\al=1$ we get the Hydrogen atom.
More generally the spectrum is bounded from below for $\al<2$,
while for the critical value $\al=2$, which is our concern here,
the prefactor $c$ becomes dimensionless and the potential is
conformally invariant. Due to scale invariance the spectrum is
constrained to be invariant under positive rescaling of the
eigenvalues. Now $c$ itself exhibits a critical value, namely
$c^*=1/4$, such that the spectrum is bounded from below and
actually non-negative only for $c \le c^*$ (since the $E=0$
solution has no nodes -- see for example \cite{LL35}). Equating
$2/(D-2)=c^*=1/4$ we arrive once more at $D^*=10$ as in
(\ref{crit-dim}).

\presub {\bf Some moduli space properties}. The merger transition
lies in finite distance in moduli space (actually, it would not
deserve to be called a ``topology change'' otherwise, since it
would require infinite ``resources'' to be reached.) The argument
was unpublished so far and here it is supplemented as appendix
\ref{finite-distance-section}. Moreover, the appearance of a
kink\footnote{More formally, a discontinuity of the tangent to the
phase line.} in the phase diagram at merger was predicted in
section (5.3) of \cite{TopChange}, and the numerics indeed seem to
exhibit some sort of a kink -- see the next section.

\subsection{Phase diagrams -- predictions and data}
\label{predicted-phase-diag}

\begin{figure}[t!] \centering \noindent
\includegraphics[width=15cm]{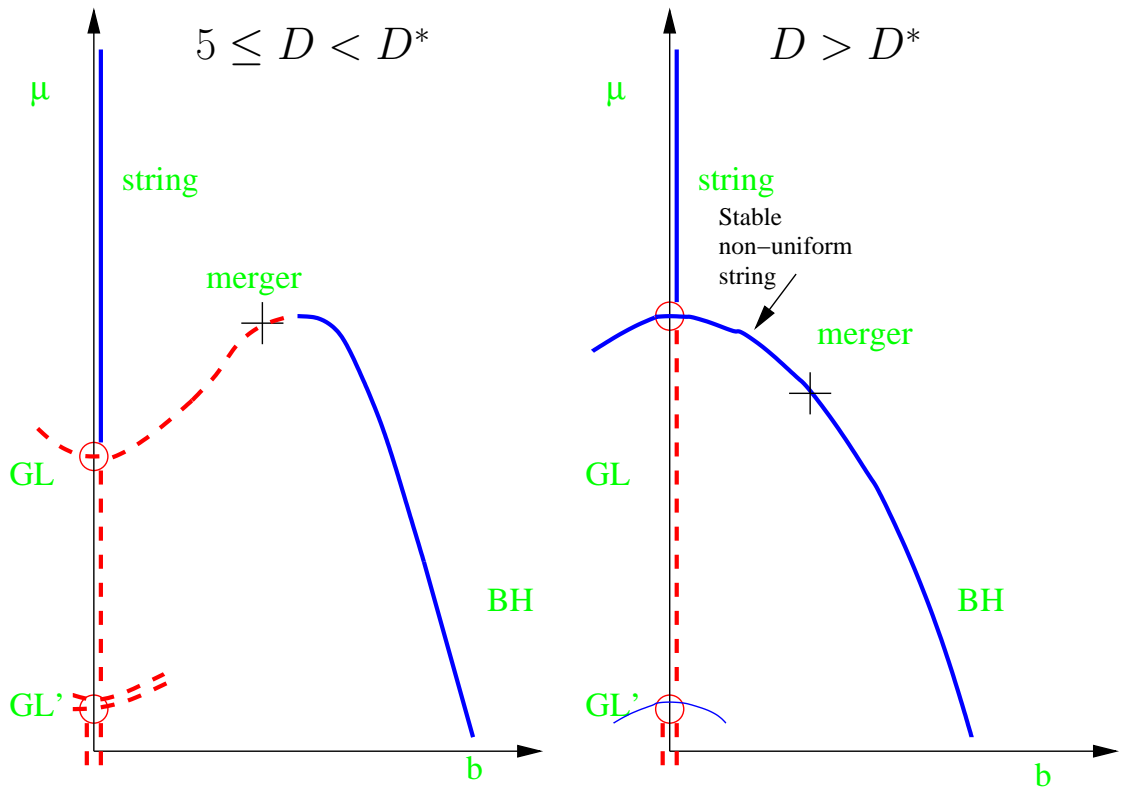}
\vskip-10.3cm
  {\LARGE $5 \le D < D^*$ \hskip4.3cm $D>
D^*$ } \vskip10.3cm \caption[]{The predicted qualitative form of
the phase diagram. The vertical axis is the control parameter
$\mu$ and the horizontal axis is the dimensionless scalar charge,
which plays the role of the order parameter (see the text for
further detail). Solid (blue) lines denote stable phases and
dashed (red) lines denote unstable phases. The ``+'' denotes the
merger point where the black-hole and black-string phases meet in
a topology changing transition (of the Euclidean solutions). Note
the dimensional dependence of the qualitative form due to the
critical dimension. The diagram on the left basically appeared in
\cite{TopChange}.} \label{PhaseDiag}
\end{figure}

We may now assemble all the theoretical input and draw the
predicted qualitative form of the phase diagram for various
spacetime dimensions -- see figure \ref{PhaseDiag}. These figures
apply with small changes to both the micro-canonical ensemble,
which is the usual physical setting where energy is conserved, and
to the canonical ensemble, namely black holes immersed in a heat
bath.

The vertical axis is the dimensionless control parameter $\mu$
defined by either the dimensionless mass $\mu_M$ (\ref{def-mu}) in
the micro-canonical ensemble or by the dimensionless inverse
temperature $\mu_\beta$ (\ref{def-mu-beta}) in the canonical. The
horizontal axis is the order parameter chosen in subsection
\ref{order-param} to be the dimensionless scalar charge (see
especially eq. (\ref{def-ab}) and the next to last paragraph in
that subsection).

The local structure around the GL point is determined by the order
of the transition as discussed in subsection
\ref{transition-order-subsection}: for $D < D^*$ the GL vertex is
first order \cite{Gubser} just like figure \ref{2tran1st}, while
for $D > D^*$ it is a second order vertex just like figure
\ref{2tran2nd}. $D^*$, the critical dimension depends on the
ensemble -- in the micro-canonical case it is $D^*=``13.5"$
\cite{SorkinD*} while in the canonical case it is $D^*=``12.5"$
\cite{KudohMiyamoto}.

Finally, we connect the black hole and black string phases, as
suggested by the central conclusion of subsection
\ref{morse-subsection}, based on Morse theory arguments and
further justified by presenting a novel topology change in
subsection \ref{merger-subsection}.

The diagrams in figure \ref{PhaseDiag} were constructed by
attempting to draw the simplest diagram consistent with the data
and assumptions in the last two paragraphs. The non-trivial nature
of these diagrams is best illustrated by the various other
possibilities that were considered in the literature, see for
example the six scenarios in \cite{HO3}, section 6.

The critical dimension $D^*_{\mbox{merger}}=10$ (\ref{crit-dim})
affecting the stability of the $\IS^m \times \IS^n$ cone in the
critical merger solution is not visible in figure \ref{PhaseDiag}.
According to the picture that emerges from \cite{scaling}, there
is a single branch of non-uniform strings, irrespective of
$D^*_{\mbox{merger}}$ (and being unstable as long as the
transition is first order), but for $D<10$ this solution
approaches a discretely self-similar (DSS) solution near the
singularity, while for $D>10$ it approaches a continuously
self-similar (CSS) solution there. While the qualitative predicted
form is similar, the kink at merger is likely to be different.

Another point to note is the critical point GL' in figure
\ref{PhaseDiag} which is there to remind us that each
(non-uniform) solution has \emph{``harmonies''} or
\emph{``copies''} gotten by trivially fitting  several cycles
inside $L$, namely replacing $L \to L/m$ for any $m \in \IZ_+$ in
the solutions. In particular the GL point has these copies.
However, these copies of the phase diagram are decoupled (except
for their connection with the uniform string) and therefore there
is no need to draw them. See also \cite{HO3}.

\presub {\bf Numerical data}. Clearly, the predicted phase
diagrams in figure \ref{PhaseDiag}, were not proven here, but
rather argued to be the simplest possibility which is consistent
with certain carefully analyzed arguments, some of them not fully
understood yet. As such it suggests to be tested by actually
obtaining these solutions. Indeed, one of the joys of this problem
is the feedback between theory\footnote{The word ``theory'' is
used here to loosely mean all the considerations that one can
apply before any exact solutions are available} and numerics,
which is in many ways like the classical feedback between
experiment and theory which we sorely miss.

\begin{figure}[t!] \centering \noindent
\includegraphics[width=10cm]{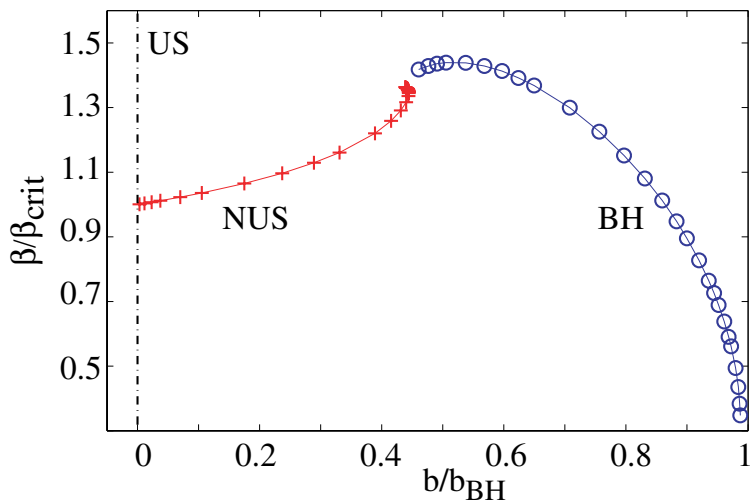}
\caption[]{Numerical data for the phase diagram in 6d. ``US'''
denotes the uniform string branch, ``NUS'' denotes the non-uniform
string branch (data from \cite{Wiseman1}) and ``BH'' is the black
hole (data from \cite{KudohWiseman2}). Axes conventions are the
same as in figure \ref{PhaseDiag} (vertical is $\mu_\beta$ and the
scalar charge is in units of mass rather than temperature).
Adapted with permission from \cite{KudohWiseman2}.}
\label{numeric-phase-diag}
\end{figure}

\begin{figure}[t!] \centering \noindent
\includegraphics[width=8cm]{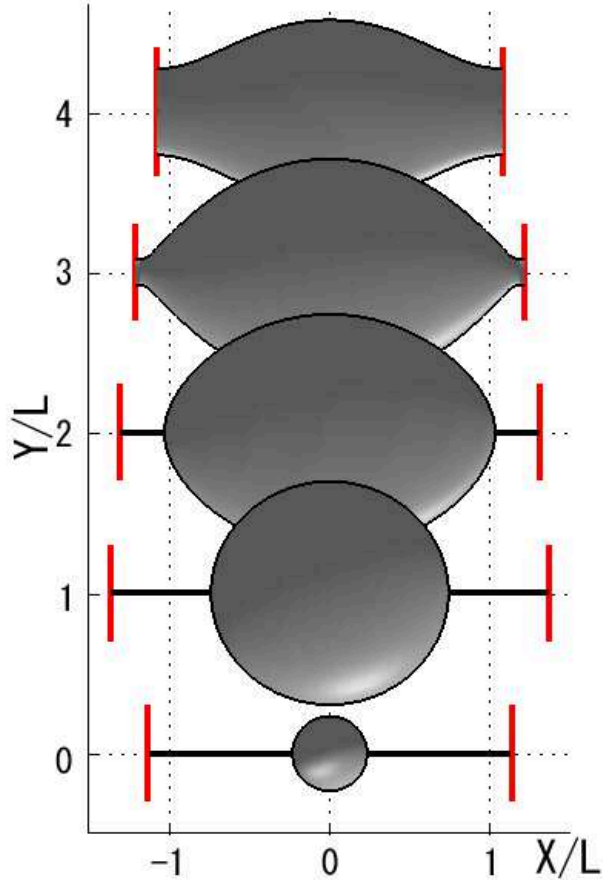}
\caption[]{Embedding diagrams for a sequence of event horizons in
6d, starting with non-uniform strings (top, data from
\cite{Wiseman1}) and passing through the merger transition to
black holes (bottom, data from \cite{KudohWiseman2}). $X$ is the
compact dimension and $Y$ is the radial direction. Note that
although the asymptotic size of the extra dimension, $L$, is kept
fixed, the ``proper'' size (for embedding purposes) changes.
Reproduced with permission from \cite{KudohWiseman2}.}
\label{visual-merger}
\end{figure}

Recently the numerical determination of the phase diagram in 6d
was all but completed \cite{KudohWiseman2} -- see figure
\ref{numeric-phase-diag}. A visualization of the merger transition
through embedding diagrams is shown in figure \ref{visual-merger}.

We see perfect agreement with the predicted diagram in figure
\ref{PhaseDiag} with $D=6$ (which basically appeared already in
\cite{TopChange}), especially regarding the prediction for a
``merger'' of the black hole and the string phases, and the
absence of a stable non-uniform phase (predicted by \cite{HM}). We
view this as a vindication of the picture presented here.
Additional interesting features of the numeric figure are a kink
at merger,\footnote{However, the geometry of the horizon and axis
seems to merge rather smoothly.} perhaps related to the predicted
kink (see the last paragraph in the previous subsection), and the
location of the merger at roughly a local maximum on the diagram.

These diagrams were obtained by combining data from two different
simulations - one for non-uniform strings \cite{Wiseman1} and one
for the black holes \cite{KudohWiseman2}, see also
\cite{KPS2,KudohWiseman1} for previous work. More details on the
challenges overcome in performing these simulations appear in the
next section.


\section{Obtaining solutions}
\label{solutions}

In the previous section we examined the qualitative features of
the phase diagram. Now we turn to the more quantitative aspects,
those required in order to obtain solutions for static black
objects.

\subsection{2d gravito-statics}
\label{gravito-statics}

{\bf Counting degrees of freedom.} The most general metric which
is static\footnotemark[16] 
 and spherically symmetric is \be \label{general-ansatz}
 ds^2 = e^{2\, A}\, dt^2 + ds^2_{(r,z)} + e^{2\, C}
 d\Omega^2_{D-3} ~,\ee
where all functions are defined on the $(r,z)$ plane,
$ds^2_{(r,z)}$ is an arbitrary metric on the plane and since the
metrics are static we might as well work with Euclidean
signature.\footnote{See also the discussion above \ref{Isum}.}

Altogether the problem is defined in the 2d $(r,z)$ plane and the
field content is the metric and two scalars $A,\, C$. That means
that we can write down a 2d action for these fields without
loosing any of the equations of motion. The action is \bea
 S &=& {\beta\, \Omega_{d-2} \over 16 \pi\, G_D}\, \int dV_2 ~e^{A+2C} \cdot \non
 &\cdot& \left[R_2 + (D-3)(D-4)\, e^{-2C} + (D-3)(D-4)\, (\del C)^2 +
 2(D-3)\,(\del A)(\del C) \right] \label{general-action} \eea
where $R_2,~ dV_2:= \sqrt{g_2}\, dr\, dz, ~\Omega_{d-2}$ are the
2d Ricci scalar, the volume element, and the area of the unit
sphere\footnote{See the definition below (\ref{rho-m}).}
$\IS^{d-2}$ (see appendix \ref{action-section} for useful formulae
to determine this and related actions).
 The total number
of fields is $3+2=5$: 3 for the 2d metric and 2 for the two
scalars. Two fields may be eliminated by a choice of coordinates
in the plane which leaves us with \emph{three fields}. As we
proceed we shall review some of the gauges that were used.

If we formally compute the number of ``dynamic'' or ``physical''
degrees of freedom we get a total of $-1+2=+1$: $-1$ is for the
metric degrees\footnote{Attributing $(d-2)(d-1)/2-1$ degrees of
freedom to gravitons in $d$ dimensions.}
 and $+2$ is for the two
scalars. So far nobody succeeded in reducing the problem to a
single field, and it is not clear whether that is possible or not,
but there is a clever ansatz due to Harmark and Obers which
reduces the problem to two fields \cite{HO1}, (see
(\ref{def-R},\ref{HO}) and thereabout) .
 Morally speaking, one may hope that the equations
for the three fields, if not reducible to a single field, could at
least be separated such that first one solves two equations for
``diffeo gauge fields'' and only then a single equation is solved
for the ``physical field''.

Note that if we relax the static requirement the number of degrees
of freedom increases  -- see  subsection \ref{time-evolution}.

\presub {\bf Constraints and boundary conditions}. Today numerical
relativists perform full 3+1 dimensional simulations with some
success. One would think that simulating a static problem, namely
``\emph{gravito-statics}'', would be well-understood by now, but
this turned out not to have been the case. The main conceptual
hurdle which was necessary to cross was the treatment of
constraints and boundary conditions. This problem was solved by
Wiseman in the 2d case \cite{Wiseman1}, as we now describe.

\presub {\it Relaxation and electro-statics}. Since Newtonian
gravito-statics is equivalent to electro-statics it is useful to
recall the method there. In electro-statics one wishes to find the
electro-static potential $\Phi(x)$ defined over some domain,
satisfying \bi
 \item  The Poisson equation \be
 \triangle \Phi = -4\, \pi\, \rho ~, \label{poisson} \ee
 where $\rho$ is the given charge density distribution
 \item Boundary conditions: Dirichlet, Neumann or some mix. \ei

A successful numerical algorithm to solve this problem is the
relaxation method. This method is very physical in the sense that
it has some similarities with the way in which an excited field
settles down or ``relaxes'' as a function of time to a static
solution. In the relaxation method one chooses a grid, consisting
of points $x_{i,j}$, and then one starts with an initial field
configuration $\Phi^{(0)}_{i,j} \equiv \Phi^{(0)}(x_{i,j})$ which
satisfies the boundary conditions. At each step $\Phi$ is modified
according to a local rule to create a sequence $\Phi^{(m)}$ which
converges to the solution as $m \to \infty$. More specifically,
once chooses a discretization of the Laplacian and solve for
$\Phi_{i,j}$. For example, if one uses a square grid with spacing
$h$ and chooses the following discretization \bea
 \( \triangle \Phi \)_{\mbox{disc}} &=& {4 \over h^2} \[ \left< \Phi \right> - \Phi \] \non
 \left< \Phi \right>_{i,j} &:=& {1 \over 4} \left[ \Phi_{i+1,j}+ \Phi_{i-1,j} + \Phi_{i,j+1} +
 \Phi_{i,j-1} \right] ~, \label{Laplace-discrete} \eea
 then after discretizing (\ref{poisson}) using (\ref{Laplace-discrete}) and solving
for $\Phi_{i,j}$ we find \be
 \Phi_{i,j}^{(m+1)}= \left< \Phi \right>_{i,j}^{(m)} - \pi \, h^2\, \rho_{i,j}
 \label{relax1} ~.
 \ee

Actually, one can go further and introduce a ``relaxation speed
parameter'', $\omega$,  defined by  \bea
 \Phi^{(m+1)} &=& \Phi^{(m)} + \omega\; \delta \Phi \non
  \delta \Phi &=& \left< \Phi \right>_{i,j}^{(m)} - \Phi^{(m)}- \pi \, h^2\, \rho_{i,j} \label{omega-def} \eea
such that $\omega=1$ corresponds to the rule (\ref{relax1}), while
$\omega>1$ is called ``over-relaxation'' and $\omega<1$ is called
``under-relaxation''. Clearly, the solution is a fixed point of
the process, irrespective of the value of $\omega$. Its importance
lies in changing the convergence properties. For some interval of
$\omega$ containing $\omega=1$ the process is guaranteed to
converge since at each step the energy is reduced and the solution
is a unique and global minimum of the energy. In this range
$\omega$ may be adjusted for convergence speed.

\presub {\it Gravito-statics}. Similarly to electro-statics,
General Relativity allows for relaxation. In our case there are 5
equations of motion, and after fixing the gauge the equations are
split to 3 equations of motion and 2 constraints (``gauge fixing
constraints''). A convenient and quite natural gauge choice is the
``conformal'' gauge \be
 ds^2 = e^{2 \, A}\, dt^2+ e^{2\, B}(dr^2+dz^2)+e^{2\,
 C}\,d\Omega^{~2}_{d-2}
\label{conformal-gauge} ~. \ee
 The action and constraints in this gauge are as follows. The action is given by \be
 S ={\beta\, \Omega_{d-2} \over 16 \pi \, G_D}\,  \int dr\,
 dz\, e^{\Psi} \[ K_{\al \beta} ~ \del_i \Phi^\al ~ \del_i \Phi^\beta +
 (D-3)(D-4)~ e^{2(B-C)}\] ~,\ee
where \bea
 \Psi &:=& A + (D-3)C \non
 \Phi^\al &:=& \[ \begin{tabular}{c}
  A \\ B \\ C  \end{tabular} \] \non 
 K_{\al \beta} &:=&  \[ \begin{tabular}{ccc}
  0 & 1 & D-3 \\
  1 & 0 & D-3 \\
  D-3 & D-3 & (D-3)(D-4) \end{tabular} \]  ~.\eea
To express the constraints compactly it is convenient to define
\be
 (Cnsr)_{ij}:=\del_i \del_j \Psi + \del_i \Psi ~ \del_j \Psi
 -K_{\al \beta} ~ \del_i \Phi^\al ~ \del_j \Phi^\beta ~, \ee
in terms of which the constraints are \bea
 0 &=& \(\sigma^1\)^{ij}\, (Cnsr)_{ij} \non
 0 &=& \(\sigma^3\)^{ij}\, (Cnsr)_{ij} ~,\eea
where $\sigma^{1,3}$ are the usual Pauli $\sigma$-matrices \be
\sigma^1 :=
\[ \begin{tabular}{cc}
  0 & 1 \\ 1 & 0 \end{tabular} \]
 ~~~~~~~~
\sigma^3 := \[ \begin{tabular}{cc}
  1 & 0 \\ 0 & -1 \end{tabular} \]~.
 \ee

In this gauge Wiseman \cite{Wiseman1} was able to formulate
gravito-statics as a relaxation problem. The 3 equations of motion
are elliptic of the form \be
 \triangle X = Src \ee
where $X$ is any of the fields $A,B,C$,
$\triangle=\del_{zz}+\del_{rr}$ is the flat space Laplacian, and
$Src$ are some non-linear functions of the fields. As such we
would like to subject them to relaxation with some b.c. at
infinity and at the horizon. Normally 3 elliptic equations require
three boundary functions as data. However, the analogy with
electro-statics, where one needs to specify only the
electro-static potential on the boundary, leads us to expect the
b.c. to consist of a single function. 
So the question is how to determine the b.c. Another problem is
how to guarantee that the remaining two constraints, which are
hyperbolic in this case, are satisfied as well.

The problems of b.c. and constraints are solved simultaneously by
noting the constraints' Bianchi identities. In this specific
ansatz the constraints are $G_{rz}$ and $G_{rr}-G_{zz}$, where $G$
is the Einstein tensor. The Bianchi identities are \bea
 \del_r (\sqrt{g}\ G^r_z) + \del_z \( {\sqrt{g} \over 2}(G^r_r-G^z_z)\)&=&0 \non
 \del_z (\sqrt{g}\ G^r_z) - \del_r \( {\sqrt{g} \over 2}(G^r_r-G^z_z)\)&=&0 ~,\eea
where $g=\det g_{\mu\nu}$ and one uses the Einstein tensor with
mixed upper and lower indices. These are Cauchy-Riemann equations
and hence \be
 G := \sqrt{g}\, \( G^r_z - {i \over 2}(G^r_r-G^z_z)\, \) \ee
 is \emph{analytic in the $r+i\, z$ variable}. For an analytic
function to vanish in a domain, we know that it is exactly enough
to impose that its real part vanishes on the boundary and that its
imaginary part vanishes at a point. Alternatively, one may choose
and arbitrary $\al$ and impose the boundary constraint
$\mbox{Re}(e^{i \al}\, G)=0$ together with $\mbox{Im}(e^{i \al}\,
G)=0$ at a single boundary point. It turns out that \emph{these
boundary conditions chosen by the constraints' Bianchi
identities}, do not only \emph{guarantee the constraints} but are
also exactly \emph{sufficient for the elliptic problem}. Note that
altogether the b.c. consist of a single function, as expected from
the analogy with electro-statics (plus a function at one point),
and that there is some freedom in specifying the b.c. (such as
choosing $\al$) which is analogous to the choice of Dirichlet/
Neumann.

Unlike electro-statics the action is neither positive-definite nor
is it quadratic in the fields and therefore convergence is not
guaranteed. In practice, however, this method performs quite well
(see also the ``convergence'' part in subsection
\ref{numerical-issues}).

Since the Cauchy-Riemann equations are special to 2d it is not
obvious how to generalize this structure to higher dimensional
gravito-statics, which is required in order to solve for black
holes in backgrounds with higher dimensional compact manifolds
($p>1$). One may speculate though, that the general rule is the
same as the last emphasized sentence.

Another issue of boundary conditions, which is dealt with in the
conformal gauge (\ref{conformal-gauge}), is fixing the boundary.
In GR, due to the dynamic nature of the metric one cannot
anticipate the final location of the horizon. However, it is
normally best for numerics to fix the location of the horizon in
coordinate space. The conformal gauge allows for analytic
coordinate change as residual gauge. It was shown that this
freedom exactly suffices in order to fix the horizon to be a
circle in the $(r,z)$ plane, by writing down elliptic equations
for the coordinate transformation \cite{KudohTanakaNakamura}. The
circle's radius, $\rho_h$, is a free parameter that determines the
size of the black hole (``the holomorphic invariant of the
domain'').

\presub {\bf Harmark-Obers coordinates}

By choosing the coordinates with some physical logic Harmark and
Obers found an ansatz with 2 rather than 3 fields \cite{HO1}. The
ansatz was constructed for black hole solutions, it explicitly
fits the uniform string, in \cite{KudohMiyamoto} it was used for
the perturbative analysis around the GL point, but it was not used
in a relaxation algorithm so far.

\begin{figure}[t!] \centering \noindent
\includegraphics[width=7cm]{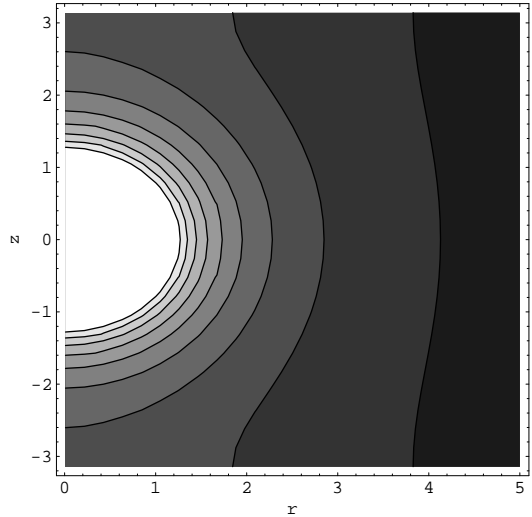}
\caption[]{The equipotential lines of the Newtonian potential
around a point-like source in 5d drawn in the $(r,z)$ plane. Note
the change in their topology from spheres to cylinders.}
 \label{Newtonian-figure}
\end{figure}

\cite{HO1} start by defining orthogonal coordinates $(R,v)$ over
the $(r,z)$ plane such that they interpolate between spherical
coordinates near the origin and cylindrical coordinates
asymptotically. To that purpose $R$ was defined through
$\Phi_N(r,z)$, the Newtonian potential of a point source at the
origin \be
 \Phi_N := \rho_{0}^{D-3}\sum_{n=-\infty}^{\infty} \frac{1}
{\left(r^{2}+(z+nL)^{2}\right)^{\frac{D-3}{2}}},
\label{thephi}\,~\ee
 (see figure \ref{Newtonian-figure})
 \footnote{\cite{HO1} find an explicit expression for $\Phi_N$ by separating
variables and expressing the radial functions in terms of modified
Bessel functions of the second kind.} ,
 by\footnote{ \cite{HO1} normalize $R$ such that $R \simeq 2 \pi\, r/L$ for $r \gg L$.} \be
 \({R_0 \over R}\)^{d-3} \propto \Phi_N ~. \label{def-R} \ee
$v$ is defined to be an orthogonal coordinate of period $2\pi$ and
it can be parameterized such that near the origin is approaches
$\chi$ (see figure \ref{coordinates}) while asympototically it
approaches $z$. This is achieved by the system \bea
 \del_r v &=& k\, r^{d-2}\,  \del_z \Phi_N \non
 \del_z v &=& -k\, r^{d-2}\,  \del_r \Phi_N ~,\eea
 where $k$ is some constant and the system is integrable since $\Phi_N$
is harmonic. We see here that this construction is special to 2d,
and it is not clear whether it can generalized to higher
dimensions.

\begin{figure}[t!] \centering \noindent
\includegraphics[width=10cm]{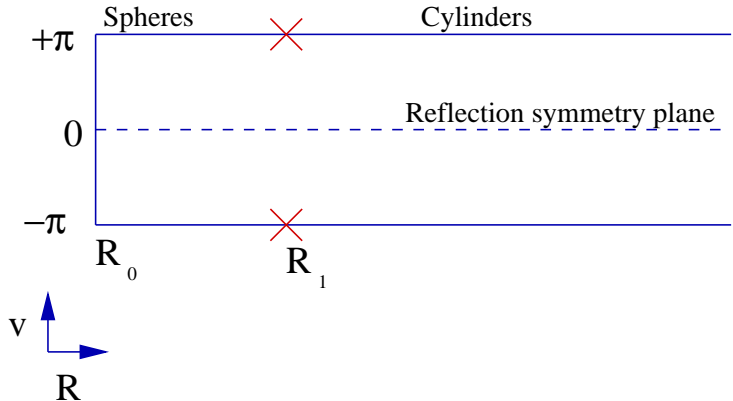}
\caption[]{The Harmark-Obers coordinates $R,v$ \cite{HO1}. $R$ is
a function of the Newtonian potential and $v$ is periodic with
period $2\pi$. The $x$ marks the point $(R,v)=(R_1,\pi)$ which is
a singular point of the coordinate transformation. For $R<R_1$
equal-$R$ surfaces are spheres while for $R<R_1$ they are
cylinders. The size of the black hole is changed by adjusting
$R_0$.} \label{HOcoord-figure}
\end{figure}

The $(R,v)$ plane is drawn in figure \ref{HOcoord-figure}. It is a
semi-infinite cylinder with one marked point (denoted by x), which
is $(r,z)=(0,L/2)$, where the equipotential surfaces turn from
spheres to cylinders and the differential of the transformation
from $(r,z)$ to $(R,v)$ vanishes. In these coordinates $R_0$, the
location of the horizon, is a free parameter that determines the
size of the black hole, while the location of the marked point
remains fixed at some $R_1$. As the black hole grows $R_0$ reaches
$R_1$ and from then on the same equations should generate string
solutions rather than BHs (for some appropriate boundary
conditions).

In the $(R,v)$ coordinates it turns out that two fields $A,\, K$
suffice (rather than three), according to the following ansatz \be
\label{HO}
 ds_{d+1}^2 = f\, dt^2 +
 \({L \over 2\pi}\)^2 \left[ f^{-1}\, A\, dR^2
+ \frac{A}{K^{d-2}}\, dv^2 + K\, R^2\, d\Omega_{d-2}^2 \right] \ee
where \be
 \label{def-f} f = 1 - \frac{R_0^{d-3}}{R^{d-3}} \ . \ee
The justification for this ansatz was demonstrated (in 6d)
\cite{Wiseman2} by showing that for any solution the equations for
a change of coordinates to the ansatz (\ref{HO}) are
self-consistent. The proof was then generalized in \cite{HO3} to
any dimension. Another way to see that reduction is to write a
fully general ansatz by adding a third field $C$, say as the
coefficient of $dv^2$, and then one finds that the $G_{tt}$
constraint gives an algebraic relation between the three fields
\cite{KolMartinez-unpublished}. Yet, it is not clear how the
reduction of fields could have been anticipated and whether there
are other ansatzs with this same property.

A further reduction is possible, expressing $A$ in terms of $K$
and its second derivatives (eq (6.6) of \cite{HO1}). However, this
step makes the equations of order higher than second, and hence
unamenable to relaxation (actually even the equations for both
$A,K$ were not brought to relaxation form so far).

\subsection{Numerical issues}
\label{numerical-issues}

We turn to a brief overview of the numerical implementation issues
of our gravito-statics problem.  The first simulation of the
system was \cite{Wiseman1} which found the correct prescription
for b.c. and constraints and applied it to obtain solutions for
non-uniform strings including highly non-uniform ones very close
to the merger point. That work benefitted from insights gained
from works on the ``black hole on a brane'' problem such as
\cite{Wiseman0}. A simulated black hole on a brane appeared in
\cite{KudohTanakaNakamura} while the first simulation for caged
BHs appeared simultaneously in \cite{KPS2} (5d) and
\cite{KudohWiseman1} (6d). However, due to convergence problems
large black holes were not possible to obtain. These problems were
largely overcome in \cite{KudohWiseman2} (5d \& 6d) whose figures
\ref{numeric-phase-diag},\ref{visual-merger} summarize the state
of the art.

\presub {\bf Implementation}. The most important decisions for
implementing the numerics are the choice of coordinates, fields
and grid. So far {\it the coordinates and fields} were chosen in
accordance with the conformal gauge (\ref{conformal-gauge}), with
variations intended to extract the singular and asymptotic
behavior for smoother numerics. For example, the ansatz used in
\cite{Wiseman1} for 6d strings was \be
 ds^2 = {r^2 \over G_5 m + r^2}\, e^{2 \, A}\, dt^2+ e^{2\,
 B}(dr^2+dz^2)+e^{2\, C}\,(G_5 m+r^2)\, d\Omega^{~2}_3 ~.\label{Wiseman1-ansatz} \ee
In \cite{KPS2} the implementation proved to be very sensitive to a
redefinition of fields, for some unclear reason, the main obstacle
being a simple linear redefinition of fields of the form
$B'=B+C,\, C'=B-C$. The more efficient Harmark-Obers coordinates
were not implemented so far and it would be interesting to do so.

When choosing a grid one should consider the following factors:
the number of nearest neighbors for each point, the density of
points and multi-grid issues. Hexagonal lattices are particulary
suited for the solution of the Laplace equation in 2d, because one
can first use some relaxation method just to distribute the grid
points according to a prescribed grid density (to allow for
variable grid spacing), and then weights for the discretized
Laplacian are uniquely determined from the points' location.
However, for convenience all simulations so far involved square
grids and the density of points was sometimes adjusted by using
the mapping from grid space to coordinate space. For instance in
\cite{KPS2} two grids were used: in the near zone the grid was
evenly spaced in $\rho$ and $\cos(\chi)$ (see figure
\ref{coordinates} for coordinate definition) while in the
asymptotic zone it was evenly spaced in $\log(r),z$. The need for
two grids came from the need for different densities of points in
the two regions. The price to pay is in the complexities of grid
matching. Another grid tool employed in \cite{KPS2} was the
``multi-grid'' where the problem was relaxed successively on finer
and finer grids, in order to accelerate convergence.

Another, less critical choice, is the choice of discretization of
the differential operators (such as (\ref{Laplace-discrete})).

\presub {\bf Convergence}. Having implemented the numerical
scheme, one typically presses ``enter'' and prays for convergence.
To achieve it one is free to use the ``convergence speed''
parameter $\omega$ (\ref{omega-def}). Often this does not suffice
and we enter the domain of black magic. One difficult problem that
arose from an innocent redefinition of the fields was already
described.

Another issue involves non-convergence as a result of problems
near the exposed $r=0$ axis (so it does not apply to strings).
This problem is known to be solved effectively by the following
``Wiseman trick'' \cite{Wiseman0}. The field $B$ appears in source
terms which are sensitive to errors near the exposed axis and
generate instabilities. To cure that one replaces $B$ by a
numerically distinct variant, $B2$, which is meant to be more
accurate than $B$ near the axis and is gotten by integrating a
constraint from the axis outward.

More generally, one expects to have a relation between physical
stability of a solution and convergence. However, the method
converges nicely for non-uniform strings which are believed to be
unstable. The convergence is the result of boundary conditions at
the horizon which force the solution to be non-uniform and which
perhaps can be described as exerting some ``pressure'' which
stabilizes the string also in the physical sense. It would be nice
to understand this issue better.

\presub {\bf Tests}. The results of the simulations must be
tested. The following tests were employed \bi
 \item Convergence.
 \item Constraints.
 \item Integrated first law.

This law (\ref{integrated-first}) relates $m,\,  \tau$ which are
measured at infinity with $\beta,\, S$ measured at the horizon. As
such it provides a strong overall test of the numerics, and it is
satisfying to find that it holds. However, note that the first law
does not test the constraints \cite[appendix C]{KudohWiseman1}.

 \item Comparison with analytic results.

\begin{figure}[t!] \centering \noindent
\includegraphics[width=7cm]{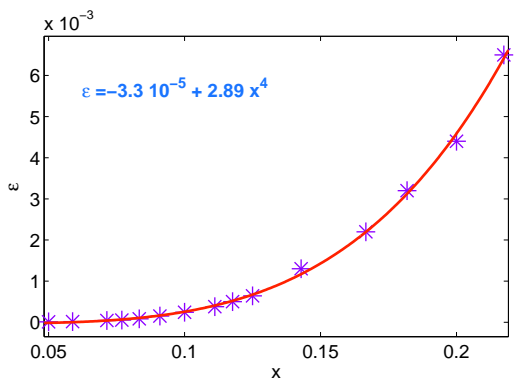}
\caption[]{5d Black hole eccentricity as a function of black hole
size. Simulations and perturbative analytic expressions agree. The
eccentricity $\eps$ is defined in (\ref{def-eps}). Reproduced from
\cite{KPS2}.} \label{eccentricity-figrue}
\end{figure}

\begin{figure}[t!] \centering \noindent
\includegraphics[width=12cm]{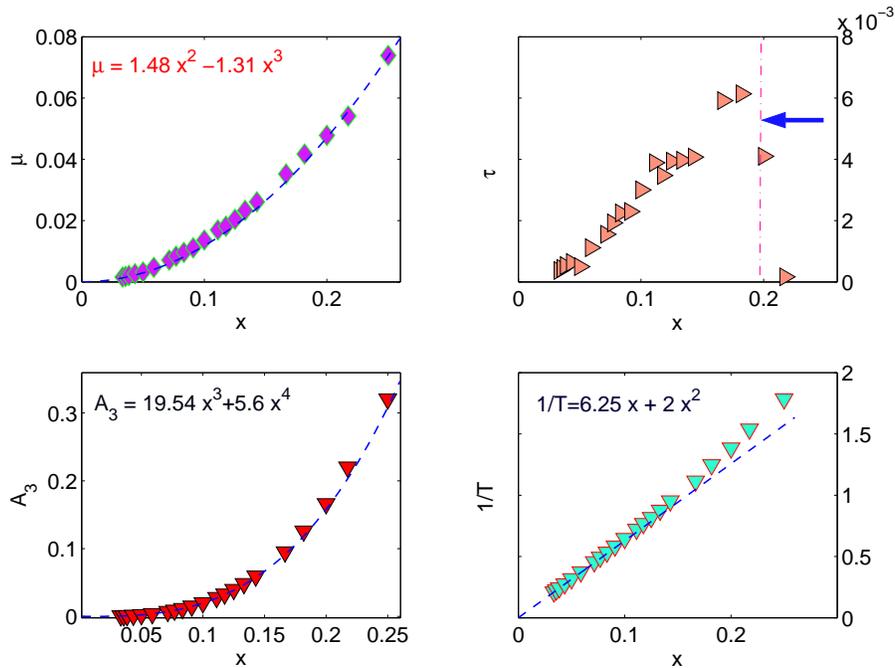}
\caption[]{Thermodynamic quantities, $\mu,\, \tau,\, A_3,\, 1/T$
which are the mass, tension, area and inverse temperature,
respectively, of a 5d black hole as a function of its size.
Simulations and perturbative analytic expressions agree, except
for the tension results which are less reliable. Reproduced from
\cite{KPS2}.} \label{BHthermodynamics-figure}
\end{figure}

\begin{figure}[t!] \centering \noindent
\includegraphics[width=7cm]{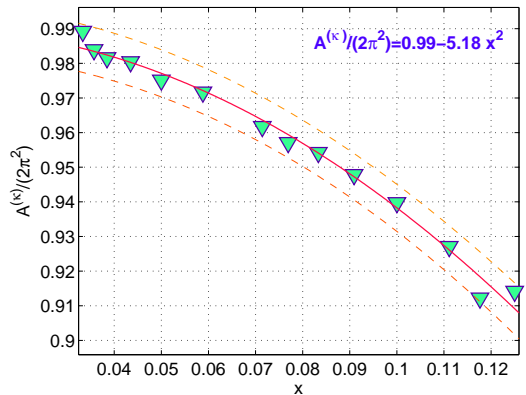}
\caption[]{The area in units of the surface gravity, $A^{\kappa}$,
as a function of 5d black hole size. This quantity measures a
correction to the temperature due to the non-zero potential at the
origin from the image BHs. Simulation and perturbative analytic
expression agree. Reproduced from \cite{KPS2}.}
\label{Akappa-figure}
\end{figure}

\begin{figure}[t!] \centering \noindent
\includegraphics[width=7cm]{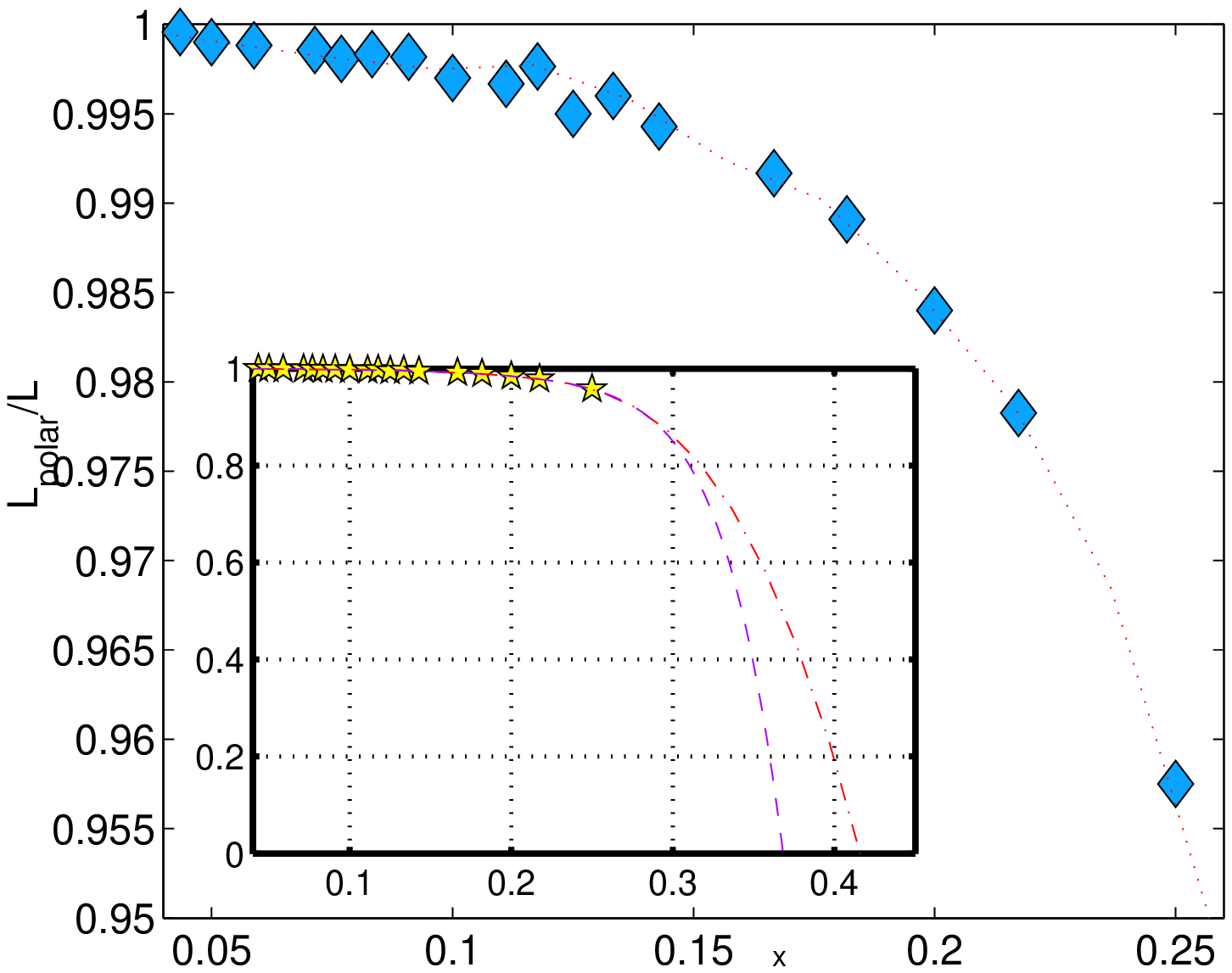}
\caption[]{The inter-polar distance (going around the compact
dimension), defined in figure \ref{Lpoles-fig}, as a function of
black hole size in 5d. The line denotes a best-fit curve. Its
quartic nature is explained in \cite{dialogue2}, though a
theoretical prediction of the prefactor is not presently
available. Reproduced from \cite{KPS2}.} \label{Lpoles-figure}
\end{figure}

Some analytic results for small black holes (see subsection
\ref{dialogue}) allow testing of the numerical data. \emph{The
satisfying agreement is demonstrated in figures
\ref{eccentricity-figrue}, \ref{BHthermodynamics-figure},
\ref{Akappa-figure}, \ref{Lpoles-figure}} taken from \cite{KPS2}.
In all figures the size of the BH is measured by $x:=2\, \rho_h/L$
where $\rho_h$ is the horizon radius in conformal coordinates and
this ``numerical'' definition of $x$, valid in this subsection,
should not be confused with a different ``analytic'' definition of
the small parameters $x$ valid in subsection \ref{dialogue}.
Stars, diamonds or triangles represent actual simulation results
while the curve represent smooth interpolations which coincide
with the theoretical predictions of \cite{dialogue} within the
simulation precision. \ei

\subsection{Time evolution}
\label{time-evolution}

As remarked in subsection \ref{issues}, ``issues'', the time
evolution of the unstable system raises deep questions regarding
singularities. Thus there is ample motivation for a dynamical
numerical simulation. On the other hand, we expect any such
simulation to crash before the singularity is reached. So in order
to achieve progress it is probably necessary to have a theoretical
local model for the time evolution, which could then be tested and
supported by a numerical simulation.

\begin{figure}[t!] \centering \noindent
\includegraphics[width=6cm]{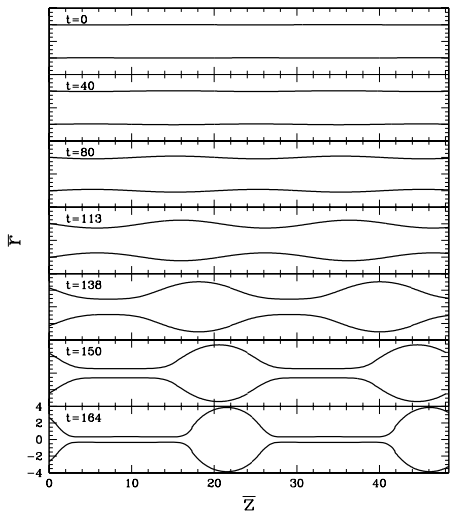}
\caption[]{A time sequence of embedding diagrams for the event
horizon of a decaying black string in 5d, from a numerical time
evolution. Initially, the string is nearly uniform (top), and by
the time the simulation comes to a stop due to grid stretching
(bottom), the horizon becomes that of a black hole with a thin
pipe connecting the poles (going around the extra dimension).
Reproduced with permission from \cite{CLOPPV}.}
\label{time-evolve-fig}
\end{figure}

Thus far a single time evolution simulation was performed:
\cite{CLOPPV} simulated the decay of a sub-GL uniform string in
5d. The main features of the evolving spacetime can be read from
figure \ref{time-evolve-fig}. The initial configuration is an
unstable string of radius $r_0$ on top of which the unstable
perturbation grows in time. The region where the horizon moves
radially inward (the ``waist'') collapses fast, stretching at the
same time in the $z$ direction (growing $g_{zz}$) until the
minimal (areal) radius reaches about $0.15\, r_0$ when the grid
stretching is so large that the simulation cannot proceed. In the
region where the horizon grows (the ``hip'') the metric approaches
the metric of a 5d BH with a comparable mass, so that the maximal
(areal) radius at the horizon is about $2\, r_0$.

Opinions on the interpretation of these results differed. The
author of this review interprets them as \emph{strong evidence
against the formation of a stable non-uniform string end-state}.
The authors of \cite{CLOPPV} are much more careful and would only
say that the results ``are not inconsistent with Gregory and
Laflamme's conjecture that the solution bifurcates into a sequence
of black holes ... [On the other hand it is] not necessarily
inconsistent with \cite{HM}... At the same time, a continuation of
the observed trend would argue against achieving a stationary
solution with a mild dependence on the string dimension''.

Later \cite{GLP}  presented a re-analysis of the metric obtained
in \cite{CLOPPV} which somewhat clarifies the picture. \cite{GLP}
observed that the affine parameter on the horizon grows extremely
fast, faster than the exponential of asymptotic time, and
suggested that the horizon might pinch off in infinite affine
parameter. This possibility was mentioned in \cite{HM} but argued
against.

In light of the critical dimension $D^*=``13.5"$ it would be very
interesting to run the simulation again for a range of dimensions,
and to test/ confirm that for $D \ge 14$ the unstable string
settles down quickly to a slightly non-uniform string.

{\bf Method}. Let us count the number of degrees of freedom, which
is of course larger than in the static case.
 The domain is the 3d space parameterized by $(r,z,t)$ and the fields
are the metric and the scalar which is roughly $g_{\theta\theta}$
or $C$ in (\ref{general-ansatz}). The total number of fields is 7.
After choosing some gauge in the 3d domain the number of fields
reduces to 4 (compared to 3 in the static case). Finally, the
``physical'' number of degrees of freedom is a total of $0+1=1$
(just like the static case): none for the graviton and +1 for the
scalar.

The most general time-dependent metric is
\begin{eqnarray}\label{metric}
ds^2=(-\alpha^2+\gamma_{AB} \beta^A \beta^B) dt^2 + 2 \gamma_{AB}
\beta^A dx^B dt + \gamma_{AB} dx^A dx^B + \gamma_{\Omega}
d\Omega^2 ~,
\end{eqnarray}
where $x^A=(r,z)$, and $d\Omega^2$ is the 2-spherical line element
with coordinates chosen orthogonal to the $t=$ constant
congruences (hence there is no shift corresponding to angular
directions). All metric components in (\ref{metric}) are defined
in the 3d domain. \cite{CLOPPV} choose the gauge \bea
 \al &=& \al_{St} \non
 \beta^z &=& 0 \non
 g_{\theta\theta} &=& r^2 ~,\eea
where the string metric function is read from the ingoing
Eddington-Finkelstein form  \be \label{st_metric}
 ds^2_{St}= -(1-2M/r)\,
d\tilde{t}^2 + 4\, M/r\, dr\, d\tilde{t} + (1+2M/r)\, dr^2 + dz^2
+ r^2\, d\Omega^2 \, . \ee They also comment that ``In a
preliminary version of our code, we also required that
$\beta^r=\beta_{\rm BS}^r$. This, however, caused a coordinate
pathology to develop at late times during the evolution of
unstable strings---specifically, some regions of the horizons
approached a zero coordinate-radius, while maintaining {\em
finite} proper radius.'' Thus the condition on $\beta^z$ was
replaced by one on $g_{\theta \theta}$.

Another hurdle turned out to be the boundary conditions at
infinity. To eliminate cut-off problems spatial infinity was
brought to a finite point by replacing $r$ by $r/(r+1)$.

\subsection{Analytic perturbation method}
\label{dialogue}

While we do not know to write the black hole metric in closed
form, metrics for small black holes can be well-approximated
everywhere: for $\rho \ll L$ the $D$ dimensional black hole is a
good approximation, while for $\rho \gg \rho_0$ ($\rho_0$ is the
\Schw radius) the Newtonian approximation is good. Moreover the
two approximations have an arbitrarily large overlap in the small
black hole limit. Therefore, one expects that the black hole
metric can be systematically expanded in a perturbation series
with a small parameter being $x:=\rho_0/L$.\footnote{This
``analytic'' definition of $x$, valid in this subsection, should
not be confused with a different ``numerical'' definition of the
small parameters $x$ valid in subsection \ref{numerical-issues}.}
 Such a perturbative method was studied in several papers:
 the general procedure and first order results were given in
\cite{dialogue} (pre-announced in \cite{KPS1}), the full second
order in 5d was obtained in \cite{KSSW1} and for a general
dimension the Post-Newtonian order was performed in
\cite{dialogue2}. A different, one zone approximation was given in
\cite{Harmark4}.

The objectives here are to \bi
 \item Provide tests for numerics.
 \item Extrapolate to the phase transition region. \ei

\begin{figure}[t!] \centering \noindent
\includegraphics[width=8cm]{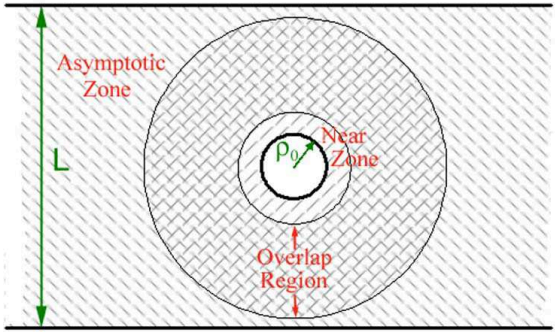}
\caption[]{The metric for small black holes can be obtained
through a ``dialogue of multipoles'' matched asymptotic expansion.
The two zones are the asymptotic zone $\rho \gg \rho_0$ and the
near zone $\rho \ll L$. The smaller the black hole the larger is
the overlap region between the two zones.} \label{zones}
\end{figure}

\begin{figure}[t!] \centering \noindent
\includegraphics[width=12cm]{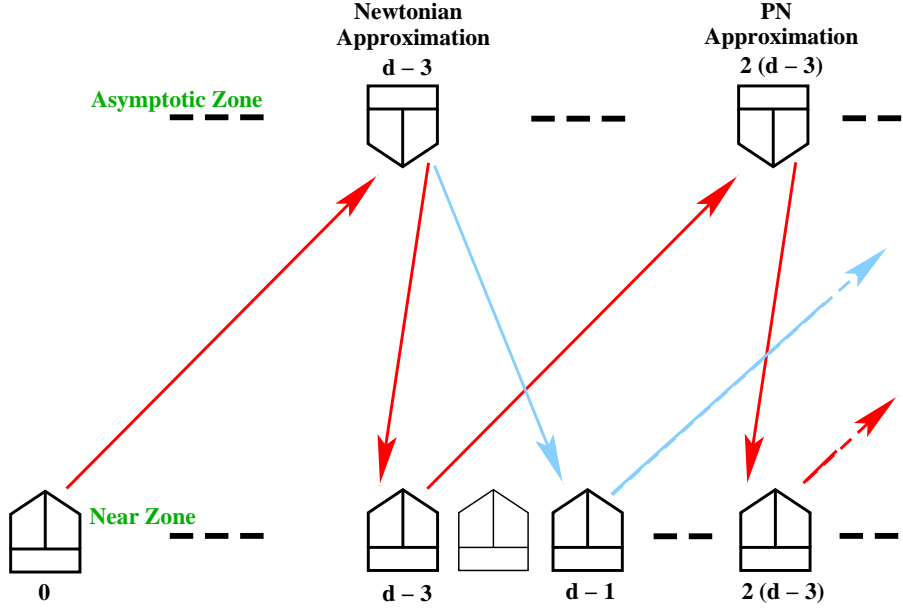}
\caption[]{The ``perturbation ladder'' for the ``dialogue of
multipoles'' matched asymptotic expansion. The upper row depicts
the asymptotic zone and the lower row the near zone. Each box
denotes a certain order in the perturbation series -- in the
asymptotic zone one counts orders of $\rho_0$ and in the near zone
orders of $L^{-1}$. The arrows denote the flow of information
between the two zones, in supplying boundary conditions through
matching. The figure shows the general pattern for an arbitrary
spacetime dimension $D$ ($d$ in the figure is denoted by $D$ in
this review). Reproduced from \cite{dialogue}.} \label{ladder}
\end{figure}

\presub {\bf Method.} It is not possible to use here the ``usual''
perturbation method, the one where a ``zeroth order'' solution is
deformed order by order to follow the deformation of a small
parameter of the problem, since here the domain of coordinates
changes with $x$, and since we do not have a zeroth order
solution. However, one can use the well-known technique of
``matched asymptotic expansion''. In \cite{dialogue} this
technique was applied by defining two zones: an ``asymptotic
zone'' and a ``near horizon zone'' (see figure \ref{zones}), which
have a large overlap in the limit $x \to 0$. The metric is solved
perturbatively in each zone, with boundary conditions coming from
matching with the other zone. The need for matching produces an
intricate (and dimension dependent) pattern of crossings between
the various orders in the two zones -- ``the perturbation ladder''
-- see figure \ref{ladder}. Effectively the gravitational field
produced by the images changes the shape of the BH, or its mass
multipoles, and that in turn back-reacts and changes the field
multipoles. This procedure is much simpler than the usual dynamic
matched asymptotic expansion on account of being static, and was
therefore given a special name  \emph{``a dialogue of
multipoles''} \cite{dialogue}.

{\it The zeroth order}.  In the asymptotic zone the zeroth order
metric is flat space with a compact dimension and the origin
removed, while in the near zone it is the \Schw black hole, with
the periodicity $L$ being far away and invisible.

{\it The Newtonian approximation}. The first order analysis in the
asymptotic zone is easy -- it is just the Newtonian approximation.
Here one considers the metric in the weak field limit \be
 g_{\mu\nu} = \eta_{\mu\nu}+h_{\mu\nu}\, , \ee
 where $\eta_{\mu\nu}$ is the flat space (Minkowski) metric and $h_{\mu\nu} \ll
 1$ is a small correction. One defines  \be
\bh_{\mu\nu}:= h_{\mu\nu}-\frac{1}{2}
h^{\alpha}_{\alpha}\,\eta_{\mu\nu}\, , \label{hbar-h} \ee
 in terms of which the harmonic gauge\footnote{Also known as the De-Donder or Lorentz gauge.}
is chosen \be
 \del^\mu \bh_{\mu\nu}=0 \,. \ee
The linearized field equations read
\begin{equation}
\frac{1}{2}\, \Box \bar{h}_{\mu\nu} = G_{\mu\nu} = 8
 \pi\, G_D\, T_{\mu\nu} \,,
\end{equation}
where $\Box$ is the flat space D'alambertian. Solving
$\bh_{\mu\nu}$ for weak and slowly moving sources where the only
non-zero component of $T_{\mu\nu}$ is taken to be $T_{00}$, and
inverting (\ref{hbar-h}) through \be
 h_{\mu\nu} =
\bh_{\mu\nu} -\frac{1}{D-2} \bh^\alpha_\alpha\, \eta_{\mu\nu}\,,
\label{h-hbar} \ee
 we find
\begin{eqnarray}
h_{tt}&=&\Phi_N, \non
h_{ij}&=&\frac{1}{D-3}\,\Phi_N\,\delta_{ij}, \label{Newton2}
\end{eqnarray}
where the Latin indices stand for the spatial components and
$\Phi_N$ is a the Newtonian potential
(\ref{thephi}).\footnote{Possibly up to a constant pre-factor,
depending on conventions.}
 In the $D=5$ case there is a useful way to
perform the summation in the definition of $\Phi_N$
(\ref{thephi}), yielding
\begin{equation}
\Phi_N=\rho_{0}^2\, {\pi \over L\, r}\,
\frac{\sinh\left(\frac{2\,\pi\,r}{L}\right)}
{\cosh\left(\frac{2\,\pi\,r}{L}\right)-\cos\left(\frac{2\,\pi\,z}{L}\right)}\,.
\end{equation}

{\it Linear perturbations around \Schw}. The first correction to
the near zone requires more work than the Newtonian approximation
in the asymptotic zone -- it is a generalization to higher $D$ of
the well-known paper by Regge and Wheeler \cite{ReggeWheeler}. The
metric is given in terms of a single function $E=E(\rho,\chi)$
which determines the metric. After separation of variables,
$E_l(\rho)$ satisfies a ``master equation'' \be
 \frac{D^{2}\,E}{dX^{2}}
 +\left(\frac{2}{X}+\frac{1}{X-1}-\frac{1}{X-w}\right)\,\frac{d\,E}{dX}
 -p\,(1+p)\frac{X+(D-4)\,w}{X\,(X-1)\,(X-w)}\,E=0
 \ee where $l$ is the angular momentum and \bea
  X &:=& (\rho/\rho_0)^{D-3} \non
  w_{l,D} &:=& -\frac{D-2}{(l-1)\,(l+D-2)} \non
 p_{l,D} &:=& \frac{l}{D-3}~. \eea
 There are four regular singularities in the complex plane including $\infty$,\footnote{Recall
 that $x_0$ is a ``regular singularity'' of the
differential equation $\alpha(x)\, y''+\beta(x)\, y' + \gamma(x)\,
y=0$ when after normalizing the equation by overall multiplication
such that $\alpha(x_0) \ne 0,\infty$ either $\beta(x)$ or
$\gamma(x)$ has a pole at $x_0$, but the order of the $\beta$-pole
is at most one and the order of the pole in $\gamma$ is at most
two.}
 so it is a ``Heun equation'', by definition.\footnote{For a short discussion
  of the Heun equation see appendix
A in \cite{dialogue}. More information can be found in
\cite{Heun}.}
 The singularities are at $0,\rho_0,\infty$ and at $w$, but the latter
can be eliminated by a field re-definition (re-defining $E$) and
in that sense is ``non-physical''. Since there are 3 remaining
singularities the solutions can be expressed in terms of
hyper-geometric functions. These results were first obtained in
\cite{IshibashiKodama} and then in \cite{dialogue} each using
somewhat different methods.

{\it Regulation of divergences at Post-Newtonian order}. At the
next, Post-Newtonian order divergences were identified and
regulated in \cite{dialogue2}. The divergences originate from
integration over the non-compact overlap region (which is the
neighborhood of the origin for the asymptotic zone). The
regularization can be defined through a ``cut-off and match''
method where one places a cut-off on the integral, matches with
the other zone, and then sends the cut-off away to remain with a
finite solution. This is equivalent to a regularization known as
Hadamard's partie finie, and it is closely related to the concept
of ``subtracting the self-energy''. This regularization allowed to
obtain the thermodynamic quantities at Post-Newtonian order.

In \cite{Harmark4} an alternative method was given employing a
single patch, and using the efficiency of Harmark-Obers
coordinates \cite{HO1}. There a ``first order'' approximation is
given,\footnote{The first order in the method of \cite{Harmark4}
actually includes also the second order results of the dialogue of
multipoles \cite{dialogue2}.} and I expect that the method could
be developed to a full perturbation series by successively
improving a suitably chosen initial guess. It has the advantage of
using only a single zone, and doing away with matching. However,
the method depends on the choice of initial guess, and unlike the
previous method, the differential operator to be inverted would
probably change at each order.

\presub {\bf Results}. In 5d the second order metric was obtained
in \cite{KSSW1} from which the following thermodynamics were
deduced \bea
 S &=& { \pi^2\, L^3 \over 2\, G_5}\, \teps^{3/2}\, \( 1+{\pi^2\,
 \teps \over 8} + {\pi^2\, \teps^2 \over 384} \) \non
 T &=& { 1 \over 2 \pi\, \mu}\, \teps^{3/2}\, \( 1-{5 \pi^2
 \teps \over 24} + {43 \pi^4\, \teps^2 \over 1152} \) \non
\tau\, L/M &=& {\pi^2
 \teps \over 6}  - {\pi^4\, \teps^2 \over 36}
 ~,\eea
 where the small parameter is defined through $\teps=8\, G_5\, M/(3 \pi\,
 L^2)=\rho_0^2/L^2$.

For arbitrary $D$ the best available determinations of the
thermodynamics quantities \cite{dialogue,Harmark4,dialogue2} are
  \bea
 S &=& {\Omega_{D-3} \over 4}\, \rho_0^{~D-2} \left[ 1+ {D-2
 \over D-3}\, \delta \right] \non
 T &=& {D-3 \over 4\pi\, \rho_0} \left[1-(D-2)\, \delta \right] \non
 \tau\, L/M &=& {D-3 \over 2}\,
 \delta \non
 M &=& {(D-2)\, \Omega_{D-2} \over 16 \pi\, G_N}\, \rho_0^{~D-3}\left[1
 + {1 \over 2}\, \delta  \right] \label{thermo}
 ~,\eea where \be
 \delta := \zeta(D-3)\, \({\rho_0 \over L}\)^{D-3} ~,\ee
and $\zeta(s):=\sum_{n=1}^{\infty} 1/n^s$ is Riemann's zeta
function, which appears here due to the sum over black hole
images. We note that the black hole tension (\ref{thermo})
vanishes to lowest order, namely $O(M)$ and the leading order
result can be explained by the Newtonian potential between the
black hole and its images \cite{dialogue2}.

When one considers results beyond the leading order, such as some
of the results above, their ``scheme dependence'' should be borne
in mind. ``Scheme dependence'' is used here to mean the freedom to
re-parameterize $\rho_0$ and through it the branch of small black
hole solutions: while to leading order all definitions of $\rho_0$
coincide (up to a multiplicative constant), there is no unique or
natural definition for the subleading corrections, and therefore
results should be accompanied by a specification of the scheme in
which they were obtained. In particular, the definition used in
\cite{dialogue2} is specified in \cite{dialogue}, subsection
3.3.2.

\begin{figure}[t!] \centering \noindent
\includegraphics[width=8cm]{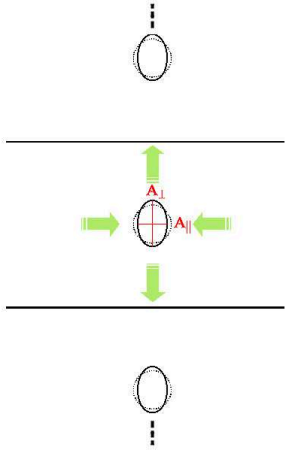}
\caption[]{The first non-spherical deformation of the horizon is a
quadruple moment, making the black hole prolate, namely elongated
along the $z$ axis. In order to measure the eccentricity we define
$A_\parallel$ the area of the equator sphere and $A_\perp$ the
area of a ``polar'' sphere.} \label{ecc-fig}
\end{figure}

\begin{figure}[t!] \centering \noindent
\includegraphics[width=8cm]{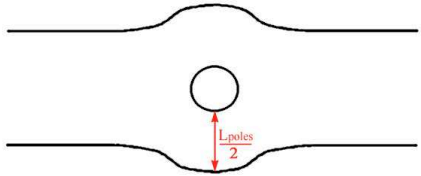}
\caption[]{The inter-polar distance is defined to be the proper
distance between the poles, going around the compact dimension.
The figure depicts the tendency of mass to ``make room for
itself'', an phenomenon termed ``the black hole Archimedes
effect'' in \cite{KPS2}.} \label{Lpoles-fig}
\end{figure}

Some nice geometric quantities can be measured as well. The
leading deviation from spherical shape is such that the BH becomes
elongated along the $z$ axis, namely prolate -- see figure
\ref{ecc-fig}. Its eccentricity can be measured by \be
 \eps:={A_\perp \over A_\parallel}-1 ={(D-3)^4\, \Gamma^2(2+{2 \over
 D-3})\, \zeta(D-1) \over 8(D-2)\, \Gamma({4 \over D-3})} \({\rho_0
 \over L}\)^{D-1} ~. \label{def-eps} \ee
 This result actually comes from beyond first order. The inter-polar
 distance (going around the compact dimension, see figure \ref{Lpoles-fig}) is given by \bea
 L_{\mbox{poles}} &=& L - 2^{D-5 \over D-3}\, \rho_0\, I_D + \dots \non
 I_D &=& 4^{1 \over D-3}\, \sqrt{\pi} {\Gamma({D-4 \over D-3})
 \over \Gamma(\half -{1 \over D-3})} ~.\eea
 where the ellipsis is order $o\( (\rho_0/L)^{D-2}\) \cdot L$.
 Since $I_5=0$ in 5d the black hole makes room for itself, exactly
compensating its size, and for that reason it was called
\emph{``the black hole Archimedes effect''} \cite{KPS2}. More
precisely, in 5d $L_{\mbox{poles}}/L=o\((\rho_0/L)^3\)$, and
numerical results indicate that indeed the next order is non-zero:
$L_{\mbox{poles}}/L=O\((\rho_0/L)^4\)$. For higher $D$,
$~0<I_D<1$, the effect is milder, and in addition, the order of
the ellipsis depends on a choice of scheme defined by eq. (B.15)
of \cite{dialogue2}.

\presub {\bf Impact on objectives}. The ability to produce results
for small black holes from two quite different methods,
perturbation theory and numerical analysis, allowed for a healthy
feedback between the two, testing and perfecting each one of them
-- see subsection \ref{numerical-issues} and especially figures
\ref{eccentricity-figrue}, \ref{BHthermodynamics-figure},
\ref{Akappa-figure}, \ref{Lpoles-figure}. While the perturbative
method does not apply directly to the phase transition region, it
provided important tests for the numerics which {\it are} capable
to study that region.

{\it Implications for the phase diagram}. Even though the phase
transition region is outside the domain of validity of the
perturbative method, the results above can be extrapolated to
provide evidence \cite{dialogue2} for the critical dimension
separating first and second order behavior, as we now proceed to
explain (of course chronologically $D^*$ was discovered first by
\cite{SorkinD*}). By extrapolating the result for the tension of
the small black hole (\ref{thermo}) one obtains an extrapolated
straight line in the phase diagram whose intersection with the
$b=0$ axis, called $\mu_X$, should be compared with $\mu_{GL}$ the
critical Gregory-Laflamme mass. While for small $D$ numerically
one finds that $\mu_X>\mu_{GL}$, for large $D$ examination of the
asymptotic expansions show that $\mu_X \ll \mu_{GL}$, which means
that for large $D$ the black hole phase is not available yet at
the onset of the GL instability and therefore the string must
decay to a different phase, which is evidence for the smooth
second order decay into the stable non-uniform string.


\section{Related work}
\label{related}

Here we mention some related work.

\begin{figure}[t!] \centering \noindent
\includegraphics[width=10cm]{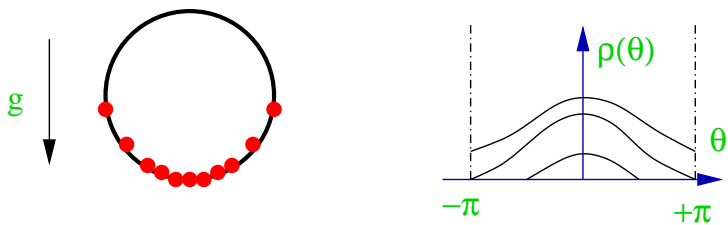}
\caption[]{In the Gross-Witten phase transition \cite{GrossWitten}
the dynamics of the eigenvalues of a unitary matrix in gauge
theory is mapped to a system of particles in a horizontal loop in
the field of gravity. In large $N$ (the number of particles) the
particle density becomes a continuous function $\rho(\theta)$. The
phase transition is between localized to non-uniform and then to
uniform distributions. Susskind (1997) identified this phenomenon
as the gauge theory dual of the black-hole black-string
transition.} \label{GW-figure}
\end{figure}

\presub {\bf Relation with gauge theory \& charged black holes}.
The gravitational phase transition is closely related to a gauge
theory phase transition, the so-called ``Gross-Witten'' transition
\cite{GrossWitten} where the eigenvalues of some unitary matrix
(originally a Wilson line over a plaquette, in a later application
\cite{AMMW}, a Wilson line around a compact dimension) change from
a clumped distribution to a uniform one as the temperature is
raised (see figure \ref{GW-figure}). This correspondence was first
noted by Susskind (1997) \cite{Susskind-GW} on a somewhat
qualitative level and was fully analyzed quantitatively by
Aharony-Marsano-Minwalla-Wiseman in \cite{AMMW}. The
correspondence with gauge theory enriches the system by adding
another parameter to the problem -- the coupling.

In \cite{HObranes} the authors showed that the phase diagram of
black holes/strings directly maps onto the phase diagram of
near-extremal black branes, and hence via holography onto the
phase diagram of the dual non-gravitational theories
(Super-Yang-Mills and ``Little String Theory''). More
specifically, they present the phases of the non-gravitational
theories and analytical results for the thermodynamics of the
localized phase and the new non-uniform phase (connected to the
``image'' of the GL point). In fact part of the early motivation
of these authors for studying black holes in Kaluza-Klein
backgrounds (see \cite{HO1}) has been precisely that, namely to
relate it to gauge theories. See also the brief review
\cite{HOrev}.

In \cite{LehnerSarbach} the GL analysis was generalized to
``point-like'' charged strings\footnote{In 5d a string may carry
either point-like or string-like charge under a Maxwell field.},
and the instability was found to persist for all charges, and
moreover $k_{GL}$ diverges as extremality is approached.

For ideas about the relation of this phase transition to the
Hagedorn phase transition in string theory see
\cite{RabinoviciBarbon0704} and references therein.

\presub {\bf Bubbles}. Taking the metric of the black string
(\ref{string-metric}) and performing a double analytic
continuation in $z,t$ (namely $t'=iz,\, z'=it$) one gets an
unstable metric called the ``bubble''. In the Euclidean setting
the analog would be to exchange $ \beta \leftrightarrow L$ thereby
inverting $\mu_\beta$, which is not that interesting. From the
Lorentzian perspective this is another phase that could be put on
the same phase diagram. It has the special property $\tau\, L =
(D-3) M$. Thus the bubble exactly saturates the Strong Energy
Condition upper bound on tension (see subsection
\ref{order-param}), which motivated \cite{EHObubbles} to study
this phase as well as combinations of several bubbles and BHs. So
far bubble phases were not seen to connect with the phases
discussed in this review, so it is conceivable that the two issues
are decoupled. See also the brief review \cite{HOrev}.

\presub {\bf Braneworld black holes}. The topic of black holes on
braneworlds is closely related to the subject of this review.
There the dimensionless parameter is the ratio of the black hole
size and the size of AdS. A comprehensive bibliography is beyond
the scope of this paper, but we shall only mention braneworld work
\cite{Wiseman0,KSSW0,KudohTanakaNakamura,Kudoh-braneworld1} which
was followed by work on compact dimensions.

\section{Summary and Open questions}
\label{summary}

{\bf Why review}.
 We start with a description of some of the major recent developments in
the field from a personal perspective, leading to the decision
that the field is ripe for this review to be written.

The Gregory-Laflamme instability and the whole black-hole
black-string transition have been attracting constant attention in
the String Theory community as well as in General Relativity since
its discovery (1993).\footnote{with a steady flux of about 20
citations/year.}
 In may 2001 the interest rose further due to the provoking ideas in \cite{HM}.
A year later I suggested in \cite{TopChange} a qualitative form
for the phase diagram which proposed the merger of the black hole
and black string phases, and rested on Morse theory and a topology
change analysis. Even today that paper includes most of what I
know about the system, but back then many felt that the evidence
was not convincing enough, and the paper was even refused for
publication by two leading journals. The non-trivial nature of
these predictions is best illustrated by the various other
possibilities that were considered in the literature, see for
example the six scenarios in \cite{HO3}, section 6.

Now,\footnote{Refers to November 2004 when the first version of
this review appeared in the archives.} after more than two years
and more than 30 papers, we have a much more detailed knowledge of
the phase diagram, culminating in the numeric results of
\cite{KudohWiseman2} which combine new numerical data for black
holes with older data for non-uniform strings into a full phase
diagram (figure \ref{numeric-phase-diag}). This full, actual-data
phase diagram exhibits the merger of the phases, vindicating
\cite{TopChange}, concluding most of the uncertainty around
\cite{HM,TopChange} as we discuss further below, and providing a
natural point in time to write a review.

\vspace{.5cm}

Below we summarize the results and open questions. The results
include two big surprises: critical dimensions and topology
change, but we remind the reader that while there was much
progress in understanding the static phase diagram, there was
practically no progress towards the deeper issues regarding time
evolution and singularities.

\subsection{Results}
\label{results}

\bi
\item Confirmation of emerging phase diagram.

The current understanding of the qualitative features of the phase
diagram\footnote{Again, we stress that this is not actual data,
but rather what seems to be the simplest possibility given the
available knowledge, some of which was obtained through research
designed to obtain these qualitative features.} is summarized in
figure \ref{PhaseDiag}, where the $D \le 13$ case originates in
\cite{TopChange}. We note that all diagrams include a merger
point.

For $D \le 13$ the end-point of decay is seen to be a black hole,
and not a stable black string which was predicted in \cite{HM}.
The evidence for that includes a numerical determination of
practically the whole branch of the non-uniform string which
emerges from the GL point and was found to have higher mass than
the critical string and thus cannot serve as an end-point for
decay \cite{Wiseman1}; the continued collapse of a time evolution
simulation \cite{CLOPPV}, coming to a stop due to grid stretching,
rather than the predicted stabilization; theoretical difficulties
in proposing a phase diagram which includes the predicted phase
and satisfy the Morse theory constraint \cite{TopChange}; and
finally the fact that such a phase was not found neither
analytically (see the attempt of \cite{DeSmet}) nor numerically.

For $D \ge 14$ the end-point of (smooth) decay does turn out to be
a stable non-uniform string (as a result of \cite{SorkinD*}), thus
partially vindicating \cite{HM}. However, I claim that in a deeper
sense, this case does not validate the arguments of \cite{HM}.
First, the arguments were independent of dimension, and second,
even for $D \ge 14$, horizon pinching, which \cite{HM} finds to be
forbidden, must happen after the non-uniform string evaporates
enough.

Thus while the claim of \cite{HM} stimulated much of the research
reported here, and while there is no doubt regarding the
calculations presented there, we find strong (actually,
overwhelming in my opinion) evidence against it. One should
therefore reconcile the derivation in \cite{HM} with the results
presented above. However, it is clear that if something went wrong
in the arguments of \cite{HM} it is rather deep and there are
lessons to be learned from it. Following the analysis of the
numerical time evolution in \cite{GLP} it seems that while indeed
pinching does not occur in finite horizon affine parameter,
consistent with the findings of \cite{HM}, it could well happen
with infinite horizon affine parameter and finite asymptotic time
(see also \cite{Marolf2005}). This possibility was indeed
discussed in \cite{HM}, and finally rejected, but with less
certainty. In \cite{Horowitz2005} this point of view seems to have
been accepted: ``It was conjectured by Kol [14] that the
nonuniform black strings should meet the squashed black holes at a
point corresponding to a static solution with a singular horizon,
and this appears to be the case [15]. ...the black string horizon
might pinch off in infinite affine parameter...  ... [the]
suggestion ... that the black string will break up into spherical
black holes might still be correct.'' Altogether, it looks like a
reasonable consensus arises where the central calculation of
\cite{HM} holds and only the arguments against pinching in
infinite time had some loop-hole,  which invalidates, however,
their bottom-line conclusion regarding the end-point of the decay
into stable non-uniform strings.

Another possibility for a flaw in the assumptions of \cite{HM}
should be mentioned, namely the reliance on the increasing area of
the event horizon. Even though the authors were careful to use a
version of the theorem valid even if there is a singularity on the
horizon, it is still quite possible that in the present context
the whole notion of the event horizon is ill-defined and/or that
the singularity completely leaves the horizon. The problem with
the definition of the event horizon is that if a singular shock
wave emerges from a naked singularity and reaches null infinity,
then we do not have the asymptotically Minkowski null infinity
which is normally used in the definition. Clearly, this issue,
like others involving the time evolution, is not well understood
yet.

 \item Critical dimensions.

The stability of cones, which have a role in the local model for
merger, exhibits a critical dimension $D^*_{\mbox{merger}}=10$
\cite{TopChange}, see subsection \ref{merger-subsection}. At a
different point in the phase diagram, the GL instability changes
from explosive (first order) to smooth (second order) for
$D>D^*_{GL}=13.5$, as was shown by Sorkin \cite{SorkinD*}. For the
canonical ensemble this happens at $D>D^*_{GL,c}=12.5$
\cite{KudohMiyamoto}.

These critical dimensions are fascinating, but it is not known
whether they have a deeper, wider meaning. The critical dimension
of the BKL theory of the approach to a space-like singularity,
$D^*=10$, is closely related. On the one hand one cannot help
wondering whether there could be a connection with the critical
dimension for super-strings, 10d, but on the other hand the
derivations and issues involved differ widely, for example, in the
phase transition no supersymmetry is involved while it is central
for the super-string.

 \item The ``merger'' topology change \cite{TopChange}.

The spaces of metrics for two different topologies, the Euclidean
black string and the black hole (in $D \ge 5$) are glued together
(subsection \ref{merger-subsection}) and a line of Ricci flat
solutions connects them, see subsection \ref{merger-subsection}
(in finite distance, see appendix \ref{finite-distance-section}).
\ei

 \presub  Additional results include \bi

 \item The use of Morse theory in GR \cite{TopChange},
  described in subsection \ref{morse-subsection}.
 \item The role of tension and the first law for this system
 \cite{HO2,KPS1}, subsection \ref{order-param}.
 \item Formulating Gravito-statics by relaxation (in 2d)
 \cite{Wiseman1}, subsection \ref{gravito-statics}.
 \item Numerical solutions: strings \cite{Wiseman1} and
 black holes \cite{KPS2,KudohWiseman1,KudohWiseman2}, subsection
 \ref{numerical-issues},
  and a dynamic time-evolution \cite{CLOPPV}, subsection \ref{time-evolution}.
 \item Developing an analytic perturbation method for small caged
 black holes -- ``a dialogue of multipoles'' \cite{dialogue,KSSW1}. See \cite{Harmark4} for a
 different analytic method. Described in subsection
 \ref{dialogue}.
\ei

\subsection{Open questions}
\label{open-questions}

\bi
 \item The deepest questions remain: are there a naked singularity and
 perhaps, a singular shock wave in the time
 evolution? -- see the discussion in subsection \ref{issues}. This
 problem could benefit from both a theoretical approach and from numerical
 analysis.

 \item Obtain solutions with discrete self-similarity (DSS) which
 were conjectured in \cite{scaling} to describe the merger
 solution locally near the pinch point (in progress).

 \item Gubser and Mitra \cite{GubserMitra} conjectured already in 2000 a connection between
 perturbative and thermodynamic instabilities of strings which is closely related to the
 issues of this review. It found confirmation in several cases,
 some recent ones being \cite{GubserD0-D2,RossWiseman,FriessGubserMitra}
 (in the last case what appeared
to be counter-examples became consistent with the conjecture after
appropriately generalizing it to allow for scalar fields). It
would be interesting to settle this conjecture.

 \item Formulate gravito-statics in more than 2d.

\item Numerically trace the phase diagram also for additional dimensions in the ranges
$10 < D \le 13$ and $D \ge 14$.

 \item Run a time evolution for $D \ge 14$ to observe the second
 order transition.
 \ei

Two items present when the first version of this review appeared
in the archives, are by now mostly resolved \bi
 \item Determine the critical dimension for torus
 compactifications.

-- torus compactifications do not reduce the critical dimension
\cite{torus}.

 \item Stability analysis in the micro-canonical ensemble. In the
canonical ensemble the Morse analysis, which identifies the
thermodynamic potential $F$ with a Morse function, allows us to
read the stability of phases off the phase diagram. It is
plausible that a generalization to other ensembles exists. For
instance, that would allow to settle the conjectured
(micro-canonical) instability of the non-uniform string (for
$D<14$).

-- Poincar\'{e}'s method (see \cite{KatzThermo} for a review)
teaches us that indeed the stability in the micro-canonical
ensemble can be determined from the form of the phase diagram just
like in the canonical ensemble, since stability changes can occur
only at turning points or at vertices where several phases meet.

 \ei

\presub {\bf  The ring}. We list some questions regarding the
physics of the sister system of rotating rings, even though they
are not part of the main subject of this review. \bi
 \item Obtain the full phase diagram: stability of phases, order of
 transition, critical points.
 \item Solutions in $D>5$?
 \ei

\vspace{0.5cm} \noindent {\bf Acknowledgements}

It is a pleasure to thank Toby Wiseman, Niels Obers and especially
an anonymous referee for reading the paper and making comments on
it, the authors of
\cite{KudohWiseman2,SorkinD*,CLOPPV,LargeD-GL,dialogue} for
permission to reproduce their figures and especially for adapting
them, in some cases, for this review; my collaborators Dan
Gorbonos, Tsvi Piran, Evgeny Sorkin and Toby Wiseman with whom I
worked on this topic; Hideaki Kudoh, Luis Lehner and Niels Obers
for discussions and correspondence; John Bahcall, Gary Gibbons,
Gary Horowitz, and Lenny Susskind for discussions and inspiration;
Roberto Emparan, Steve Gubser, Akihiro Ishibashi, Igor Klebanov,
Hermann Nicolai, Mukund Rangamani, Edward Witten for some specific
discussions; and finally Shmu'el Elitzur, Amit Giveon, and Eliezer
Rabinovici my group partners in Jerusalem for their assistance
during these two first years of mine here. I also wish to thank
the following institutions for their hospitality during the course
of the work reviewed here: Max-Planck Institute at Golm, Humboldt
University Berlin, Cambridge University, Amsterdam University, the
Perimeter Institute, MIT and Harvard University.

BK is supported in part by The Israel Science Foundation (grant no
228/02) and by the Binational Science Foundation BSF-2002160.

\newpage
\appendix
\section{Formulae for action manipulation}
\label{action-section}

Here we collect some formulae which are useful for manipulation of
actions.

The Ricci scalar in the presence of a general fibration \bea
 ds^2 &=& ds^2_X + \sum_i\, e^{2\, F_i}\, ds^2_{Yi} \Rightarrow \non
 R &=& R_X + \sum_i\, [ e^{-2\, F_i}\, R_{Yi} - 2\, d_i\,
 \wttriangle (F_i) - d_i (\del F_i)^2] -\sum_{i,j}\, d_i\, d_j\,
 (\del F_i \cdot \del F_j) \label{Rfibration} \eea
 where the fibration fields depend only on the $x$ coordinates $F_i=F_i(x)$,
 $R_X, ~R_{Yi}$ are the Ricci scalars of the spaces $X,\, Y_i$,
 $d_i$ are the dimensions $\mbox{dim}(Y_i)$,
 and the Laplacian ($\wttriangle$) and grad-squared
($\del \cdot \del$) are evaluated in the X space.

The Ricci scalar of a conformally transformed metric (see for
example \cite{Wald-ConformalRicci}) \bea
 \widetilde{ds}^2  &=& e^{2\, w}\, ds^2 \Rightarrow \non
 \widetilde{R} &=& e^{-2\, w}\, [ R -2\, (\hat{d}-1)\, \triangle w-(\hat{d}-1)(\hat{d}-2)\, (\del w)^2]
 \label{Rconformal} \eea
 where $\hat{d}$ is the dimension of the space and the Laplacian
and grad-squared are evaluated in the non-tilded metric.

\section{Topology change is a finite distance away}
\label{finite-distance-section}

Scaling down a smoothed cone gives a family of metrics which
approaches the singular cone, as discussed is subsection
\ref{merger-subsection} (see figure \ref{ScaledCones}). In this
appendix we wish to show that the singular cone is at a finite
distance in moduli space, just like the conifold.

Let us denote a smooth cone metric with some specific length scale
(of the smoothed tip) by \be
  \widetilde{ds}^2 = d\trho^2 + e^{2\, a(\trho)}\, d\Omega^2_{\IS^m} +
 e^{2\, b(\trho)}\, d\Omega^2_{\IS^n} ~.\ee
The family of rescaled cone metrics is defined by \be
 ds^2= e^{-2 \sigma}\, \widetilde{ds}^2 ~,\ee
and we wish to compute the distance in moduli space from
$\sigma=0$ to $\sigma=\infty$ ($\sigma=\infty$ is the non-smooth
cone).

The metric on  moduli space can be found by adding an auxiliary
coordinate $t$, making $\sigma$ $t$-dependent, $\sigma=\sigma(t)$,
and evaluating its kinetic term. Since one wants to hold the
asymptotic form of the cone fixed it is useful to introduce
$\rho=e^{-\sig}\, \trho$. Then the rescaled cone metric is
$ds^2=d\rho^2+e^{2\,(a(\rho e^\sigma)-\sigma)}\, d\Omega_m +
e^{2\,(b(\rho e^\sigma)-\sigma)}\, d\Omega_n$ , where
$d\Omega_m=d\Omega^2_{\IS^m}$. Adding $t$-dependence and
substituting back to $\trho$ we get \be
 ds^2=dt^2 + e^{-2 \sigma}\, \left[ (d\trho - \dot{\sigma}\, \trho\, dt)^2
  +  e^{2\, a}\, d\Omega_m +
 e^{2\, b}\, d\Omega_n \right] \ee

Using (\ref{Rfibration}) the Ricci scalar is \bea
 R &=& -m(m+1)\, (\del(a-\sig))^2 - 2\, m\, \triangle(a-\sig) \non
   & & -n(n+1)\, (\del(b-\sig))^2 - 2\, n\, \triangle(b-\sig) \non
   & & - 2\, m\, n\, \del(a-\sig)\, \del(b-\sig) ~,\eea
where the differential operators are in the $(\trho,t)$ plane. All
terms which vanish when $\dot{\sigma}=0$ must cancel since the
resolved cone is Ricci-flat, and one finds \bea
 R &=& -2 \left[ (m\, a' + n\, b') \trho - (D-1) \right]\,
 \ddot{\sigma} \\
 & +& \dot{\sigma}^2 \left[ -\trho^2(m(m-1)\,e^{-2\, a} +n(n-1)\,e^{-2\,
 b}) + 2\, \trho\,(D-1)(m\, a' + n\, b') - D(D-1) \right] \nonumber \eea
A useful test for this expression is that it must vanish when
substituting the singular cone metric rather than the resolved
cone.

The resulting bulk action after multiplying by $\sqrt{g}=e^{m\, a
+ n\, b-D\, \sig}$ and eliminating the second derivatives
 using integration by parts is \bea
\label{bulk_sigma}
 S &=& \int dt\, \dot{\sigma}^2\, e^{-D\, \sigma} \int d\trho\,  e^{m\, a + n\,
 b}\, \\
 &\cdot& \left[-\trho^2(m(m-1)\, e^{-2\, a} + n(n-1)\, e^{-2\,
 b})-2\, \trho (m\, a' + n\, b') + D(D-1) \right] \nonumber \eea

This action should be regularized by comparison with the singular
cone. This can be done by introducing a large-distance cut-off
$\rho= \Lambda$ and therefore $\trho_\Lambda=\Lambda\, e^\sig$ and
subtracting for the cone. We are interested in large $\sig$ and
hence $\trho_\Lambda \to \infty$. In this limit the integral seems
to be dominated by the large $\trho$ behavior of the integrand:
the large $\trho$ behavior of $a,\, b$ compared to the singular
cone is given by the linearized perturbations
(\ref{linearized-exponents}) \be
 \delta a,\, \delta b \sim \rho^s, ~~ \mbox{Re}(s)=-(D-2)/2 \ee
 The large $\trho_\Lambda$ behavior of the kinetic term in (\ref{bulk_sigma}) is
 \be e^{-D\, \sigma}\,  \int^{\Lambda\, e^\sig} d\trho\, \trho^{D-1}\,
 \trho^{-(D-2)} \sim  e^{-D\,
\sigma} (\Lambda^2 \, e^{2\, \sigma})  ~~.\ee
 Hence the metric on moduli space in the large $\sigma$ limit is
 \be
 ds^2_{\cal M} = e^{-(D-2)\, \sig}\, d\sig^2 \ee and the distance
is finite  \be
  \int^{+\infty} \exp(-{D-2 \over 2}\, \sigma)\, d\sigma < \infty \ee
 for $D > 2$, namely always, since we were only
interested in $D \ge 5$.

\newpage
\bibliographystyle{JHEP}
\bibliography{revbib}

\providecommand{\href}[2]{#2}\begingroup\raggedright\begin{thebibliography}{10}

\bibitem{CosmoBilliard}
T.~Damour, M.~Henneaux, and H.~Nicolai, {\it Cosmological billiards},  {\em
  Class. Quant. Grav.} {\bf 20} (2003) R145--R200,
  [\href{http://xxx.lanl.gov/abs/hep-th/0212256}{{\tt hep-th/0212256}}].

\bibitem{Mazur}
P.~O. Mazur, {\it Black hole uniqueness theorems},
  \href{http://xxx.lanl.gov/abs/hep-th/0101012}{{\tt hep-th/0101012}}. p.10-11.

\bibitem{Emparan-with-hair}
R.~Emparan, {\it Rotating circular strings, and infinite non-uniqueness of
  black rings},  {\em JHEP} {\bf 03} (2004) 064,
  [\href{http://xxx.lanl.gov/abs/hep-th/0402149}{{\tt hep-th/0402149}}].

\bibitem{BenaWarnerRing}
I.~Bena and N.~P. Warner, {\it One ring to rule them all ... and in the
  darkness bind them?},  \href{http://xxx.lanl.gov/abs/hep-th/0408106}{{\tt
  hep-th/0408106}}.

\bibitem{EEMRdipoleRing}
H.~Elvang, R.~Emparan, D.~Mateos, and H.~S. Reall, {\it Supersymmetric black
  rings and three-charge supertubes},  {\em Phys. Rev.} {\bf D71} (2005)
  024033, [\href{http://xxx.lanl.gov/abs/hep-th/0408120}{{\tt
  hep-th/0408120}}].

\bibitem{uniqueness}
B.~Kol, {\it Speculative generalization of black hole uniqueness to higher
  dimensions},  \href{http://xxx.lanl.gov/abs/hep-th/0208056}{{\tt
  hep-th/0208056}}.

\bibitem{HelfgottOzYanay}
C.~Helfgott, Y.~Oz, and Y.~Yanay, {\it On the topology of black hole event
  horizons in higher dimensions},
  \href{http://xxx.lanl.gov/abs/hep-th/0509013}{{\tt hep-th/0509013}}.

\bibitem{GallowaySchoen}
G.~J. Galloway and R.~Schoen, {\it A generalization of {H}awking's black hole
  topology theorem to higher dimensions},
  \href{http://xxx.lanl.gov/abs/gr-qc/0509107}{{\tt gr-qc/0509107}}.

\bibitem{HawkingStewart}
S.~W. Hawking and J.~M. Stewart, {\it Naked and thunderbolt singularities in
  black hole evaporation},  {\em Nucl. Phys.} {\bf B400} (1993) 393--415,
  [\href{http://xxx.lanl.gov/abs/hep-th/9207105}{{\tt hep-th/9207105}}].

\bibitem{MyersPerry}
R.~C. Myers and M.~J. Perry, {\it Black holes in higher dimensional
  space-times},  {\em Ann. Phys.} {\bf 172} (1986) 304.

\bibitem{EmparanReall-ring}
R.~Emparan and H.~S. Reall, {\it A rotating black ring in five dimensions},
  {\em Phys. Rev. Lett.} {\bf 88} (2002) 101101,
  [\href{http://xxx.lanl.gov/abs/hep-th/0110260}{{\tt hep-th/0110260}}].

\bibitem{EEMRsusyRing}
H.~Elvang, R.~Emparan, D.~Mateos, and H.~S. Reall, {\it A supersymmetric black
  ring},  {\em Phys. Rev. Lett.} {\bf 93} (2004) 211302,
  [\href{http://xxx.lanl.gov/abs/hep-th/0407065}{{\tt hep-th/0407065}}].

\bibitem{ArcioniLozano}
G.~Arcioni and E.~Lozano-Tellechea, {\it Stability and thermodynamics of black
  rings},  \href{http://xxx.lanl.gov/abs/hep-th/0502121}{{\tt hep-th/0502121}}.

\bibitem{ConcentricRing1}
J.~P. Gauntlett and J.~B. Gutowski, {\it Concentric black rings},  {\em Phys.
  Rev.} {\bf D71} (2005) 025013,
  [\href{http://xxx.lanl.gov/abs/hep-th/0408010}{{\tt hep-th/0408010}}].

\bibitem{ConcentricRing2}
J.~P. Gauntlett and J.~B. Gutowski, {\it General concentric black rings},  {\em
  Phys. Rev.} {\bf D71} (2005) 045002,
  [\href{http://xxx.lanl.gov/abs/hep-th/0408122}{{\tt hep-th/0408122}}].

\bibitem{ElvangEmparanFigueras}
H.~Elvang, R.~Emparan, and P.~Figueras, {\it Non-supersymmetric black rings as
  thermally excited supertubes},  {\em JHEP} {\bf 02} (2005) 031,
  [\href{http://xxx.lanl.gov/abs/hep-th/0412130}{{\tt hep-th/0412130}}].

\bibitem{LargeD-GL}
B.~Kol and E.~Sorkin, {\it On black-brane instability in an arbitrary
  dimension},  {\em Class. Quant. Grav.} {\bf 21} (2004) 4793--4804,
  [\href{http://xxx.lanl.gov/abs/gr-qc/0407058}{{\tt gr-qc/0407058}}].

\bibitem{torus}
B.~Kol and E.~Sorkin. to appear.

\bibitem{Myers-4dcaged}
R.~C. Myers, {\it Higher dimensional black holes in compactified space- times},
   {\em Phys. Rev.} {\bf D35} (1987) 455.

\bibitem{Korotkin_Nicolai}
D.~Korotkin and H.~Nicolai, {\it A periodic analog of the {Schwarzschild}
  solution},  \href{http://xxx.lanl.gov/abs/gr-qc/9403029}{{\tt
  gr-qc/9403029}}.

\bibitem{FrolovFrolov}
A.~V. Frolov and V.~P. Frolov, {\it Black holes in a compactified spacetime},
  {\em Phys. Rev.} {\bf D67} (2003) 124025,
  [\href{http://xxx.lanl.gov/abs/hep-th/0302085}{{\tt hep-th/0302085}}].

\bibitem{IMSY}
N.~Itzhaki, J.~M. Maldacena, J.~Sonnenschein, and S.~Yankielowicz, {\it
  Supergravity and the large n limit of theories with sixteen supercharges},
  {\em Phys. Rev.} {\bf D58} (1998) 046004,
  [\href{http://xxx.lanl.gov/abs/hep-th/9802042}{{\tt hep-th/9802042}}].

\bibitem{ABKR}
S.~A. Abel, J.~L.~F. Barbon, I.~I. Kogan, and E.~Rabinovici, {\it String
  thermodynamics in {D}-brane backgrounds},  {\em JHEP} {\bf 04} (1999) 015,
  [\href{http://xxx.lanl.gov/abs/hep-th/9902058}{{\tt hep-th/9902058}}].

\bibitem{CHR}
A.~Chamblin, S.~W. Hawking, and H.~S. Reall, {\it Brane-world black holes},
  {\em Phys. Rev.} {\bf D61} (2000) 065007,
  [\href{http://xxx.lanl.gov/abs/hep-th/9909205}{{\tt hep-th/9909205}}].

\bibitem{EHM}
R.~Emparan, G.~T. Horowitz, and R.~C. Myers, {\it Exact description of black
  holes on branes},  {\em JHEP} {\bf 01} (2000) 007,
  [\href{http://xxx.lanl.gov/abs/hep-th/9911043}{{\tt hep-th/9911043}}].

\bibitem{Tangherlini}
F.~R. Tangherlini, {\it Schwarzschild field in $n$ dimensions and the
  dimensionality of space problem},  {\em Nuovo Cim.} {\bf 27} (1963) 636.

\bibitem{Harmark4}
T.~Harmark, {\it Small black holes on cylinders},  {\em Phys. Rev.} {\bf D69}
  (2004) 104015, [\href{http://xxx.lanl.gov/abs/hep-th/0310259}{{\tt
  hep-th/0310259}}].

\bibitem{dialogue}
D.~Gorbonos and B.~Kol, {\it A dialogue of multipoles: Matched asymptotic
  expansion for caged black holes},  {\em JHEP} {\bf 06} (2004) 053,
  [\href{http://xxx.lanl.gov/abs/hep-th/0406002}{{\tt hep-th/0406002}}].

\bibitem{KSSW1}
D.~Karasik, C.~Sahabandu, P.~Suranyi, and L.~C.~R. Wijewardhana, {\it Analytic
  approximation to 5 dimensional black holes with one compact dimension},  {\em
  Phys. Rev.} {\bf D71} (2005) 024024,
  [\href{http://xxx.lanl.gov/abs/hep-th/0410078}{{\tt hep-th/0410078}}].

\bibitem{dialogue2}
D.~Gorbonos and B.~Kol, {\it Matched asymptotic expansion for caged black
  holes: Regularization of the post-{Newtonian} order},  {\em Class. Quant.
  Grav.} {\bf 22} (2005) 3935--3959,
  [\href{http://xxx.lanl.gov/abs/hep-th/0505009}{{\tt hep-th/0505009}}].

\bibitem{KPS1}
B.~Kol, E.~Sorkin, and T.~Piran, {\it Caged black holes: Black holes in
  compactified spacetimes. {I}: Theory},  {\em Phys. Rev.} {\bf D69} (2004)
  064031, [\href{http://xxx.lanl.gov/abs/hep-th/0309190}{{\tt
  hep-th/0309190}}].

\bibitem{KPS2}
E.~Sorkin, B.~Kol, and T.~Piran, {\it Caged black holes: Black holes in
  compactified spacetimes. {II}: 5d numerical implementation},  {\em Phys.
  Rev.} {\bf D69} (2004) 064032,
  [\href{http://xxx.lanl.gov/abs/hep-th/0310096}{{\tt hep-th/0310096}}].

\bibitem{KudohWiseman1}
H.~Kudoh and T.~Wiseman, {\it Properties of {Kaluza-Klein} black holes},  {\em
  Prog. Theor. Phys.} {\bf 111} (2004) 475--507,
  [\href{http://xxx.lanl.gov/abs/hep-th/0310104}{{\tt hep-th/0310104}}].

\bibitem{KudohWiseman2}
H.~Kudoh and T.~Wiseman, {\it Connecting black holes and black strings},  {\em
  Phys. Rev. Lett.} {\bf 94} (2005) 161102,
  [\href{http://xxx.lanl.gov/abs/hep-th/0409111}{{\tt hep-th/0409111}}].

\bibitem{GL1}
R.~Gregory and R.~Laflamme, {\it Black strings and p-branes are unstable},
  {\em Phys. Rev. Lett.} {\bf 70} (1993) 2837--2840,
  [\href{http://xxx.lanl.gov/abs/hep-th/9301052}{{\tt hep-th/9301052}}].

\bibitem{GubserMitra}
S.~S. Gubser and I.~Mitra, {\it Instability of charged black holes in anti-de
  {Sitter} space},  \href{http://xxx.lanl.gov/abs/hep-th/0009126}{{\tt
  hep-th/0009126}}.

\bibitem{GubserMitra-detail}
S.~S. Gubser and I.~Mitra, {\it The evolution of unstable black holes in
  anti-de {Sitter} space},  {\em JHEP} {\bf 08} (2001) 018,
  [\href{http://xxx.lanl.gov/abs/hep-th/0011127}{{\tt hep-th/0011127}}].

\bibitem{Reall-onGM}
H.~S. Reall, {\it Classical and thermodynamic stability of black branes},  {\em
  Phys. Rev.} {\bf D64} (2001) 044005,
  [\href{http://xxx.lanl.gov/abs/hep-th/0104071}{{\tt hep-th/0104071}}].

\bibitem{GLcharged}
R.~Gregory and R.~Laflamme, {\it The instability of charged black strings and
  p-branes},  {\em Nucl. Phys.} {\bf B428} (1994) 399--434,
  [\href{http://xxx.lanl.gov/abs/hep-th/9404071}{{\tt hep-th/9404071}}].

\bibitem{GPY}
D.~J. Gross, M.~J. Perry, and L.~G. Yaffe, {\it Instability of flat space at
  finite temperature},  {\em Phys. Rev.} {\bf D25} (1982) 330--355.

\bibitem{GL0}
R.~Gregory and R.~Laflamme, {\it Hypercylindrical black holes},  {\em Phys.
  Rev.} {\bf D37} (1988) 305.

\bibitem{SorkinD*}
E.~Sorkin, {\it A critical dimension in the black-string phase transition},
  {\em Phys. Rev. Lett.} {\bf 93} (2004) 031601,
  [\href{http://xxx.lanl.gov/abs/hep-th/0402216}{{\tt hep-th/0402216}}].

\bibitem{HM}
G.~T. Horowitz and K.~Maeda, {\it Fate of the black string instability},  {\em
  Phys. Rev. Lett.} {\bf 87} (2001) 131301,
  [\href{http://xxx.lanl.gov/abs/hep-th/0105111}{{\tt hep-th/0105111}}].

\bibitem{TopChange}
B.~Kol, {\it Topology change in general relativity and the black-hole
  black-string transition},  \href{http://xxx.lanl.gov/abs/hep-th/0206220}{{\tt
  hep-th/0206220}}.

\bibitem{explosive}
B.~Kol, {\it Explosive black hole fission and fusion in large extra
  dimensions},  \href{http://xxx.lanl.gov/abs/hep-ph/0207037}{{\tt
  hep-ph/0207037}}.

\bibitem{HO2}
T.~Harmark and N.~A. Obers, {\it New phase diagram for black holes and strings
  on cylinders},  {\em Class. Quant. Grav.} {\bf 21} (2004) 1709,
  [\href{http://xxx.lanl.gov/abs/hep-th/0309116}{{\tt hep-th/0309116}}].

\bibitem{TraschenFoxTension}
J.~H. Traschen and D.~Fox, {\it Tension perturbations of black brane
  spacetimes},  {\em Class. Quant. Grav.} {\bf 21} (2004) 289--306,
  [\href{http://xxx.lanl.gov/abs/gr-qc/0103106}{{\tt gr-qc/0103106}}].

\bibitem{TownsendZamaklarTension}
P.~K. Townsend and M.~Zamaklar, {\it The first law of black brane mechanics},
  {\em Class. Quant. Grav.} {\bf 18} (2001) 5269--5286,
  [\href{http://xxx.lanl.gov/abs/hep-th/0107228}{{\tt hep-th/0107228}}].

\bibitem{TraschenTension>0}
J.~H. Traschen, {\it A positivity theorem for gravitational tension in brane
  spacetimes},  {\em Class. Quant. Grav.} {\bf 21} (2004) 1343--1350,
  [\href{http://xxx.lanl.gov/abs/hep-th/0308173}{{\tt hep-th/0308173}}].

\bibitem{STI-Tension>0}
T.~Shiromizu, D.~Ida, and S.~Tomizawa, {\it Kinematical bound in asymptotically
  translationally invariant spacetimes},  {\em Phys. Rev.} {\bf D69} (2004)
  027503, [\href{http://xxx.lanl.gov/abs/gr-qc/0309061}{{\tt gr-qc/0309061}}].

\bibitem{EHObubbles}
H.~Elvang, T.~Harmark, and N.~A. Obers, {\it Sequences of bubbles and holes:
  New phases of {Kaluza-Klein} black holes},  {\em JHEP} {\bf 01} (2005) 003,
  [\href{http://xxx.lanl.gov/abs/hep-th/0407050}{{\tt hep-th/0407050}}].

\bibitem{HOtension}
T.~Harmark and N.~A. Obers, {\it General definition of gravitational tension},
  {\em JHEP} {\bf 05} (2004) 043,
  [\href{http://xxx.lanl.gov/abs/hep-th/0403103}{{\tt hep-th/0403103}}].

\bibitem{GibbonsHawking-action}
G.~W. Gibbons and S.~W. Hawking, {\it Action integrals and partition functions
  in quantum gravity},  {\em Phys. Rev.} {\bf D15} (1977) 2752--2756.

\bibitem{York-action}
J.~York, James~W., {\it Role of conformal three geometry in the dynamics of
  gravitation},  {\em Phys. Rev. Lett.} {\bf 28} (1972) 1082--1085.

\bibitem{LL}
L.~Landau, {\em Statsitical Physics}.
\newblock Pergamon, 1993.

\bibitem{Gubser}
S.~S. Gubser, {\it On non-uniform black branes},  {\em Class. Quant. Grav.}
  {\bf 19} (2002) 4825--4844,
  [\href{http://xxx.lanl.gov/abs/hep-th/0110193}{{\tt hep-th/0110193}}].

\bibitem{Wiseman1}
T.~Wiseman, {\it Static axisymmetric vacuum solutions and non-uniform black
  strings},  {\em Class. Quant. Grav.} {\bf 20} (2003) 1137--1176,
  [\href{http://xxx.lanl.gov/abs/hep-th/0209051}{{\tt hep-th/0209051}}].

\bibitem{KudohMiyamoto}
H.~Kudoh and U.~Miyamoto, {\it On non-uniform smeared black branes},
  \href{http://xxx.lanl.gov/abs/hep-th/0506019}{{\tt hep-th/0506019}}.

\bibitem{Park}
M.-I. Park, {\it The final state of black strings and p-branes, and the
  {Gregory-Laflamme} instability},  {\em Class. Quant. Grav.} {\bf 22} (2005)
  2607--2614, [\href{http://xxx.lanl.gov/abs/hep-th/0405045}{{\tt
  hep-th/0405045}}].

\bibitem{KatzThermo}
J.~Katz, {\it Thermodynamics and self-gravitating systems},  {\em Found. Phys.}
  {\bf 33} (2003) 223--269,
  [\href{http://xxx.lanl.gov/abs/astro-ph/0212295}{{\tt astro-ph/0212295}}].

\bibitem{HawkingPage}
S.~W. Hawking and D.~N. Page, {\it Thermodynamics of black holes in anti-de
  {S}itter space},  {\em Commun. Math. Phys.} {\bf 87} (1983) 577.

\bibitem{KolWiseman}
B.~Kol and T.~Wiseman, {\it Evidence that highly non-uniform black strings have
  a conical waist},  {\em Class. Quant. Grav.} {\bf 20} (2003) 3493--3504,
  [\href{http://xxx.lanl.gov/abs/hep-th/0304070}{{\tt hep-th/0304070}}].

\bibitem{scaling}
B.~Kol, {\it Choptuik scaling and the merger transition},
  \href{http://xxx.lanl.gov/abs/hep-th/0502033}{{\tt hep-th/0502033}}.

\bibitem{KlebanovRangamaniWitten-private}
I.~Klebanov, M.~Rangamani, and E.~Witten. private communication.

\bibitem{LL35}
L.~D. Landau and E.~M. Lifshitz, {\em Quantum mechanics}.
\newblock Pergamon, 1977.
\newblock \S35.

\bibitem{HO3}
T.~Harmark and N.~A. Obers, {\it Phase structure of black holes and strings on
  cylinders},  {\em Nucl. Phys.} {\bf B684} (2004) 183--208,
  [\href{http://xxx.lanl.gov/abs/hep-th/0309230}{{\tt hep-th/0309230}}].

\bibitem{HO1}
T.~Harmark and N.~A. Obers, {\it Black holes on cylinders},  {\em JHEP} {\bf
  05} (2002) 032, [\href{http://xxx.lanl.gov/abs/hep-th/0204047}{{\tt
  hep-th/0204047}}].

\bibitem{KudohTanakaNakamura}
H.~Kudoh, T.~Tanaka, and T.~Nakamura, {\it Small localized black holes in
  braneworld: Formulation and numerical method},  {\em Phys. Rev.} {\bf D68}
  (2003) 024035, [\href{http://xxx.lanl.gov/abs/gr-qc/0301089}{{\tt
  gr-qc/0301089}}].

\bibitem{Wiseman2}
T.~Wiseman, {\it From black strings to black holes},  {\em Class. Quant. Grav.}
  {\bf 20} (2003) 1177--1186,
  [\href{http://xxx.lanl.gov/abs/hep-th/0211028}{{\tt hep-th/0211028}}].

\bibitem{KolMartinez-unpublished}
B.~Kol and M.~Rodriguez-Martinez. unpublished.

\bibitem{Wiseman0}
T.~Wiseman, {\it Relativistic stars in {Randall-Sundrum} gravity},  {\em Phys.
  Rev.} {\bf D65} (2002) 124007,
  [\href{http://xxx.lanl.gov/abs/hep-th/0111057}{{\tt hep-th/0111057}}].

\bibitem{CLOPPV}
M.~W. Choptuik {\em et.~al.}, {\it Towards the final fate of an unstable black
  string},  {\em Phys. Rev.} {\bf D68} (2003) 044001,
  [\href{http://xxx.lanl.gov/abs/gr-qc/0304085}{{\tt gr-qc/0304085}}].

\bibitem{GLP}
D.~Garfinkle, L.~Lehner, and F.~Pretorius, {\it A numerical examination of an
  evolving black string horizon},  {\em Phys. Rev.} {\bf D71} (2005) 064009,
  [\href{http://xxx.lanl.gov/abs/gr-qc/0412014}{{\tt gr-qc/0412014}}].

\bibitem{ReggeWheeler}
T.~Regge and J.~A. Wheeler, {\it Stability of a {Schwarzschild} singularity},
  {\em Phys. Rev.} {\bf 108} (1957) 1063--1069.

\bibitem{Heun}
A.~Ronveaux, {\em Heun's differential equations}.
\newblock University press, New York, 1995.

\bibitem{IshibashiKodama}
H.~Kodama and A.~Ishibashi, {\it A master equation for gravitational
  perturbations of maximally symmetric black holes in higher dimensions},  {\em
  Prog. Theor. Phys.} {\bf 110} (2003) 701--722,
  [\href{http://xxx.lanl.gov/abs/hep-th/0305147}{{\tt hep-th/0305147}}].

\bibitem{GrossWitten}
D.~J. Gross and E.~Witten, {\it Possible third order phase transition in the
  large {N} lattice gauge theory},  {\em Phys. Rev.} {\bf D21} (1980) 446--453.

\bibitem{AMMW}
O.~Aharony, J.~Marsano, S.~Minwalla, and T.~Wiseman, {\it Black hole - black
  string phase transitions in thermal 1+1 dimensional supersymmetric
  {Yang-Mills} theory on a circle},  {\em Class. Quant. Grav.} {\bf 21} (2004)
  5169--5192, [\href{http://xxx.lanl.gov/abs/hep-th/0406210}{{\tt
  hep-th/0406210}}].

\bibitem{Susskind-GW}
L.~Susskind, {\it Matrix theory black holes and the {Gross Witten} transition},
   \href{http://xxx.lanl.gov/abs/hep-th/9805115}{{\tt hep-th/9805115}}.

\bibitem{HObranes}
T.~Harmark and N.~A. Obers, {\it New phases of near-extremal branes on a
  circle},  {\em JHEP} {\bf 09} (2004) 022,
  [\href{http://xxx.lanl.gov/abs/hep-th/0407094}{{\tt hep-th/0407094}}].

\bibitem{HOrev}
T.~Harmark and N.~A. Obers, {\it Phases of {Kaluza-Klein} black holes: A brief
  review},  \href{http://xxx.lanl.gov/abs/hep-th/0503020}{{\tt
  hep-th/0503020}}.

\bibitem{LehnerSarbach}
O.~Sarbach and L.~Lehner, {\it Critical bubbles and implications for critical
  black strings},  {\em Phys. Rev.} {\bf D71} (2005) 026002,
  [\href{http://xxx.lanl.gov/abs/hep-th/0407265}{{\tt hep-th/0407265}}].

\bibitem{RabinoviciBarbon0704}
J.~L.~F. Barbon and E.~Rabinovici, {\it Touring the {Hagedorn} ridge},
  \href{http://xxx.lanl.gov/abs/hep-th/0407236}{{\tt hep-th/0407236}}.

\bibitem{KSSW0}
D.~Karasik, C.~Sahabandu, P.~Suranyi, and L.~C.~R. Wijewardhana, {\it Small
  (1-{TeV}) black holes in {Randall-Sundrum I} scenario},  {\em Phys. Rev.}
  {\bf D69} (2004) 064022, [\href{http://xxx.lanl.gov/abs/gr-qc/0309076}{{\tt
  gr-qc/0309076}}].

\bibitem{Kudoh-braneworld1}
H.~Kudoh, {\it Thermodynamical properties of small localized black hole},  {\em
  Prog. Theor. Phys.} {\bf 110} (2004) 1059--1069,
  [\href{http://xxx.lanl.gov/abs/hep-th/0306067}{{\tt hep-th/0306067}}].

\bibitem{DeSmet}
P.-J. De~Smet, {\it Black holes on cylinders are not algebraically special},
  {\em Class. Quant. Grav.} {\bf 19} (2002) 4877--4896,
  [\href{http://xxx.lanl.gov/abs/hep-th/0206106}{{\tt hep-th/0206106}}].

\bibitem{Marolf2005}
D.~Marolf, {\it On the fate of black string instabilities: An observation},
  {\em Phys. Rev.} {\bf D71} (2005) 127504,
  [\href{http://xxx.lanl.gov/abs/hep-th/0504045}{{\tt hep-th/0504045}}].

\bibitem{Horowitz2005}
G.~T. Horowitz, {\it Higher dimensional generalizations of the {Kerr} black
  hole},  \href{http://xxx.lanl.gov/abs/gr-qc/0507080}{{\tt gr-qc/0507080}}.

\bibitem{GubserD0-D2}
S.~S. Gubser, {\it The {Gregory-Laflamme} instability for the {D2-D0} bound
  state},  {\em JHEP} {\bf 02} (2005) 040,
  [\href{http://xxx.lanl.gov/abs/hep-th/0411257}{{\tt hep-th/0411257}}].

\bibitem{RossWiseman}
S.~F. Ross and T.~Wiseman, {\it Smeared {D0} charge and the {Gubser-Mitra}
  conjecture},  {\em Class. Quant. Grav.} {\bf 22} (2005) 2933--2946,
  [\href{http://xxx.lanl.gov/abs/hep-th/0503152}{{\tt hep-th/0503152}}].

\bibitem{FriessGubserMitra}
J.~J. Friess, S.~S. Gubser, and I.~Mitra, {\it Counter-examples to the
  correlated stability conjecture},
  \href{http://xxx.lanl.gov/abs/hep-th/0508220}{{\tt hep-th/0508220}}.

\bibitem{Wald-ConformalRicci}
R.~M. Wald, {\em General Relativity}.
\newblock The University of Chicago Press, 1984.
\newblock appendix D.

\end{thebibliography}\endgroup

\end{document}